\documentclass[12pt, reqno, a4paper]{amsart} 
\usepackage{latexsym}
\usepackage[english]{babel}
\usepackage{fancyhdr}
\usepackage[mathscr]{eucal}
\usepackage{amsfonts}
\usepackage{amssymb}
\usepackage{bbm}
\usepackage{latexsym}
\usepackage{color}
\usepackage[dvipsnames]{xcolor}
 \usepackage{comment} 

\usepackage{amsmath,amsthm}

\theoremstyle{plain}
\newtheorem{theorem}{Theorem}[section]
\newtheorem{remark}[theorem]{Remark}
\newtheorem{lemma}[theorem]{Lemma}
\newtheorem{corollary}[theorem]{Corollary}
\newtheorem{proposition}[theorem]{Proposition}

\theoremstyle{definition}
\newtheorem{definition}[theorem]{Definition}

\newcommand {\absleq} {{\leq_{|\, \cdot\, |}\, }}

\def\tK {{\tilde{K}}}

\def\Bg{{\mathcal B}}
\def\Cg{{\mathcal C}}
\def\Dg{{\mathcal D}}
\def\Fg{{\mathcal F}}
\def\Gg{{\mathcal G}}
\def\Hg {{\mathcal H}}
\def\Kg {{\mathcal K}}
\def\Lg {{\mathcal L}}
\def\Mg{{\mathcal M}}
\def\tMg{\tilde{\mathcal M}}
\def\Ng{{\mathcal N}}

\def\Og{{\mathcal O}}
\def\Qg {{\mathcal Q}}

\def\Rg {{\mathcal R}}

\def\Wg {{\mathcal W}}
\def\Xg {{\mathcal X}}
\def\Yg {{\mathcal Y}}

\def\tG {{\tilde{G}}}
\def\tI {{\tilde{I}}}
\def\tK {{\tilde{K}}}

\def\tM {{\tilde{M}}}

\def\tT {{\tilde{T}}}

\numberwithin{equation}{section}
\def\HL {{L^2(\R^2)}}

\def\te{{\tilde{e}}}

\def\tm{{\tilde{\mu}}}
\def\tOm{{\tilde{\Omega}}}

\def\loc{{\rm loc}}

\def\ben{\begin{enumerate}}
\def\een{\end{enumerate}}
\def\bgdf{\begin{definition}}
\def\eddf{\end{definition}}
\def\bglm{\begin{lemma}}
\def\edlm{\end{lemma}}
\def\bgpf{\begin{proof}}
\def\edpf{\end{proof}}
\def\bgth{\begin{theorem}}
\def\edth{\end{theorem}}
\def\bgcor{\begin{corollary}}
\def\edcor{\end{corollary}}
\def\bgprop{\begin{proposition}}
\def\edprop{\end{proposition}}
\def\bgrm{\begin{remark}}
\def\edrm{\end{remark}}
\def\bgal{\begin{align}}
\def\edal{\end{align}}
\def\bgala{\begin{align*}}
\def\edala{\end{align*}}
\def\lbeq(#1){\label{eqn:#1}}
\def\refeq(#1){{\rm (\ref{eqn:#1})}}
\def\refeqs(#1,#2){{\rm (\ref{eqn:#1}) and (\ref{eqn:#2})}}
\def\refeqss(#1,#2,#3){{\rm (\ref{eqn:#1}),\ (\ref{eqn:#2}) and (\ref{eqn:#3})}}
\def\refeqsss(#1,#2,#3,#4){{\rm (\ref{eqn:#1}),\ (\ref{eqn:#2}),\ 
(\ref{eqn:#3}) and (\ref{eqn:#4})}}
\def\lbth(#1){\label{th:#1}}
\def\refth(#1){{\rm Theorem \ref{th:#1}}}
\def\refths(#1,#2){{\rm Theorems \ref{th:#1} and \ref{th:#2}}}
\def\refthb(#1){{\bf Theorem \ref{th:#1}}}
\def\lblm(#1){\label{lm:#1}}
\def\reflm(#1){{\rm Lemma \ref{lm:#1}}}
\def\reflms(#1,#2){{\rm Lemmas \ref{lm:#1} and \ref{lm:#2}}}
\def\reflmss(#1,#2,#3){{\rm Lemmas \ref{lm:#1}, \ref{lm:#2} and \ref{lm:#3}}}
\def\reflmsss(#1,#2,#3,#4){{\rm Lemmas \ref{lm:#1},\, \ref{lm:#2},\, \ref{lm:#3} and \ref{lm:#4}}}
\def\reflmb(#1){{\bf Lemma \ref{lm:#1}}}
\def\lbprop(#1){\label{prp:#1}}
\def\refpropb(#1){{\bf Proposition \ref{prp:#1}}}
\def\refprop(#1){{\rm Proposition \ref{prp:#1}}}
\def\refprops(#1,#2,#3,#4){{\rm Propositions \ref{prp:#1},\, \ref{prp:#2},
\, \ref{prp:#3} \, and \ref{prp:#4}}}
\def\refpropb(#1){{\bf Proposition \ref{prp:#1}.}}
\def\lbcor(#1){\label{cor:#1}}
\def\refcor(#1){{\rm Corollary \ref{cor:#1}}}
\def\refcors(#1,#2){{\rm Corollaries \ref{cor:#1} and \ref{cor:#2}}}
\def\lbrm(#1){\label{rm:#1}}
\def\refrm(#1){{\rm Remark \ref{rm:#1}}}
\def\lbass(#1){\label{ass:#1}}
\def\refass(#1){{\rm Assumption \ref{ass:#1}}}
\def\lbdf(#1){\label{df:#1}}
\def\refdf(#1){{\rm Definition \ref{df:#1}}}
\def\refdfs(#1,#2){{\rm Definitions \ref{def:#1} and \ref{def:#2}}}
\def\lbsec(#1){\label{s:#1}}
\def\refsec(#1){{\rm \S\ref{s:#1}}}
\def\lbsubsec(#1){\label{ss:#1}}
\def\refsubsec(#1){{\rm \S\ref{ss:#1}}}

\newcommand{\lam}{\lambda}

\def\Bb{{\bf B}}

\def\ph{{\varphi}}

\def\bqn{\begin{equation}}
\def\eqn{\end{equation}}
\def\C{{\mathbb C}}
\def\N{{\mathbb N}}

 \def\Cb{{\overline{\mathbb C}}}

\def\R{{\mathbb R}}
\def\a{\alpha}
\def\b{\beta}
\def\c{\gamma}
\def\Ga{\Gamma}
\def\d{\delta}

\def\Hs {{\mathscr H}}
\def\Ls {{\mathscr L}}

\def\Sg{{\mathcal S}}

\def\p{\psi}

\def\ep{\varepsilon}

\def\th{\theta}
\def\k{\kappa}
\def\m{\mu}

\def\r{\rho}
\def\s{\sigma}
\def\t{\tau}
\def\w{\omega}
\def\W{\Omega}
\def\la{\langle}
\def\ra{\rangle}

\def\lap{\Delta}
\def\ax{{\la x \ra}}
\def\ay{{\la y \ra}}

\def\pa{{\partial}} 

\def\tv{\tilde{v}}
\def\te{\tilde{e}}

\def\br{\begin{array}}
\def\er{\end{array}}

\def\Ker{\rm Ker\,}

\def\tW{\tilde{\W}}

\begin{document}
\allowdisplaybreaks

\title[$L^p$-boundedness of wave operators for $\lap^2 + V$ on $\R^2$]
{The $L^p$-boundedness of wave operators for 4-th order  
Schr\"odinger operators on $\R^2$, I. \\
Regular case} 
\footnote{{\it 2020 Mathematics Subject. Classification}  
Primary 47A40, Secondary 81Q10. \\
Keywords: $L^p$-spaces, Wave operators, 
Fourth order Schr\"odinger operators in $\R^2$}

\author[A.~Galtbayar]{Artbazar Galtbayar}
\address{Center of Mathematics for Applications
and Department of Applied Mathematics \\
National University of Mongolia \\
University Street 3, Ulaanbaatar (Mongolia). \\ 
\footnote{Supported by project P2024-4835 funded by National University of Mongolia}}
\email{galtbayar@num.edu.mn}
\author[K.~Yajima]{Kenji Yajima}
\address{Department of Mathematics \\ Gakushuin University 
\\ 1-5-1 Mejiro \\ Toshima-ku \\ Tokyo 171-8588 (Japan). \\ 
\footnote{Supported by JSPS grant in aid for scientific research No. 23K03187}}
\email{kenji.yajima@gakushuin.ac.jp}

\begin{abstract} 
We prove that wave operators of scattering theory for fourth
order Schr\"odinger operators $H = \lap^2 + V (x)$ on $\R^2$ 
with real potentials $V(x)$ 
such that $\ax^3 V(x) \in  L^{\frac43}(\R^2)$ and 
$\la x \ra^{10+\ep} V(x) \in L^1 (\R^2)$ for an $\ep>0$, 
$\ax=(1+|x|^2)^{\frac12}$,  are bounded in $L^p (\R^2)$  for all $1<p<\infty$ 
if  $H$ is regular at zero in the sense that there are no non-trivial solutions 
to $(\lap^2 + V(x))u(x)=0$ such that $\ax^{-1} u(x) \in L^\infty(\R^2)$  and if 
positive eigenvalues are absent from $H$. This reduces $L^p$-mapping 
properties of functions $f(H)$ of $H$ to those of Fourier multipliers $f(\lap^2)$. 
\end{abstract}

\maketitle 

\section{Introduction, Theorems}

We consider fourth order Schr\"odinger operators in $\R^2$:
\[
H = \lap^2 + V (x), \quad  \lap = \pa^2/\pa x_1^2 + \pa^2/\pa x_2^2, \quad 
x = (x_1 , x_2 ) \in \R^2
\]
with potentials $V (x)$ which are real valued and satisfy for a $q > 1$ that 
\bqn \lbeq(1-1)
N_q (V )(x) : =\left(\int_{|x-y|<1} |V(y)|^q dy \right)^\frac{1}{q}\in  L^1 (\R^2 ) .
\eqn 
Condition \refeq(1-1)  implies  $N_q (V ) \in L^r (\R^2 )$ for all $1\leq r \leq \infty$;  
multiplication 
with $|V(x)|^\frac12$ is relatively compact with respect to $-\lap$ (\cite{IS}), hence to
$\lap^2$ in the Hilbert space $\Hs \colon = L^2 (\R^2 )$;  the quadratic form
\[
q(u, v) : = \int_{\R^2}(\lap u(x) \overline{\lap v(x)} + V (x)u(x)\overline{v(x)})dx,  
\quad u, v \in H^2 (\R^2 )
\]
is closed and bounded from below in $\Hs$ and it defines the self-adjoint operator
$H$,  $H^2 (\R^2)$ being the Sobolev space of order two. Moreover, the
argument of Ionescu and Schlag (\cite{IS}) shows that  
\begin{itemize}
\item[$\bullet$] the spectrum of $H$ consists of the absolutely continuous (AC
for short) part $[0, \infty)$ and the bounded set of eigenvalues which
are discrete in $\R \setminus \{0\}$ and accumulate possibly to zero;
\item[$\bullet$] the wave operators $W_\pm$ defined by the strong limits in $\Hs$:
\[
W_\pm  = \lim_{t\to \pm \infty} e^{itH} e^{-itH_0} ,
\quad  H_0 = \lap^2 
\]
exist and are complete in the sense that ${\rm Image}\, W_\pm  = \Hs_{ac} (H)$,
$\Hs_{ac} (H)$ being the AC subspace of  $\Hs$ for H. Let  $P_{ac} (H)$ be the 
projection to $\Hs_{ac} (H)$.
Then, 
\bqn \lbeq(inter) 
f(H)P_{ac}(H) = W_\pm f(H_0) W_{\pm}^\ast 
\eqn 
for Borel functions $f$ on $\R$ (the intertwining property of $W_\pm$).
\end{itemize}

The wave operators $W_\pm$ are partial isometries on $\Hs$ and, 
a fortiori, are bounded in $\HL$. In this paper we are concerned with 
whether they are bounded in $L^p(\R^2)$ for some 
$1\leq p \leq \infty$ other than $p=2$. Note that, 
since $H$ and $H_0$ are real operators, $W_{+}$ 
and $W_{-}$ are complex conjugate of each other 
and they are simultaneously bounded or unbounded in 
$L^p(\R^2)$. 

If $W_\pm$ are bounded in $L^p(\R^2)$ for $p$ in a subset $I$ of 
$[1,\infty)$, then it follows from the intertwining property \refeq(inter) 
that 
\bqn \lbeq(inter-estimate)
\|f(H)P_{ac}(H)\|_{\Bb(L^{q'},L^p)}
\leq C_{pq} \|f(H_0)\|_{\Bb(L^{q'},L^p)}
\eqn  
for $p\in I$ and $q'\in I^\ast=\{q/(q-1)\colon q\in I\}$, 
which reduces some mapping properties between 
$L^p$-spaces of $f(H)P_{ac}(H)$, 
the AC-part of $f(H)$, to those of $f(H_0)$ which is a 
Fourier multiplier. 
We write $\ax=(1+|x|^2)^{1/2}$ for $x \in \R^d$, $d\in \N$. 

Because of this property the $L^p$-boundedness of wave operators 
has attracted interest of many authors and various results have been obtained 
for ordinary Schr\"odinger operators $H=-\lap +V(x)$ as well as for $H=\lap^2+ V(x)$   
on $\R^d$, $d\geq 1$,  under various assumptions. The results depend on $d$ and the spectral 
property of $H$ at the threshold. We first list some of the results for $-\lap+ V(x)$: 
When $d=1$, $W_\pm$ are bounded in $L^p$  for 
$1<p<\infty$ if $V$ satisfies $\ax V\in L^1(\R)$  
(\cite{Weder-2, DF, Weder-1,GY}). For $d\geq 2$, the range of 
$p$ for which $W_\pm$ are bounded in $L^p(\R^d)$ depends on the structure of 
the space of the zero energy resonances:  
\[
\Ng_\infty^{(2)}(H)=\{\ph \colon |\ph(x)|\leq C \ax^{2-d}, \ (-\lap + V(x))\ph(x)=0\}. 
\]
If $\Ng^{(2)}_\infty(H)=\{0\}$, then 
$W_\pm$ are bounded in $L^p(\R^d)$  for $1<p<\infty$ 
if $d=2$ (\cite{Y-2dim,JY}) and for all $1\leq p \leq \infty$ 
if $d \geq 3$ (\cite{BS,Y-3dim}). 
If $\Ng_\infty^{(2)}(H)\not=\{0\}$, the range of $p$ 
for which $W_\pm$ are bounded in $L^p$ shrinks and it is determined by the 
maximal rate of decay $\c_c$ of $\ph \in \Ng_{\infty}^{(2)}(H)\setminus \{0\}$: 
\[
\c_c= \sup_{\ph \in \Ng_\infty^{(2)}\setminus \{0\}} 
\{\c \colon \ax^{\c} |\ph(x)| \in L^\infty(\R^d)\}
\]
irrespective of potentials provided that they decay fast enough as 
$|x|\to \infty$. It takes too much space to recall the results for this case 
and, for more information, we refer to the introduction of 
\cite{Ya-2dim-new,Ya-4dim-new} and the references therein, 
\cite{Y-odd-sing,FY,GG,GG-4,KY,EGG,Y-3d-sing} among others. 

For wave operators for $H=\lap^2 + V(x)$ the investigation 
started only recently and the following results have been obtained 
under suitable conditions on the decay at infinity and the smoothness 
of $V(x)$ in addition to the absence of positive eigenvalues of $H$. 
When $d=1$, $W_\pm$ are bounded in $L^p(\R^1)$ 
for $1<p<\infty$ but not for $p=1$ and $p=\infty$; they are bounded 
from the Hardy space $H^1$ to $L^1$ and from $L^1$ to $L^1_{w}$
(\cite{MWY}); if $d=3$ and 
$\Ng_\infty(H)\colon =\{u \in L^\infty(\R^3): (\lap^2+ V)u=0\}=0$ 
then $W_\pm$ are bounded in $L^p(\R^3)$ for $1<p<\infty$ (\cite{GG-16}); 
if $d \geq 5$ and 
$\Ng_\infty(H)
=\cap_{\ep>0}\{u \in \ax^{-\frac{d}{2}+2+\ep}L^2(\R^d): (\lap^2 + V)u=0\}=0$, 
then they are bounded in $L^p(\R^d)$ for all $1\leq p \leq \infty$ 
(\cite{EG,EG-1,EGG-2}). 
There have been no results for $\lap^2 +V(x)$ in $\R^2$ so far,  
however, an extensive study has been made by \cite{LSY} on dispersive estimates 
for time dependent equations $i\pa_t u= (\lap^2+V) u$ on $\R^2$, 
from which we borrow some results. 

In two dimensions, as in higher dimensions, the $L^p$-boundedness of  $W_\pm$ 
depends on the nature of the space of zero energy resonances
\bqn \lbeq(reso-space)
\Ng_\infty (H) \colon = \{\ph \in \la x \ra L^\infty (\R^2) : (\lap^2 + V (x))\ph(x) = 0\}.  
\eqn 
Note that zero energy eigenfunctions of $H$ are included in $\Ng_\infty(H)$.

\bgdf  We say that the operator $H$ is {\it regular} (at zero)
if $\Ng_\infty (H) = \{0\}$ and is {\it singular} otherwise.
\eddf 

This is the first of the set of two papers on the $L^p$-boundedness 
of wave operators for $H= \lap^2 + V(x)$ in $\R^2$ and we prove here 
that $W_\pm$ 
are bounded in $L^p (\R^2 )$ for all $1 < p <\infty$ if $H$ is regular at zero. 
We plan to study the singular case in a forthcoming paper.

We use the following terminology and notation:

\bgdf  We say an operator is a {\it good operator} (GOP in short)  
if it is bounded in  $L^p (\R^2)$ for all  $1 <p < \infty$.
\eddf 

For Banach
spaces $\Xg$ and  $\Yg$,  $\Bb(\Xg , \Yg)$ is the Banach space of bounded operators
from $\Xg$ to $\Yg$ and  $\Bb(\Xg ) = \Bb(\Xg , \Xg )$;
the Fourier transform of $u$ is defined by 
\[
\hat{u}(\xi) = (\Fg u)(\xi) = (2\pi)^{-1} \int_{\R^2}e^{-ix\xi} u(x)dx.
\]
For Borel functions $f (\lam)$ on $[0, \infty)$,  
$f (|D|)$ is the Fourier multiplier defined by  $f (|\xi|)$:
\[
f (|D|)u(x) = \frac1{2\pi}\int_{\R^2}e^{ix\xi} f (|\xi|)\hat{u}(\xi)d\xi.
\]
For  $a >0$,  $\chi_{\leq a}(\lam)$ and  $\chi_{\geq a}(\lam)$ are smooth functions 
of $\lam\in [0, \infty)$ such that
\[
\chi_{\leq a}(\lam) = 
\left\{ \br{ll} 1,    &  \lam \leq a, \\  0 , & \lam \geq 2a \er \right. 
\quad 
\chi_{\geq a}(\lam) = 
\left\{ \br{ll} 1,    &  \lam \geq a, \\  0 , & \lam \leq a/2. 
\er \right. 
\]
Operators $\chi_{\geq a} (|D|)$ and  $\chi_{\leq a}(|D|)$ are GOPs;  they are cut-off 
functions to 
the high and the low energy parts respectively and $W_\pm \chi_{\geq a}(|D|) $ 
and  $W_\pm \chi_{\leq a}(|D|)$ are called 
the high and the low energy parts of $W_\pm$. 

\bgdf We say a function $\m(\lam)$ on $(0,\infty)$ is a {\it good multiplier } 
(GMU for short) if it is of class $C^2$ and satisfies 
$|\pa^j \m(\lam)|\leq C \lam^{-j}$ for $0<\lam<\infty$ and   
$0\leq j \leq 2$. 
\eddf 
\noindent 
If $\m(\lam)$ is a GMU, then $\m(|D|)$ is a GOP by 
H\"ormander's theorem (\cite{Hor}).

We have following two theorems for high energy parts $W_\pm \chi_{\geq a}(|D|)$. 
They hold irrespective of whether $H$ is regular or singular and under much weaker 
assumptions on the decay at infinity of $V$ than for the  low energy parts. For $1 \leq  p \leq \infty$,   
$L^p_{{\rm loc,u}}$ is the uniform localization of $L^p(\R^2)$:
\[
L^p_{\rm loc,u} \colon = \{u :  \|u\|_{L^p_{\rm loc,u}} : = \sup_{x\in \R^2} N_p (u) (x)< \infty\}.
\]

\bgth \lbth(small) Suppose $N_q(V) \in L^1(\R^2)$ for a $q>1$ 
and $\la \log |x|\ra^{2} V \in L^1(\R^2)$. Let $a>0$.   
Then, there exists $c_0>0$ such that 
$W_{\pm} \chi_{\geq{a}}(|D|)$ are GOPs whenever 
$\|V\|_{L^{q}_{\rm loc,u}}+ \|\la \log |x|\ra^{2} V\|_{L^1}\leq c_0$. 
\edth 

In the next theorem we assume $H$ has no positive eigenvalues. 
We remark that $H$ may have 
positive eigenvalues even when $V \in C_0^\infty(\R^2)$ 
(\cite{FSW,MWY}).   This is a sharp contrast  to the case of ordinary 
Schr\"odinger operators $-\lap + V$ which have no positive eigenvalues 
for a large class of short-range potentials (\cite{IJ,KT}). 
When $V$ is small as in \refth(small), $H$ has no positive eigenvalues. 
Note $N_q(V) \in L^1(\R^2)$ for any $1<q\leq 4/3$ if $\ax^3 V(x) \in  L^{\frac43}(\R^2)$. 

\bgth \lbth(high) Suppose that  
$\ax^3 V(x) \in  L^{\frac43}(\R^2)$ and 
that $H$ has no positive eigenvalues. Then, 
$W_{\pm}\chi_{\geq a}(|D|)$ is GOP for any $a>0$.
\edth 

For low energy parts we assume faster decay of $V(x)$ as $|x|\to \infty$.  

\bgth \lbth(regular-case)  Let 
$\ax^3 V(x) \in  L^{\frac43}(\R^2)$ and
$\la x \ra^{10+\ep} V (x) \in L^1 (\R^2 )$  for an  $\ep > 0$. 
Suppose that $H$ is regular at zero and $H$ has no positive
eigenvalues. Then, $W_\pm$ are GOP.
\edth

\bgcor 
If conditions of \refth(regular-case) are satisfied and if $p, q \in 
(1, \infty)$ then, there exists a constant $C_{pq} \geq 1$ such that
\[
C_{pq}^{-1} \|f (H_0)\|_{\Bb(L^q ,L^p)} \leq \|f (H)P_{ac} (H)\|_{\Bb(L^q ,L^p )} 
\leq  C_{pq} \|f (H_0)\|_{\Bb(L^q ,L^p )}
\]
for any Borel function $f$ of  $\lam \in [0, \infty)$.
\edcor

The rest of the paper is devoted to proving Theorems.
We assume throughout the paper that $H$ has no positive eigenvalues.
Since complex conjugation $\Cg u(x) = \overline{u(x)}$ changes the direction of
time,  $\Cg e^{-itH} \Cg = e^{itH}$ and, hence, $W_+ =\Cg W_{-}\Cg$,  we shall prove Theorems 
only for $W_{-}$.  We use the following notation and conventions: 
We denote $L^2(\R^2)$ by $\Hs$, $\|u\|=\|u\|_2$ and $(u,v)$ is 
the inner product of $\Hs$; the notation 
\[
(u,v)= \int_{\R^2} u(x)\overline{v(x)}dx \ \ \mbox{and} \ \  
\la u,v\ra= \int_{\R^2} \overline{u(x)}v(x)dx
\]
will be used whenever the integral makes sense, e.g. for 
$u \in \Sg(\R^2)$ and $v \in \Sg'(\R^2)$. 
Various (unimportant) constants are denoted by $C$ and it may 
differ from one place to the other.
For  $j = 0, 1, \cdots$,  $f^{(j)}(\lam) = (d^j f /d \lam^j )(\lam)$;
\[
\Dg_\ast : = \Fg^{-1} C_0^\infty(\R^2 \setminus \{0\}); 
\]
for functions $F(x)$ of $x \in \R^2$, $M_F$ is the multiplication with $F(x)$:
\[
M_F u(x) = F(x)u(x). 
\] 
For the potential $V(x)$ of $H$ we set 
\[
U(x)={\rm sign}\, V(x) \colon= 
\left\{ \br{ll} 1,  & \ V(x)\geq 0 ,  
\\ -1,    &  \ V(x) < 0,  
\er 
\right.  \quad  v(x) = |V(x)|^{1/2}
\]
and $v$ is reserved for this notation.

Resolvents of $H_0$ and $H$ are denoted by $R_0(z)=(H_0-z)^{-1}$ and $R(z)= (H-z)^{-1}$ 
respectively and for $\lam>0$ 
\[
R_0^\pm(\lam)= \lim_{\ep \to 0}(H_0-\lam \mp i\ep)^{-1} \ \mbox{and} \ 
R^\pm(\lam)= \lim_{\ep \to 0}(H-\lam \mp i\ep)  
\] 
are the boundary values at $\lam>0$. 
It is well known (cf. \cite{IS})  that under the assumption \refeq(1-1)  
\bqn  \lbeq(Mg+)
\Mg^{\pm}(\lam)\colon = M_U + M_v R_0^{\pm}(\lam^4) M_v , \quad \lam>0,  
\eqn 
is bounded in $\Hs$ and that the bounded inverses
$\Mg^{\pm}(\lam)^{-1}$ exist if and only if $\lam^4$ is {\it not} an eigenvalue of $H$, 
in which case $R^\pm(\lam^4)$ are given by the symmetric resolvent equations  
\[
R^{\pm}(\lam^4) = R_0^\pm(\lam^4) - R_0^\pm(\lam^4) M_v \Mg^\pm(\lam)^{-1}
M_v R_0^\pm(\lam^4) .
\]
Moreover, if we define 
\bqn \lbeq(Qgv)
\Qg_v(\lam) = M_v \Mg^{+}(\lam)^{-1} M_v,  \quad \lam>0, 
\eqn 
then the wave operator $W_{-}$ may be represented via the integral: 
\bqn \lbeq(sta)
W_{-}u(x) = u(x) - \int_0^\infty (R_0^{+}(\lam^4)\Qg_v(\lam) 
\Pi(\lam) u)(x) \lam^3 d\lam, \ u \in \Dg_\ast, 
\eqn 
where $\Pi(\lam)=(2/\pi{i})(R_0^{+}(\lam^4)-R_0^{+}(\lam^4))$ is the spectral projection 
for $H_0$. The representation formula \refeq(sta) is the starting point of our analysis.  
The integral in \refeq(sta) is absolutely convergent for all $x \in \R^2$.  Replacing $\Qg_v(\lam) $ 
by an operator valued function $T(\lam)$, we define 
\bqn \lbeq(Omega-T-int)
\tOm(T(\lam)) u(x)\colon = \int_0^\infty (R_0^{+}(\lam^4)T(\lam)\Pi(\lam)u)(x) \lam^3 d\lam.
\eqn 
When $T(\lam)$ is independent of $\lam$, we write $\W(T)$ for $\tOm(T(\lam))$. 

\bgdf  \lbdf(GPR)
We say that $T(\lam)$ is a {\it good producer}  (GPR for short) 
if $\tOm(T(\lam))$ is a GOP.  It is a {\it GPR for low or high energy}  
if  $\tOm_{\leq a}(T(\lam))\colon = \tOm(T(\lam))\chi_{\leq a}(|D|)$ 
or $\tOm_{\geq a}(T(\lam))\colon =\tOm(T(\lam))\chi_{\geq a}(|D|)$ 
is a GOP for a $a>0$ respectively. 
\eddf 
\noindent 

Integral operators $T$ and their integral kernels $T(x,y)$ will be often identified and we 
shall often say integral operator $T(x,y)$; we denote 
\[
\Lg^1= L^1(\R^2 \times \R^2) ,
\] 
which will also denote the space of integral operators with integral kernels $T(x,y) \in \Lg^1$.

 In \S 2, we collect  various  mostly well known results in scattering theory, 
prove some results on the free resolvent $R_0(z)$ and show
that $\lam^2 \k(\lam) T(\lam)$ 
is a GPR for low energy if $T(\lam) \in L^1_{\loc}([0,\infty),\Lg^1)$ and  $\k$ is a GMU, 
which will play an important role in \S 5.  We prove 
\refth(small) and \refth(high) in \S 3. 
In \S 4, we study the asymptotic expansion (up to a finite order) of $\Qg_v(\lam)$ as 
$\lam \to +0$  and show in \S 5 that most terms 
in the expansion are 
GPRs by using  results of \S 2.   \refth(regular-case) will be 
proved in \S 6 by showing via integration by parts that the remaining 
operator in the expansion, which is an oscillatory integral operator, is 
also a GPR for low energy.  Estimates on some integral operators 
and Fourier multipliers are postponed to the appendix 
to avoid interrupting the main line of the argument.  
In what follows $a \absleq b$ means $|a|\leq |b|$.

\section{Preliminaries}

\subsection{Integral representation of free resolvent} 
Let $\C^{++}=\{z\in \C \colon \Re z>0, \Im z>0\}$, 
$\Cb^{++}$ its closure and, for $z \in \C^{++}$, 
$R_0(z^4)= (H_0 -z^4)^{-1}$ and $G_0(z^2)= (-\lap - z^2)^{-1}$. 
A little algebra shows that
\bqn 
R_0(z^4)= \frac{1}{2z^2}(G_0(z^2)- G_0(-z^2)). \lbeq(resol)
\eqn
Let $\Rg(z,x)$ and $\Gg(z,x)$ be 
the convolution kernels of $R_0(z^4)$ and 
$G_0(z^2)$ respectively: For $u \in \Dg_\ast$
\begin{align*}
& R_0(z^4)u(x) = \int_{\R^2}\Rg(z,x-y) u(y) dy, \\   
& G_0(z^2)u(x) = \int_{\R^2}\Gg(z,x-y) u(y) dy.
\end{align*}
It follows from \refeq(resol) that  
\bqn 
\Rg(z,x) = \frac1{2z^2}(\Gg(z,x) - \Gg(iz, x)).  \lbeq(res-ker)
\eqn 
Let $\R_{\pm}= \{\lam\in \R \colon \pm \lam>0\}$. 
When $z \in \C^{++}$ approaches $i\R^{+}$, $z^2$ does 
$\R^{-}$ and $z^4$  the lower edge $\R^{+}$ of $\C\setminus [0,\infty)$. 
It is well-known that $\Gg(z,x)$ for $z \in \C^{+}$ may be expressed  
via the Hankel function of the first kind (p.469 of \cite{CH} or DLMF 10.9.10): 
\bqn 
\Gg(z,x)= \frac{i}{4}H^{(1)}_0 (z|x|), \quad 
H^{(1)}_0(z)= \frac2{i\pi}\int_0^\infty e^{iz\cosh{t}}dt.
\lbeq(Green) 
\eqn  
Note that $iz \in \C^{+}$ if $z\in \C^{++}$. 
Change of variable $t \to \cosh t -1$ yields 
\begin{equation}  
H_0^{(1)}(z)= 
\frac{2e^{iz}}{i\pi}
\int_0^\infty e^{iz{t}}t^{-\frac12}
(2+t)^{-\frac12}{dt}. \lbeq(int-10)
\end{equation}

The following representation of  $\Gg(z,x)$ is more convenient for us. 
\bglm 
For $z\in \C^{+}$ 
\bqn 
\Gg(z,x)= \frac{1}{2\pi} {e^{iz|x|}}
\int_0^\infty t^{-\frac12} e^{-{t}}\left({t}
-2iz|x|\right)^{-\frac12}dt, \lbeq(Gg) 
\eqn  
where the branch of square root is such that  
$\left({t} -2iz|x|\right)^{-\frac12}>0$ when $z\in i\R_{+}$. 
For every $x\not=0$, $\C^{+} \ni z \mapsto \Gg(z,x)$ may be analytically  
continued through $\R_{\pm}$ to $\C \setminus {i}(-\infty, 0]$. 
The representation formula \refeq(Gg) holds with $iz$ in place 
of $z$ when $\Re z\geq 0$. 
\edlm 
\bgpf  
Let $z \in \C^{++}$  and 
$\Ga_z= \{t \in \C\setminus\{0\} \colon 0<\arg z + \arg t <\pi\}$. 
Then, the function $f(t) \colon = e^{iz{t}}t^{-\frac12} (2+t)^{-\frac12}$ 
on $\R_{+}$ may be analytically extended to  
$\Ga_z$ and, on closed sub-sectors of $\Ga_z$, 
it decays exponentially as $|t|\to \infty $. 
Thus, the contour of integration of \refeq(int-10) may be rotated to 
$\Cg_z= \{t=e^{i\th}r/|z| \colon r>0\}$, $\th=\pi/2-\arg{z}$. 
On $\Cg_z$ we have  
\[
\frac{dt}{t^{\frac12} (2+t)^{\frac12}} = \frac{dr} {r^{\frac12}(r-2iz)^{\frac12}} 
\]
where the branch of square root is such that $(r-2iz)^{-\frac12}\in \C^{++}$ 
and 
\bqn \lbeq(int-1a)
H_0^{(1)}(z)
= \frac{2e^{iz}}{i\pi} \int_0^\infty  
e^{-r}r^{-\frac12} 
\left({r}-2iz\right)^{-\frac12}{dr}, \quad z \in \C^{++}
\eqn 
(cf.  \cite{CH}, p. 525).  It is clear that \refeq(int-1a) can be analytically extended to $\C^{+}$ 
and we obtain \refeq(Gg)  for $z \in \C \setminus i(-\infty, 0]$. 
\edpf   
\subsection{Some estimates on free resolvent} 
\bglm \lblm(Green) Let $j=0,1, \dots$.  We have following estimates:
\ben 
\item[\textrm{(1)}]  Let $z \in \Cb^{+}\setminus \{0\}$. Then,   
for $|z||x|\geq 1/2$   
\bqn \lbeq(large)
|\pa_z^j \Gg({z},x)|
\leq C e^{-(\Im z)|x|}|z|^{-\frac12}|x|^{j-\frac12} 
\leq C e^{-(\Im z)|x|}|x|^{j}   
\eqn 
and for $|z||x|\leq 1/2$  
\bqn \lbeq(small-lam) 
|\pa_z^j \Gg(z,x)| \leq C \left\{\br{ll} |z|^{-j}, & \quad j \geq 1, \\
\la \log (|z||x|)\ra, & \quad j=0. \er \right.
\eqn 
\item[\textrm{(2)}] For $\Re{z}\geq 0$ with $z\not =0$,  $\Gg(iz, x)$ 
satisfies estimates \refeq(large) and \refeq(small-lam) 
with $iz$ in place of $z$.  
\een 
 (The series expansion of $H^{(1)}_0(z)$ 
given below provides more detailed information on 
$\pa_z^j \Gg(z,x)$ for small $|z||x|\leq 1/2$.) 
\edlm 
\bgpf Statement (2) follows from (1) immediately and we prove 
the latter only. We write \refeq(Gg) in the form 
$\Gg(z,x)=\frac{1}{2\pi} {e^{iz|x|}}N(z|x|)$ where 
the definition of $N(z)$ should be obvious. 
Then, Leibniz' formula and the chain rule imply 
$\pa_z^j \Gg({z},x)= \sum_{k=0}^j C_{jk} e^{iz|x|}N^{(k)}(z|x|) |x|^j$.  
For $z \in \Cb^{+}$ we have   
$|t-2iz|\geq (2/\sqrt{5})(t +|z|)$ and  
\bqn   
|N^{(k)}(z)| 
\leq C_k \int_0^\infty e^{-t}t^{-\frac12}(t+|z|)^{-k-\frac12}dt, 
\quad k=0, \dots, j. 
\lbeq(F-estimate)
\eqn 
Hence, $|N^{(k)}(z)|\leq C_k |z|^{-k-\frac12}$ and  \refeq(large) follows. 
When $|z|\leq 1$,  the right of 
\refeq(F-estimate) is bounded by a constant times 
\[
\int_0^{|z|}t^{-\frac12}|z|^{-k-\frac12}dt + 
\int_{|z|}^1 t^{-k-1}dt + C  
\leq C_k \left\{\br{l} |z|^{-k}, \quad k \geq 1, \\ \la \log |z|\ra, \quad k=0. 
\er \right. 
\]
This implies \refeq(small-lam)
\edpf 

We omit the proof of the following lemma which is evident from \refeq(resol) 
and \reflm(Green). 

\bglm \lblm(R-prop)   The convolution kernel $\Rg(z,x)$ of $(H_0-z^4)^{-1}$, 
$z \in \C^{++}$ can be extended to 
a smooth function of $z$ on the punched 
closure $\Cb^{++}\setminus \{0\} $ where 
it satisfies  
\bqn \lbeq(R-prop) 
|\pa_z ^j\Rg(z,x) |\leq C  \left\{ 
\br{ll} |z|^{-2-j }\la \log (z|x|) \ra,  & |z||x| \leq 1, \\
 |z|^{-2-\frac12} |x|^{j-\frac12} \leq |z|^{-2}|x|^j, &  |z||x|\geq 1. \er \right. 
\eqn 
\edlm   

We need some more notation;  for a closable operator $T$ on $\Hs$, 
$[T]$ is its closure. When $[T]$ is bounded, we abuse notation and say 
$T$ is bounded and denote $[T]$ simply by $T$.   
The space of Hilbert-Schmidt operators from Hilbert space $\Kg_1$ to $\Kg_2$  
is denoted by $\Hg_2(\Kg_1, \Kg_2)$; $\Hg_2= \Hg_2(\Hs, \Hs)$. 
For $T\in \Hg_2$, $T (x,y) \in L^2(\R^2 \times \R^2)$ and $M_v T M_v\in \Lg^1$. 

\bglm \lblm(pre-final)  Let $1<q\leq \infty$ and $a>0$. 

\vspace{0.2cm}
\noindent 
{\rm (1)} If $V\in L^1(\R^2) \cap L^q_{\loc, u}(\R^2)$,  then   
$M_v R_0(z^4)M_v$ is an $\Hg_2$-valued holomorphic 
function of $z\in\C^{++}$, it can be continuously extended to 
the punched closure $\Cb^{++}\setminus\{0\}$ and is bounded 
for $|z|\geq a$ for any $a>0$.

\vspace{0.2cm}
\noindent 
{\rm (2)} For $j=0,1, \dots$ and on $\Cb^{++}\setminus\{0\}$, 
$M_v R_0(z^4)M_v$  satisfies  
\bqn  \lbeq(pre-final) 
\|\pa_{z}^j M_v R_0(z^4)M_v \|_{\Hg_2}\leq 
C |z|^{-2}(\|V\|_{L^q_{u,loc}(\R^2)}+ \|\ax^{(2j-1)_{+}} V\|_{1}).
\eqn 
\edlm 
\bgpf  We write $\|\pa_{z}^j M_v R_0(z^4)M_v \|_{\Hg_2}^2$ as the sum 
\[
\left(
	\int_{|z||x-y|\geq 1}+ 
	\int_{|z||x-y|\leq 1}
	\right) 
	|V(x)||\Rg^{(j)}(z, x-y)|^2 |V(y)| dx dy. 
\]
It follows from \refeq(R-prop) for  $|z|>a$ that 
\bqn 
\int_{|z||x-y|\geq 1} |V(x)||\Rg^{(j)}(|z|, x-y)|^2 |V(y)| dx dy 
\leq C|z|^{-5} \|\ax^{(2j-1)_{+}} V\|_1^2 ,
\lbeq(1-a)
\eqn 
H\"older's inequality implies that for any $1<q\leq \infty$   
\begin{align*}
& C \iint_{|z||x-y|\leq 1}\la 
\log(|z||x-y|)\ra^2 |V(x)V(y)| dxdy  \notag \\
& \leq  C |z|^{-2/q'} \|\log |x|\|_{L^{q'}(|x|<1)}^2
\sup_{y\in \R^2} \|V(x)\|_{L^q(|x-y|\leq 1)} \|V(y)\|_{L^1},
\end{align*}
where $q'= q/(q-1)$ is the H\"older conjugate exponent.  Hence 
\begin{align}
\int_{|z||x-y|\leq 1}&  |V(x)||\Rg^{(j)}(|z|, x-y)|^2 |V(y)| dx dy \notag
\\
& \leq C|z|^{-2(2+j)-2/q'} \sup_{y\in \R^2} \|V(x)\|_{L^q(|x-y|\leq 1)} \|V(y)\|_{L^1}.
\lbeq(1-b)
\end{align} 
By combining \refeqs(1-a,1-b) 
\[
\|\pa_{z}^j M_v R_0(z^4)M_v \|_{\Hg_2}\leq 
C |z|^{-\min(\frac52, 3+j-\frac1{q})}(\|V\|_{L^q_{u,loc}(\R^2)}+ \|\ax^{(2j-1)_{+}} V\|_{1}), 
\]
and we obtain \refeq(pre-final).  Since $\Rg(z,x)$ is holomorphic in $z\in \Cb^{++}$ 
and is continuous in $z\in \C^{++}\setminus \{0\}$,  
the dominated convergence theorem implies 
that $M_v R_0(z^4)M_v$ enjoys the same property as an $\Hg_2$-valued function.  
\edpf

\subsection{Series expansion of the free resolvent} 
The well-known series expansion of the Hankel function (\cite{DLMF})  yields    
\bqn  
\Gg(z,x) = 
\sum_{n=0}^\infty \left( g(z|x|) + \frac{c_n}{2\pi} 
\right) \frac{(-z^2 |x|^2/4)^n}{(n!)^2}, \lbeq(Green-b) 
\eqn 
where $c_n= 1/(2(n+1))+ \sum_{j=1}^n 1/j$ and 
\bqn 
g(z)= -\frac1{2\pi}\log\left(\frac{z}{2}\right)-\frac{\c}{2\pi}, \quad 
\mbox{$\c$ being Euler's number}. 
\lbeq(g)
\eqn 
It follows that $\Rg(z,x)$ has the series expansion: 
\begin{gather}  
\frac{i}{8z^2}\sum_{n\geq 0, even} 
\frac{(z^2 |x|^2)^n}{4^n (n!)^2}
- \frac1{z^2}\sum_{n\geq 1, odd} {g}_n(z|x|)
\frac{(z^2|x|^2)^n}{4^n(n!)^2},  \lbeq(Rg0-rev-a) \\
{g}_n(z) = g(z)+\frac{c_n}{2\pi}-\frac{i}{8}.
\lbeq(Rg0-rev-ag)
\end{gather}
When reordered in the descending order as $z \to 0$ \refeq(Rg0-rev-a) becomes 
\begin{align}
\Rg(\lam, x) & =\frac{i}{8\lam^2} - \frac{g_1(\lam)}{4}|x|^2 
+ \frac{(\log |x|)|x|^2}{4\cdot 2\pi} 
+ \frac{i\lam^2}{8 \cdot 4^2\cdot 2^2 }|x|^4 
\notag \\
& \hspace{1cm} - \frac{\lam^4 g_3(\lam)}{4^3(3!)^2}|x|^6 
+\frac{\lam^4}{4^3 \cdot 2 \pi (3!)^2} |x|^6\log |x|+ \cdots 
\lbeq(descend-a) \\
& =\lam^{-2}\Gg_{0}(x)+ g_1(\lam) \Gg_{2}(x) + \Gg_{2,l}(x) 
+ \lam^2 \Gg_4(x)\notag  \\
& \hspace{1cm} + \lam^4 g_3(\lam)\Gg_6(x)+ \lam^4 \Gg_{6,l}(x) + \cdots\,, 
\lbeq(descend-b)
\end{align} 
where \refeq(descend-b) is the definition of 
$\Gg_0(x), \Gg_2(x), \Gg_{2,l}(x), \dots$. Note that, for even $n$,  
$\Gg_{2n,l}(x)=0$ and $g_{n}(\lam)$ does not appear 
in front of $\Gg_{2n}(x)$.  

Let $G_{2n}$ and $G_{2n,l}$ respectively be the convolution operators with 
$\Gg_{2n}(x)$ and $\Gg_{2n,l}(x)$:
\[
G_{2n}u(x) = \int_{\R^2}\Gg_{2n}(x-y)u(y) dy, \quad 
G_{2n,l}u(x) = \int_{\R^2}\Gg_{2n,l}(x-y)u(y) dy
\]
and, when sandwiched by $M_v$, we denote 
\[
G_{2n}^{(v)}= M_v G_{2n} M_v, \ \ G_{2n,l}^{(v)}= M_v G_{2n,l} M_v, \ n=1,2,\dots\,.
\]
Then, at least formally, $\Mg^{+}(\lam^4)=M_U + M_v R_0^{+}(\lam^4)M_v$ 
has the following expansion as $\lam \to 0$: 
\begin{align}  
\Mg^{+}(\lam^4)& = 
(i/8\lam^{2})\|V\|_1 P + 
g_1(\lam) G_{2}^{(v)} + (M_U + G_{2,l}^{(v)}) \notag \\ 
& \hspace{1cm} + \lam^2 G_4^{(v)} 
+ \lam^4 (g_3(\lam)G_6^{(v)}+ G_{6,l}^{(v)}) + \cdots ,
\lbeq(Uplus)
\end{align}
where $P$ is the orthogonal projection 
\[
P= \colon \frac{v\otimes v}{\|V\|_1} = \tv \otimes \tv, \quad \tv= \frac{v}{\|v\|_2}.  
\]
We denote $Q=1-P$.  The precise meaning of the expansion \refeq(Uplus) will be made 
clear in what follows.

\reflm(pre-final) and the localization of Kato's theorem (cf. \cite{RS4}, Theorem XIII.30) 
imply that $M_v$ is $H_0$-smooth in the sense of Kato on $[a,\infty)$ for any $a>0$. 
Moreover, we learn from \cite{IS} that  the boundary values 
on $\R_{+}=(0,\infty)$, $\Mg^{\pm}(\lam^4)=\Mg(\lam^4 \pm i0)$, of 
\bqn 
\Mg(z^4)\colon = M_U + M_v R_0(z^4) M_v , \quad z \in \C^{++} \lbeq(Mdef) 
\eqn 
have inverses in $\Bb(\Hs)$ if and only if $\lam^4$ is not an eigenvalue 
of $H$; if $H$ has no positive eigenvalues, $\Mg^{+}(\lam^4)^{-1}$ 
is $\Bb(\Hs)$-valued locally H\"older continuous in $\lam>0$; we have  
for $z \in \Cb^{++} \setminus\{0\}$ that 
\bqn  
R(z^4) = R_0(z^4)- R_0(z^4)M_v \Mg(z^4)^{-1} M_v R_0(z^4)
\eqn 
and $M_v$ is locally $H$-smooth in the sense of Kato. We should mention that these results  
are known by \cite{Ag} and \cite{Ku} under a weaker but explicit decay condition 
at infinity: $\|M_{\ax^{1+\ep}V}\|_{\Bb(H^2, \Hs)} \leq C $,  $\ep>0$.  
 
\subsection{Stationary representation formula} 
Let $E_{H_0}(d\lam)$ be the spectral measure for $H_0$. 
Let for $u \in \Dg_\ast$,   
\bqn \lbeq(spect-proj-def)
\Pi(\lam)u(x)= \frac{2}{\pi{i}}\lim_{\ep \downarrow 0}
(R_0(\lam^4+i\ep)-R_0(\lam^4-i\ep))u(x).
\eqn 
Then, Stone's theorem implies for any  Borel set $I \subset [0,\infty)$
\[
E_{H_0}(I)u(x)= \int_{\lam^4 \in I} \Pi(\lam) u(x) \lam^3 d\lam. 
\] 
Let, for $a\in \R^2$, $\t_a$ be the translation by $a$: $\t_a{u}(x)=u(x-a)$. 

\bglm \lblm(s-proj) 
Let $u \in \Dg_\ast$. Then,  $\Pi(\lam)u(x)$ is a smooth function of 
$(\lam,x) \in \R_{+}\times \R^2$ and 
\bqn 
\Pi(\lam)u(x) = \frac1{2\pi \lam^2}
\int_{{\mathbb S}}
e^{{i\lam}{x\w}}\hat{u}(\lam \w)d\w 
= (\Pi(\lam)\tau_{-x}u)(0)\,.  \lbeq(spect-proj)
\eqn  
\ben 
\item[(1)] There exists a compact interval  $K \subset (0,\infty)$ such that 
$\Pi(\lam) u(x)= 0$ for all $x\in \R^2$ if $\lam \not\in K$.  
\item[(2)]  For a constant $C>0$ 
\bqn \lbeq(s-proj) 
|\Pi(\lam) u(x)| \leq C \ax^{-1/2}, \quad (\lam, x)\in \R_{+}\times \R^2.
\eqn 
\item[(3)] For Borel functions $f(\lam)$ of $\lam\in [0,\infty)$ we have    
\bqn \lbeq(mult)
f(\lam) \Pi(\lam)u(x) = \Pi(\lam) f(|D|)u (x).  
\eqn 
\een
\edlm 
\bgpf We prove \refeq(spect-proj) only.  Then, statements (1) and (3) are obvious 
and  (2) follows by the stationary phase method.  
By using Fourier transform, polar coordinates 
$\xi=\rho\w$  and the change of variable $\r \to \r^{1/4}$, we obtain 
\begin{align*}
\Pi(\lam)u(x) & = \frac{2}{2\pi(i\pi)}
\int_{\R^2}e^{ix\xi}
\left(\frac1{|\xi|^4- \lam^4- i\ep}- \frac1{|\xi|^4- \lam^4+ i\ep}\right)
\hat{u(\xi)}d\xi \\  
& = \frac{1}{2\pi}\frac{\ep}{\pi}
\int_{0}^{\infty}
\frac{e^{i\r^{1/4}{x}\w}}{(\r- \lam^4)^2+ \ep^2}
\left( \int_{\mathbb S}\hat{u}(\r^{\frac14}\w)d\w\right)\
\r^{-\frac12}d\r .
\end{align*}
Let $\ep \to 0$. Equation \refeq(spect-proj) follows. 
\edpf

We restate \refeq(sta) as a theorem which is the starting point of our analysis (see e.g., \cite{RS3}).

\bgth \lbth(repre-formula)
Suppose the condition of \refth(small) or \refth(high) is satisfied. Then,  
for $u \in \Dg_\ast$
\bqn \lbeq(sta-00)
W_{-}u(x) = u(x) - 
\int_0^\infty (R_0^{+}(\lam^4)\Qg_v(\lam)\Pi(\lam)u)(x) \lam^3 d\lam, 
\eqn 
where the integral converges absolutely for almost all $x \in \R^2$.
\edth 
\bgpf  By \reflms(Green,s-proj) the integral on the right of 
\refeq(sta-00) is only over a compact interval of $(0,\infty)$ 
and \reflm(pre-final) (1) implies 
$\Mg^{+}(\lam)^{-1}=  (1+ M_U M_v R^{+}_0(\lam^4)M_v)^{-1}  M_U$ is the sum of 
$M_U$ and an $\Hg_2$-valued continuous function. Hence,  \reflm(meaning)  below 
implies that the integral in \refeq(sta-00)  converges absolutely for a.e. $x\in \R^2$. 
The other part of the theorem is well known.     
\edpf 

As we shall 
exclusively deal with $W_{-}$ only, {\it we shall often 
omit the superscript $+$ from $R_0^{+}(\lam^4)$ and $\Mg^{+}(\lam^4)$.}

\subsection{Good producers}  
Recall definition \refeq(Omega-T-int) of 
${\tOm}(T(\lam))$, $\tOm_{\leq a}(T(\lam))$ and $\tOm_{\geq a}(T(\lam))$  
and also \refdf(GPR) of GPRs. Then, thanks to the representation formula \refeq(sta-00), 
the proof of Theorems amounts to 
proving that $\Qg_v(\lam)$ is a GPR for small and for large energies.  

By virtue of the property \refeq(mult) of $\Pi(\lam)$ we have  
\begin{gather}
\tOm_{\leq {a}}(T(\lam)) u(x)=
\int_0^\infty (R_0^{+}(\lam^4)T(\lam)\Pi(\lam)u)(x) \lam^3 \chi_{\leq {a}}(\lam)d\lam, \\  
\tOm_{\geq {a}}(T(\lam)) u(x)= \int_0^\infty 
(R_0^{+}(\lam^4)T(\lam)\Pi(\lam)u)(x) \lam^3 \chi_{\geq {a}}(\lam)d\lam.
\end{gather}
When $T(\lam)$ is independent of $\lam$, 
we write $\W_{\leq {a}}(T) $ and $\W_{\geq {a}}(T) $ for 
$\tOm_{\leq {a}}(T(\lam)) $ and $\tOm_{\geq {a}}(T(\lam)) $ respectively.

The next lemma shows that integral \refeq(Omega-T-int) is well-defined for a 
large class of integral operators $T(\lam,x,y)$. 

\bglm  \lblm(meaning) Let $T \in L^1_{\loc}((0,\infty), \Lg^1)$, $\m(\lam)$ a GMU 
and  $u \in \Dg_\ast$.  Then,  for almost every $x\in \R^2$,  the integral 
\bqn 
\int_0^\infty \iint_{\R^4} \Rg(\lam,x-z) T(\lam,z,y)(\Pi(\lam)u)(y) \lam^3 \m(\lam) dzdyd \lam 
\lbeq(int-1)
\eqn 
is absolutely convergent,  it is equal to $\tW(T(\lam))\m(|D|)u(x)$ and, it may be iteratively integrated 
in an arbitrary order.
\edlm  
\bgpf 
Let $B=\{x\in \R^2 \colon |x|<R\}$. By \reflm(s-proj) 
$(\Pi(\lam)u)(y)=0$ for $\lam \not\in [a,b]\Subset (0,\infty)$ and  
$|(\Pi(\lam)u)(y)|\leq C \ay^{-\frac12}$  and, by \reflm(R-prop), 
\[
\sup_{a\leq \lam \leq b, z \in \R^2} 
\int_{B} |\Rg(\lam,x-z)|dx = C <\infty .
\]
It follows that  
\begin{align*}
& \iint_{\R^4} 
\left(\int_{B \times \R_{+}}|\Rg(\lam,x-z)T(\lam, z, y)(\Pi(\lam)u)(y)\m(\lam)\lam^3|
dx d\lam \right) dy dz \\
& \qquad  \leq  C \iint_{\R^4} \left( \int_a^b |T(\lam, z, y)|d\lam\right)  \ay^{-\frac12} dz dy < \infty  
\end{align*}
and the integral on \refeq(Omega-T-int) is absolutely convergent for a.e. $x\in \R^2$.  
Then, it can be integrated iteratively in an arbitrary order by Fubini's theorem 
and is equal to $\tW(T(\lam))\m(|D|)u(x)$. 
\edpf 

\bgdf For $k\in \N \cup \{0\}$,  a Banach space $\Bg$ and 
a function $\a(\lam)$ of $0<\lam<a$, $\Og^{(k)}_{\Bg}(\a(\lam))$ 
is the space of  
$\Bg$-valued $C^k$-functions $T(\lam)$ of $0<\lam<a$ such that 
\bqn \lbeq(Ogk)
\|\pa_\lam^j T(\lam)\|_{\Bg}\leq C_j \lam^{-j}\a(\lam), \quad j=0, \dots, k. 
\eqn
We write $\Og_{\Bg}(\a(\lam))$ for $\cap_{k=0}^\infty \Og^{(k)}_{\Bg}(\a(\lam))$. 
We abuse notation and we use the same letter $\Og^{(k)}_{\Bg}(\a(\lam))$    
to denote a function in the space. 
\eddf 
It should be clear that 
\begin{itemize}
\item
If $\C$-valued function $\b(\lam)$ satisfies 
$\pa_\lam^j \b(\lam)\absleq C_j \lam^{-j}\b(\lam)$ for $j=0, \dots, k$, then $\b(\lam)\Og^{(k)}_{\Bg}(\a(\lam))=
\Og^{(k)}_{\Bg}(\b(\lam)\a(\lam)) $. 
\item
If $\Bg$ is a Banach algebra, $\Og^{(k)}_{\Bg}(\a(\lam)) \Og^{(k)}_{\Bg}(\b(\lam))
=\Og^{(k)}_{\Bg}(\a(\lam)\b(\lam))$.
\end{itemize}

We define the integral operator $K$ by 
\bqn \lbeq(Ka-def) 
Ku(x) = \int_0^\infty \Rg(\lam, x) (\Pi(\lam) u)(0)\lam^3 d\lam,  \quad u \in \Dg_\ast.
\eqn  
Since  $(\Pi(\lam) u)(0)\in C_0^\infty(\R_{+})$, integral \refeq(Ka-def) is 
integrable for all $x \not =0$. 

\bglm \lblm(WT-trans)
Suppose that $T \in \Lg^1$,  $\m\in C(\R)$ and $u \in \Dg_\ast$. 
Then,  for almost all $x \in \R^2$,  the integral on the right converges absolutely and 
\bqn 
\W(T)\m(|D|)u(x) = \iint_{\R^4} T(y,z) (\tau_{y} K \m(|D|)\tau_{-z} u)(x)  dy dz. 
\lbeq(W-t)
\eqn 
\edlm 
\bgpf By virtue of \reflm(meaning)   $\W(T)\m(|D|)u(x)$ is equal to  
\bqn \lbeq(linear)
\int_0^\infty \left(\iint_{\R^4}
\Rg(\lam,x-y)T(y,z)(\Pi(\lam)u)(z)dy dz \right) \m(\lam)\lam^3 d\lam.
\eqn 
and the integral converges absolutely for almost all $x \in \R^2$. Then, since 
translations and Fourier multipliers commute,  
\refeq(linear) is equal to 
\[
\iint_{\R^4} T(y,z) \left( \tau_y \int_0^\infty 
\Rg(\lam,x)(\Pi(\lam)\tau_{-z}\mu(|D|)u)(0)\lam^3 d\lam \right) dy dz. 
\]
The function in the parenthesis is equal to 
$(\t_{y} K \m(|D|)\tau_{-z} u)(x) $. This proves the lemma. 
\edpf 

Let 
\bqn \lbeq(pi-2)
\Pi_2 (\lam)u(x)= \frac1{2\pi} \int_{{\mathbb S}^1}e^{i\lam x \w}\hat u(\lam\w)d\w
\eqn 
so that $\Pi(\lam)u(x) =\lam^{-2}\Pi_2 (\lam)u(x)$. 
\begin{align} 
& \tK_{1} u(x)= \int_0^\infty \Gg(\lam, x) \Pi_2(\lam) u(0)\lam d\lam, \lbeq(tk1)\\
& \tK_{2} u(x)= \int_0^\infty  \Gg(i\lam,x) \Pi_2(\lam) u(0)\lam d\lam  \lbeq(tk2)
\end{align}

\bglm \lblm(K-GOP)  
{\rm (1)} Operators $\tK_1$ and $\tK_2$ are GOPs.  

\noindent
{\rm (2)} Let $\m(\lam) = \lam^2 \k(\lam)$ and $\k(\lam)$ be a GMU. Then,  
$K\m(|D|)$ is a GOP. 
\edlm 
\bgpf 
Replacing $\lam^3$ by $\lam^5$ in \refeq(Ka-def),  let, for $u \in \Dg_\ast$,
\[
\tK {u}(x) = \int_0^\infty \Rg(\lam, x) (\Pi(\lam) u)(0)\lam^5 d\lam .
\]
Then, $\tK {u} (x) = \frac12(\tK_{1}u(x) + \tK_{2} u(x))$ by \refeq(res-ker). 
We prove (1). 
Since $K \mu(|D|)= \tK \circ \k(|D|)$, statement (2) follows from (1).  
The proof patterns after that of the 
corresponding theorem in section 3 of \cite{GY-2024} 
and we shall be a little sketchy here.  Let $Mu(\r)$ be the mean of 
$u(x)$ over the circle $|x|=\r$: 
\bqn  \lbeq(sphe-mean)
Mu(\r) = \frac1{2\pi}\int_{{\mathbb S}^1} u(\r\w) dw.
\eqn 

By using Fourier transform, \refeq(spect-proj) and  polar coordinates 
$\lam \w= \eta$,  we obtain 
\begin{align}
	{\tK}_{1}u(x)& = \int_0^\infty 
	\left(\frac1{(2\pi)^3}
	\int_{\R^2\times {\mathbb{S}}}
	\frac{e^{ix\xi}\hat{u}(\lam\w)}{|\xi|^2-\lam^2- i0}d\xi{d\w}
	\right)\lam {d\lam}\notag \\
	& = \frac1{(2\pi)^4} \int_{\R^2_y}\left(
	\iint_{\R^2\times \R^2}
	\frac{e^{ix\xi-iy\eta}d\xi{d\eta}}
	{|\xi|^2-|\eta|^2- i0}
	\right)u(y) dy . \lbeq(K1-a)
\end{align}
The inner integral of \refeq(K1-a) is equal to 
\begin{align}
	& \lim_{\ep \to +0}\frac{(2\pi)^2 i}{2}\int_0^\infty e^{-\ep{t}}
	\left(\frac1{2\pi}\int_{\R^2}e^{ix\xi- it\frac{|\xi|^2}{2}}d\xi\right) 
	\left(\frac1{2\pi}\int_{\R^2}e^{-iy\eta+ it\frac{|\eta^2|}{2}}d\eta\right) dt 
	\notag \\
	& = \lim_{\ep \to +0} \frac1{2i}\int_0^\infty 
	e^{i{(|x|^2-|y|^2+i\ep)}/{2t}}\frac{dt}{t^2} 
	= \lim_{\ep \to +0}\frac{1}{|x|^2-|y|^2+i\ep} .
\end{align}
It follows that $\tK_1$ is equal to the singular integral operator 
\bqn \lbeq(K-3)
{\tK}_{1}u(x)= \lim_{\ep \to+0} 
\frac1{(2\pi)^4} \int_{\R^2_y}\frac{u(y)dy}{|x|^2-|y|^2+i\ep}.
\eqn 
Thus, ${\tK}_{1}u(x)$ is rotationally invariant and, 
if we write ${\tK}_{1}u(x)= {\tK}_{1}u(\r)$, $\r= |x|$, then   
\bqn 
{\tK}_{1}u(\sqrt{\r})= \frac1{2(2\pi)^3}
\int_0^\infty \frac{Mu(\sqrt{r})}{\r-r+i0}dr .
\eqn 
Via the $L^p$-boundedness of Hilbert transform we then obtain 
\begin{align}
	& \|{\tK}_{1}u\|_{L^p(\R^2)}^p 
	= \pi \int_0^\infty |{\tK}_{1}u(\sqrt{\r})|^p d\r 
	\leq C \int_0^\infty |Mu(\sqrt{r})|^p dr \notag \\
	& = 2C \int_0^\infty |Mu({r})|^p rdr
	\leq \frac{C}{\pi} \|u\|_p^p. \lbeq(K-4)
\end{align}
Similar computation shows that  
\begin{align}
& \tK_{2}u(x)  = \frac1{(2\pi)^2}\int_0^\infty \left(\int_{\R^2} 
\frac{e^{ix\xi} d\xi}{|\xi|^2+\lam^2}\right) 
\left(\int_{{\mathbb S}^1}\hat{u}(\lam\w)d\w\right) \lam d\lam \notag \\
& = 
\int_{\R^2}\left(\iint_{\R^4} \frac{e^{i(x\xi+y\eta)}}{\xi^2 + \eta^2}d{\xi}d{\eta}\right) u(y) dy 
= C\int_{\R^2} \frac{u(y)}{x^2+y^2}dy  \lbeq(K-5) 
\end{align} 
and  ${\tK}_{2}$ is also a GOP by \reflm(preparation).  This finishes the proof.
\edpf  

 Combining \reflm(WT-trans) and \reflm(K-GOP) (2) , we obtain the following lemma 
which will play important role for proving \refths(high,regular-case). 
The proof of the lemma is a modification of that of Lemmas 3.4 and 3.5 of \cite{Ya-2dim-new}.  

\bglm \lblm(T-K)
  Let $\mu(\lam)= \lam^2 \k(\lam)$ and $\k(\lam)$ be a GMU. 
Then:  

\noindent
{\rm (1) } If $T(x,y)\in \Ls^1$, then   
$\W(T)\m(|D|)$ is a GOP and 
\bqn \lbeq(T-K)  
\|\W(T)\m(|D|)\|_{\Bb(L^p)} \leq C_{p} \|T \|_{\Ls^1}.
\eqn 
The same holds for multiplication $M_f$ by $f(x) \in L^1(\R^2)$ 
and  
\bqn  \lbeq(T-F)
\|\W(M_f)\m(|D|)\|_{\Bb(L^p)} \leq C_{p} \|f \|_{L^1}.
\eqn  
{\rm (2)}  Suppose $T(\lam)\in C^2((0,\infty), \Ls^1)$ and it satisfies  
\bqn \lbeq(Tlam-int) 
\lim_{\lam \to 0}(\|T(\lam)\|_{\Ls^1}+ \lam \|T'(\lam)\|_{\Ls^1})=0, 
\ 
\int_0^\infty \lam \|T''(\lam)\|_{\Ls^1}d\lam<\infty.  
\eqn  
Then, ${\tOm}(T(\lam))\m(|D|)$ is a GOP and 
\bqn \lbeq(Tlam-K) 
\|{\tOm}(T(\lam))\m(|D|)\|_{\Bb(L^p)} 
\leq C_{p}\int_0^\infty \lam \|T''(\lam)\|_{\Ls^1}d\lam.
\eqn 
The same holds with obvious modifications when 
$T(\lam)$ is replaced by multiplication $M_{f(\lam)}$ 
by $f(\lam,x) \in C^2((0,\infty), L^1(\R^2))$. 
\edlm 

\bgcor   \lbcor(T-K)  Let $0<a<b$. 

\vspace{0.2cm}
\noindent 
{\rm (1)} Let $\m(\lam)$ be replaced by $\chi_{\geq a}(\lam)$. Then,  
\reflm(T-K) {\rm (1)} holds and, if $\int_0^\infty \lam \|T''(\lam)\|_{\Lg^1}d\lam<\infty$, 
then \reflm(T-K) {\rm (2)} holds.  
 
\vspace{0.2cm}
\noindent 
{\rm (2)}  Suppose $T(\lam) \in C^1([0, 2b), \Lg^1)\cap C^2((0,2b), \Lg^1)$ 
and  it satisfies \refeq(Tlam-int)  except
$\lim_{\lam\to 0}\|T(\lam)\|_{\Lg^1}=0$. Then, for $0<a<b$, 
$\lam^2\chi_{\leq a}(\lam) T(\lam)$ 
is a GPR.  

\vspace{0.2cm}
\noindent 
{\rm (3)}  Operator valued function $\chi_{\leq a}(\lam)\Og_{\Lg^1}(\lam^2 (\log \lam)^{-1})$  is a GPR. 
\edcor 
\bgpf  
(1)  We evidently have $\chi_{\geq a}(\lam)= \lam^2 \cdot (\chi_{\geq a}(\lam)/\lam^2)$ 
and $\chi_{\geq a}(\lam)/\lam^2$ is a GMU.   Statement (1) follows from \reflm(T-K) (2). 

(2) Let $R(\lam)= T(\lam)\chi_{\leq b}(\lam) - T(0)\chi_{\leq b}(\lam)$. 
Then, 
\[
\lam^2 T(\lam) \chi_{\leq a}(\lam)
=\lam^2 \chi_{\leq a}(\lam) R(\lam) + \lam^2 \chi_{\leq a}(\lam) T(0),  
\]
$R(\lam)$ satisfies \refeq(Tlam-int) and $T(0) \in \Lg^1$. Since  
$\chi_{\leq a}(\lam)$ is a GMU, (2) follows from \reflm(T-K) (1) and (2).  

(3) We have 
$\chi_{\leq a}(\lam)\Og_{\Lg^1}(\lam^2 (\log \lam)^{-1})
=\lam^2 \chi_{\leq a}(\lam)\Og_{\Lg^1}((\log \lam)^{-1})$ 
and $\Og_{\Lg^1}((\log \lam)^{-1})$ satisfies \refeq(Tlam-int). 
\reflm(T-K) (2) implies (3). 
\edpf 

\paragraph{\bf Proof of \reflmb(T-K)} The proof is almost a repetitions of that of 
Lemmas 3.3 and Proposition 3.6 of \cite{GY-2024} and we 
only outline it here, referring to \cite{GY-2024} for details. 

(1)  For $T \in \Lg^1$, \refeq(T-K) follows by  applying 
\reflm(K-GOP) (2)  and Minkowski's inequality 
to \refeq(W-t).  For  the multiplication $M_f$, we have from \refeq(W-t) that 
\bqn 
\W(M_f)\m(|D|)u(x) = \iint_{\R^4} f(y) (\tau_{y} K\circ \m(|D|)  \tau_{-y} u)(x)  dy   
\lbeq(linear-a)
\eqn 
and \refeq(T-F) follows again via Minkowski's inequality.  

(2) By integration by parts we have 
\bqn \lbeq(int-parts)
T(\lam)= \int_0^{\infty} (\r-\lam)_{+}T''(\r)d\r.
\eqn
We substitute \refeq(int-parts) for $T(\lam)$ in \refeq(Omega-T-int). 
Since  $\W(T)\mu(|D|)$ depends on $T\in \Lg^1$ continuously by \refeq(T-K), 
it follows that  
\[
{\tOm}(T(\lam))\m(|D|)u = 
	\int_0^\infty(\r-\lam)_{+} \W(T''(\r))\m(|D|)u d\r 
\] 
and \refeq(mult) implies that the right side is equal to 
\bqn \lbeq(intW)
\int_0^\infty \r \W(T''(\r))\m(|D|)(1-|D|/\r)_{+} u d\r. 
\eqn
Since $(1-|\xi|)_{+}= (1-|\xi|^2)_{+} (1+|\xi|)^{-1} \in L^1(\R^2)$ 
(\cite{Stein}, p. 389), 
$(1-|D|/\r)_{+}\in \Bb(L^p)$ for all $1\leq p\leq \infty$ 
and $\|(1-|D|/\r)_{+}\|_{\Bb(L^p)}$ is independent of $\r>0$.  
Apply Minkowski's inequality to \refeq(intW).  Estimate \refeq(T-K) implies the desired 
\refeq(Tlam-K).  
\qed

\section{Proof of \refths(small,high)} 
The strategy of the proof is similar to that of Theorems 1.1 and 1.3 of 
\cite{GY-2024}.  Via \refeq(sta-00) and \refeq(mult) , we see that 
$W_{-}\chi_{\geq{a}}(|D|)u$ is equal to  
\bqn \lbeq(sta-high)
\chi_{\geq{a}}(|D|) u - 
\int_0^\infty R_0^{+}(\lam^4)\Qg_v(\lam)\Pi(\lam)u \lam^3 
\chi_{\geq{a}}(\lam)d\lam\,.
\eqn 
Formally expanding by Neumann series, we have 
\bqn \lbeq(Neumann)
\Qg_v(\lam)= M_v(M_U + M_v R_0(\lam^4) M_v)^{-1}M_v= V - VR_0(\lam^4)V +\cdots. 
\eqn 
Substituting \refeq(Neumann) in \refeq(sta-high)  
produces the well-known Born series:  
\bqn 
W_{-}\chi_{\geq{a}}(|D|)= \chi_{\geq{a}}(|D|) - 
W_1 \chi_{\geq{a}}(|D|) + \cdots, \lbeq(Born) 
\eqn 
where $W_n \chi_{\geq{a}}(|D|)$ is given by the integral 
\bqn 
\int_0^\infty R_0(\lam^4) (M_VR_0(\lam^4))^{n-1}M_V 
\Pi(\lam) \lam^3 \chi_{\geq{a}}(\lam)d\lam\,. \lbeq(Wn-0)
\eqn   

\paragraph{\textbf{Proof \refthb(small)}} 
We assume $c_0$ is small. Then, Proposition 2.1 in \cite{Y-3dim} may be adapted 
to show that the series on the right of \refeq(Born) converges to the left-hand side 
in the operator norm of $\Bb(L^2)$. Thus, the proof will be  finished if we show that it 
converges in the operator norm of $\Bb(L^p)$ for $1<p<\infty$.   
We employ the following lemma whose proof is postponed to the Appendix 
on the Fourier multiplier $\Rg(|D|,y)$ 
with parameter $y \in \R^2$ defined via the convolution kernel $\Rg(\lam,x)$ by 
\bqn 
\Rg(|D|,y)u(x) = \frac1{2\pi}\int_{\R^2}e^{ix\xi} \Rg(|\xi|, y)\hat{u}(\xi) d\xi, \quad 
u \in \Dg_\ast.
\eqn 

\bglm \lblm(H02y) Let $a>0$ and $1<p<\infty$. Then, 
there exists a constant $C_{a,p}$ independent 
of $y \in \R^2\setminus\{0\}$ such that 
\bqn 
\|\Rg(|D|,y)\chi_{\geq {a}}(|D|)\|_{\Bb(L^p)}\
\leq C_{a,p}(1+ |\log |y||). 
\lbeq(H02y-a)
\eqn  
\edlm

It is clear that $\chi_{\geq {a}}(|D)$ is a GOP and,  by virtue of \refeq(T-K),  
\[
\| W_1 \chi_{\geq{a}}(|D|)u\|_p  = \|\W(M_V)\chi_{\geq{a}}(|D|)u\|_p 
\leq C_p \|V\|_{L^1}\|u\|_p.
\]
Under the conditions of \refth(small) the following integrals converge 
absolutely   for a.e. $x \in \R^2$ and $W_2 \chi_{\geq a}(|D|)u(x)$ is equal to the integral 
${\displaystyle \int_0^\infty \lam^3 \chi_{\geq{a}}(\lam) d\lam }$ of 
\begin{align}
& \iint_{\R^2 \times \R^2}
\Rg(\lam,x-x_1)V(x_1) \Rg(\lam, x_1-x_2) V(x_2) (\Pi (\lam)u)(x_2) dx_1 dx_2 
\notag \\
& = \iint_{\R^2 \times \R^2}
\Rg(\lam,x-y)V(y) \Rg(\lam, z)V(y-z) (\Pi (\lam)u)(y-z) dydz \notag  \\
& = \iint_{\R^2 \times \R^2}
\Rg(\lam,x-y)V_z^{(2)}(y) (\Pi (\lam) \Rg(|D|, z)\tau_z u)(y) dydz,  \lbeq(inner-1)
\end{align}
where we changed $x_1 \to y, x_2 \to y-z$ in the first step; 
in the second  we set  $V_z^{(2)}(y) = V(y) V(y-z)$ and changed 
$\Rg(\lam, z)(\Pi (\lam)u)(y-z)$ to $ (\Pi (\lam) \Rg(|D|, z)\tau_z u)(y)$ 
via \refeq(mult). 
Integrating \refeq(inner-1) with respect to $\lam$ and changing 
the order of integrals,  we obtain 
\bqn \lbeq(W2c)
W_2 \chi_{\geq a}(|D|)u=\int_{\R^2} 
\W(M_{V^{(2)}_z})\chi_{\geq a}(|D|)\Rg(|D|,z)\tau_z u dz.  
\eqn 
Apply  \refeq(T-K) and \refeq(H02y-a) to \refeq(W2c) and estimate as in the 
proof of \reflm(pre-final).  
We obtain 
\begin{align}
\|W_2 \chi_{\geq a}(|D|)u\|_p 
& \leq C \|u\|_p \int_{\R^8} |V(x)V(x-y)|(1+|\log |y||)dxdy  \notag \\
& \leq C(\|V\|_{L^{q}_{loc,u}}+ \|\la \log |x|\ra^{2} V\|_{L^1})^2 \|u\|_p 
\lbeq(W2-intro).
\end{align}
For $n \geq 3$, let $V^{(n)}_{Y_{n-1}}(x)= 
V(x) V(x-y_1) \cdots V(x-y_1-\cdots-y_{n-1})$ for  $Y_{n-1}=(y_1,\dots,y_{n-1})$.
Repeating the change of variables as in \refeq(inner-1), we have 
\begin{multline*}
	(M_V R_0(\lam^4))^{n-1} M_V u(x) \\ 
	= \iint_{\R^{4(n-1)}}
	V^{(n)}_{Y_{n-1}}(x)
	\left(\prod_{j=1}^{n-1} \Rg(\lam|y_j|)\right)\tau_{y_1+\cdots+y_{n-1}}u(x) 
	dy_1 \dots dy_{n-1} .
\end{multline*}
It follows that $W_{n}\chi_{\geq{a}}(|D|)u(x)$ is equal to 
\begin{align*}
&	\int_0^\infty \int_{\R^4} \int_{\R^{4(n-1)}}\Rg(\lam |x-y|) 
	V^{(n)}_{Y_{n-1}}(y) \left(\prod_{j=1}^{n-1} \Rg(\lam|y_j|)\right) \notag \\
& \hspace{3cm} 
	\times \Pi(\lam)\tau_{y_1+\cdots+y_{n-1}}u(y) \lam^3 \chi_{\geq{a}}(\lam)
	dy_1 \cdots dy_{n-1} dy d\lam \notag \\ 
& =\int_{\R^{4(n-1)}} \W(M_{V^{(n)}_{Y_{n-1}}})
\chi_{\geq{a}}(|D|)\prod_{j=1}^{n-1}
\Rg (|y_j||D|) \tau_{y_1+\cdots+ y_{n-1}} {u} dy_1 \dots dy_{n-1},   
\end{align*}
where we used \refeq(mult) in the second step. 
Then, Minkowski's inequality,  
\reflms(T-K,H02y) imply that $\|W_n \chi_{\geq{a}}(|D|)u\|_p$ is bounded by   
\begin{align*}
	& C_{a,p}^n 
	\int_{\R^{4(n-1)}}  
	\|V^{(n)}_{y_1,\dots,y_{n-1}}\|_{L^1(\R^4)} 
	\prod_{j=1}^{n-1} \la \log |y_j|\ra \|u\|_p dy_1 \dots dy_{n-1} 
	\notag \\
	&= C_{a,p}^n 
	\int_{\R^{4n}}  
	|V(x_0)| 
	\prod_{j=1}^{n-1} 
	|V(x_j)| \la |\log |x_{j-1}-x_j|\ra  
	\|u\|_p dx_0 \dots dx_{n-1}. 
\end{align*}
We estimate the last integral inductively by using Schwarz' and  H\"older's 
inequalities $n$-times, which yields  
\bqn \lbeq(small-effect-0)
\|W_n \chi_{\geq{a}}(|D|) u\|_p \leq C_p C_{a,p}^n 
(\|V\|_{L^{q}_{loc,u}}
+ \|\la \log |x|\ra^{2} V\|_{L^1})^n \|u\|_p 
\eqn 
with constants independent of $V$ and $n$. 
Hence the series \refeq(Born) 
converges in $\Bb(L^p)$ for $1<p<\infty$ 
if $C_{ap}(\|V\|_{L^{q}_{loc,u}}+ \|\la \log |x|\ra^{2} V\|_{L^1})<1$. 
This proves \refth(small).  

\paragraph{\textbf{Proof \refthb(high)}}
Since positive eigenvalues are absent from $H$ by the assumption,  the inverse 
$\Mg^{+}(\lam^4)^{-1}$ exists for all $\lam>0$ 
as was remarked in the introduction. We expand $\Qg_v(\lam)$ of 
\refeq(sta-high) as follows:
\begin{align}  
& \Qg_v(\lam)= V -M_V R_0(\lam^4)M_V + D_2(\lam), \lbeq(with) \\ 
& D_2(\lam)= M_v (M_w R_0(\lam^4)M_v)^2 (1+M_w R_0(\lam^4)M_v)^{-1} M_w, 
\end{align} 
where $w= U v$ so that $vw= V$. 
As was shown in the proof of \refth(small), $V$ and $M_V R_0(\lam^4)M_V$ 
are GPR's;   by virtue of \refeq(pre-final) and Leibniz' rule   
\[
\|(\pa/\pa\lam)^{j} D_2(\lam)\|_{\Ls^1} \leq 
C\lam^{-4} (\|V\|_{L^q_{u,loc}}+ \|\ax^{3} V\|_{1})^2, \quad 
j=0,1,2
\]
and \refcor(T-K) (1) implies that $\chi_{\geq a}(\lam) D_2(\lam)$ is a GPR.  
This completes the proof of \refth(high). 
\qed 

\section{Threshold expansion of  $\Mg(\lam)^{-1}$}  
In what follows we shall study the low energy part $W_{-}\chi_{\leq{a}}(|D|)$ 
and prove \refth(regular-case) assuming that $V(x)$ satisfies, in addition to \refeq(1-1), the faster decay 
at infinity:      
\bqn \lbeq(condi-1)
 \ax^{10+\ep} V(x) \in L^1(\R^2)\  \quad \mbox{for an}\ \ep>0.  
\eqn   
In view of the representation formula \refeq(sta-00) for $W_{-}$, we need to study 
$\Qg_v(\lam)$ or ${\Mg^{+}(\lam)}^{-1}$ for small $\lam >0$.   
Actually, the asymptotic expansion of $\Mg^{+}(\lam)^{-1}$  as $\lam \to 0$ 
has been obtained by \cite{LSY}, however, it does not seem to us completely correct 
and, we have decided to redo it here in a fashion suitable for our purpose. 
As we shall be exclusively concerned with small energies 
we shall {\it often omit the phrase ``for small $\lam>0$''} from various expressions, 
e.g., GPR will mean GPR for small energy etc. 

\subsection{Preparation}

 In view of \refeq(Uplus)  we set 
${\tMg}^{+}(\lam^4) \colon = c_v \lam^2 \Mg^{+}(\lam^4)$ with 
\[
c_v \colon = 8(i\|V\|_1)^{-1} 
\] 
so that 
\bqn   \lbeq(tMg)
{\Mg}^{+}(\lam^4)^{-1}= c_v \lam^{2}{\tMg}^{+}(\lam^4)^{-1}.
\eqn 
Let $T_0$ be the $0$-th order term in the series \refeq(Uplus):  
\bqn \lbeq(t0-def)
T_0 = M_U + G_{2,l}^{(v)}. 
\eqn 
The {\it tilde} symbol indicates that operators are multiplied by $c_v$ 
(except the ${\tMg}^{+}(\lam^4)$ defined above): 
\bqn  
\tT_0 \colon= c_v T_0, \ \tG_2^{(v)}\colon = c_v G_2^{(v)} \ \mbox{etc}.
\lbeq(tilde)
\eqn 
For shortening formulas we set 
 \begin{align}
& A_2(\lam)\colon = \lam^2 g_1(\lam) \tG_2^{(v)}+ \lam^2 \tT_0 ,	\lbeq(A2) \\
& \Ng_4(\lam) \colon= \lam^4 \tG_4^{(v)}+ \lam^6(g_3(\lam)\tilde{G}_6^{(v)}+ \tilde{G}_{6,l}^{(v)})
	+ \cdots, \lbeq(n4) 
\end{align} 
so that
\bqn 
{\tMg}^{+}(\lam^4)= P+ A_2(\lam)+ \Ng_4(\lam). \lbeq(M4-1)
\eqn 
Separating multiplications from $\Hg_2$-valued functions, we also write      
\bqn  
A_2(\lam) = \lam^2\tM_{U} + N_2(\lam),  \ 
N_2(\lam)= \lam^2 g_1(\lam) \tG_2^{(v)} + \lam^2 \tG_{2,l}^{(v)} .
\lbeq(A2lam) 
\eqn 
We set 
\bqn \lbeq(f-lam)
f(\lam,x) \colon = (1 + \lam^2 \tilde{U}(x))^{-1}
= 1 - \lam^2 \tilde{U}(x)+ \Og_{L^\infty(\R^2)}(\lam^4) .
\eqn  

\bglm\lblm(JN-3)   {\rm (1)}  We have following estimates:
\begin{align}
\Ng_4 (\lam) & = \Og^{(4)}_{\Hg_2}(\lam^{4}). 
\lbeq(Ng4)  \\ 
({\bf 1}+ A_2(\lam))^{-1}& = {\bf 1}-A_2(\lam)+ \Og_{\Bb(\Hs)}^{(4)}(\lam^4 (\log\lam)^2)  
\lbeq(A-inv-a) \\
\hspace{2cm}
& = M_{f}- M_f N_2 (\lam)M_f + \Og_{\Hg_2}^{(4)}(\lam^4(\log\lam)^2).
\lbeq(A-inv)
\end{align} 

\noindent 
{\rm (2)}  The operator valued function $\lam^2 M_v ({\bf 1}+ A_2(\lam))^{-1}M_v$ is a GPR. 
\edlm 
\bgpf  (1)  Let  $\ep>0$ be as in \refeq(condi-1).  Let $n(x,y) \colon = \tv(x)\tv(y)|x-y|^{4} $.  
Then, $n(x,y) \in \Hg_2$  by \refeq(condi-1).   If $\lam |x-y|\leq 1$ and $j=0,\dots, 4$, then 
\refeq(Rg0-rev-a) implies 
\bqn \lbeq(l-small)
\pa_\lam^j \Ng_4(\lam,x,y) \absleq 
C \lam^{4-j} n(x,y).
\eqn 
For $\lam |x-y|\geq 1$,  we estimate each term on the right of 
\[
\Ng_4(\lam,x,y)=c_v \lam^2 v(x)\Rg(\lam,x-y)v(y)- \tv(x)\tv(y)- N_2(\lam,x,y) 
\]
separately.   Let $j=0, \dots, 4$.  From  \reflm(Green)  
\[
|\pa_{\lam}^j (\lam^2 v(x)\Rg(\lam,x-y)v(y))|\leq 
C|x-y|^j v(x) v(y) \leq C \lam^{4-j} n(x,y);  
\] 
we evidently have 
$\tv(x)\tv(y)= n(x,y)|x-y|^{-4} \absleq \lam^{4} n(x,y)$; 
\[
\pa_\lam^j N_2(\lam,x,y) \absleq Cv(x)v(y)(\lam|x-y|)^{2+\frac{\ep}{2}-j}|x-y|^j 
\leq C \lam^{4-j} n(x,y).
\]
Combining these estimates, we obtain \refeq(Ng4).   

Estimate \refeq(A-inv-a) is evident and we  omit the proof. 
We have   
\[
{\bf 1}+A_2(\lam)= M_{f}^{-1}(1+ M_f N_2(\lam))
\]
and $\|M_f N_2(\lam)\|_{\Hg_2}\leq C \lam^2 \la \log \lam\ra$ by \refeq(A2lam).  
It follows $({\bf 1}+ A_2(\lam))^{-1} = (1+ M_f N_2(\lam))^{-1}M_f$. 
Expanding the right side yields \refeq(A-inv). 

(2)  Substitute \refeq(f-lam) in \refeq(A-inv) and sandwich the result by $M_v$. 
Then,   
$M_v(- M_f N_2 (\lam)M_f + \Og_{\Hg_2}^{(4)}(\lam^4\log^2\lam))M_v$ 
satisfies the condition of \refcor(T-K) (2) and it produces a GPR when 
multiplied by $\lam^2$.  If we write as  
$\lam^2 M_v M_{f(\lam)}M_v= \lam^2 (M_V -\lam^2 M_{V \tilde{U}} + 
M_v\Og_{L^\infty}(\lam^4)M_v )$, we can apply  \reflm(T-K) (1) to the first two 
terms and (2) to the last.  This proves statement (2). 
\edpf

\subsection{Jensen-Nenciu's lemma and Feshbach formula}

We compute $\tMg^{+}(\lam^4)^{-1}$ by using the following lemmas.  

\bglm[\cite{JN}] \lblm(JN)  Let $A$ be a closed operator in a 
Hilbert space $\Hg$ and $S$ a projection. Suppose $A+S$ has bounded 
inverse. Then, $A$ has bounded inverse if and only if 
\bqn \lbeq(JN-B)
B= S - S(A+S)^{-1}S 
\eqn 
does so in $S\Hg$ and, in this case, 
\bqn \lbeq(JN-A)
A= (A+S)^{-1}+ (A+S)^{-1}SB^{-1} S (A+S)^{-1}.
\eqn  
\edlm

\bglm[Feshbach formula] \lblm(FS) Let $\Xg= \Xg_1 \dot{+} \Xg_2$ be the 
direct sum of Banach spaces and 
\[
A= \begin{pmatrix}
a_{11} & a_{12} \\ a_{21}& a_{22}
\end{pmatrix}
\]
an operator matrix in this decomposition. 
Suppose $a_{11}$, $a_{22}$ are closed, $a_{12}$, $a_{21}$ bounded  
and bounded inverse $a_{22}^{-1}$ exists in $\Xg_2$. 
Then, $A^{-1}$ exists if and only if 
$d \colon = (a_{11}- a_{12}a_{22}^{-1}a_{21})^{-1}$ exists in $\Xg_1$. In this case we have 
\bqn \lbeq(FS-formula)
A^{-1} = \begin{pmatrix}  d & -d a_{12} a_{22}^{-1} \\
-a_{22}^{-1}a_{21} d & a_{22}^{-1}a_{21}d a_{12} a_{22}^{-1} + a_{22}^{-1}
\end{pmatrix}.
\eqn 
\edlm 

We apply \reflm(JN) to the pair $(A,S)= ({\tMg}^{+}(\lam^4), Q)$. We begin 
with the following lemma.  We assume $0<\lam<a$ for a small $a>0$. 

\bglm \lblm(JN-3a)
{\rm (1)} The inverse $({\tMg}^{+}(\lam^4)+ Q)^{-1}$ exists in $\Bb(\Hs)$ and    
\bqn \lbeq(JN-a)
({\tMg}^{+}(\lam^4)+ Q)^{-1}=({\bf 1}+A_2(\lam))^{-1}- \Ng_4(\lam)+ 
\Og_{\Hg_2}^{(4)}(\lam^6\log\lam).
\eqn 
{\rm (2))} For small energies $\lam^2 M_v (\tilde{\Mg}^{+}(\lam^4)+ Q)^{-1}M_v$ is a GPR.  
\edlm

\bgpf  
(1) Let ${\Ng}_{4,1}(\lam) \colon = \Ng_{4}(\lam)({\bf 1}+ A_2(\lam))^{-1}$.  
Then,  
\bqn \lbeq(tN4)
{\Ng}_{4,1}(\lam)= \Ng_{4}(\lam)+ \Og^{(4)}_{\Hg_2}(\lam^6\log\lam)
\eqn 
and $({\tMg}^{+}(\lam^4)+ Q)^{-1} 
= ({\bf 1}+A_2(\lam))^{-1}({\bf 1} + {\Ng}_{4,1}(\lam))^{-1}$. 
We expand $({\bf 1} + {\Ng}_{4,1}(\lam))^{-1}$ and use \refeq(A-inv-a) and \refeq(tN4). 
This gives \refeq(JN-a). 

(2) By virtue of \reflm(JN-3) (2), $\lam^2M_v (1+ A_2(\lam))^{-1}M_v$ 
is a GPR.   We may apply \reflm(T-K) (2) to 
$\lam^2 \Og_{\Hg_2}^{(4)}(\lam^4)$ 
to conclude the {proof.}   
\edpf   

We proceed following the strategy of \reflm(JN): We define 
\bqn 
B(\lam) \colon =Q-Q({\tMg}^{+}(\lam^4)+ Q)^{-1}Q  \lbeq(4-20)
\eqn 
and investigate $B(\lam)^{-1}$.  
Let 
\begin{align}   
& A_{2,m}(\lam)\colon = 
\lam^2 (g_1(\lam) \tG^{(v)}_2 + T_1(\lam)), \lbeq(r-s) \\
& T_1(\lam) = (\tM_U + \tG_{2,l}^{(v)}) - c_v^2\lam^2 M_{f(\lam)} + \lam^4 \tM_U^3 .
   \lbeq(T1-def)
\end{align}
Separating multiplication operators from those in $\Hg_2$, we also write    
\bqn \lbeq(A2m)
A_{2,m}(\lam) = \lam^2 M_{h(\lam)}+ N_2(\lam), \ \ 
h(\lam,x)=\tilde{U}(x) - c_v^2\lam^2 f(\lam,x)+\lam^4 \tilde{U}(x)^3.
\eqn 
For shortening formulas, we set 
\bqn \lbeq(AQ) 
{A}_Q (\lam)\colon 
=\lam^{-2} Q {A}_{2,m}(\lam)Q= Q(g_1(\lam) \tG_2^{(v)}+ T_1(\lam))Q .
\eqn

\bgdf We say that an operator valued function $A(\lam)$ of $\lam \in (0,a)$ 
is {\it of variable separable} if it is a finite sum $\sum_{j=1}^N f_j(\lam) A_j $ 
of products of operators $A_j$ which are $\lam$-independent and scalar functions 
$f_j(\lam)$,  $1\leq j \leq N$. 
\eddf 
 Operators of variable separable are easier to handle 
as $f_j(\lam) A$ is a GPR if $A \in \Lg^1$ and $f_j(\lam) = \lam^2 \k(\lam)$ 
with a GMU $\k(\lam)$. 

We first simplify the right of \refeq(4-20). 
 
\bglm  \lblm(BF5a)  We have 
\bqn 
B(\lam) = \lam^2 A_Q(\lam)+  F_4(\lam)+ \Og_{Q\Hg_2}^{(4)}(\lam^6(\log\lam)^3) , \lbeq(BF5) 
\eqn 
where $F_4(\lam)\in \Og_{Q\Hg_2}(\lam^4(\log\lam)^2)$ is of variable separable and is equal to 
\bqn \lbeq(BF5a)
F_4(\lam)= - Q(N_2(\lam)^2+ \lam^2 \tM_U N_2(\lam) + \lam^2 N_2(\lam) \tM_{U}-\lam^4\tG_4^{(v)} )Q. 
\eqn 
\edlm 
\bgpf 
Substitute \refeq(JN-a) for $ ({\tMg}^{+}(\lam^4)+ Q)^{-1}$ in \refeq(4-20). 
We have 
\begin{align*}  
& B(\lam) = Q- Q({\bf 1}+A_2(\lam))^{-1}Q + Q\Ng_4(\lam)Q+ 
\Og_{Q\Hg_2}^{(4)}(\lam^6\log\lam) \notag \\
& = Q(A_2(\lam) -A_2(\lam)^2  (1+ A_2(\lam))^{-1} + \lam^4 \tG_4^{(v)})Q  +
\Og_{Q\Hg_2}^{(4)}(\lam^6\log\lam). 
\end{align*} 
By using \refeq(A2lam), \refeq(A-inv)  and that $f(\lam,x) = 1 + \Og_{L^\infty}(\lam^2)$ 
we see that $-Q(A_2 (\lam)^2 (1+ A_2(\lam))^{-1}- \lam^4 \tG_4^{(v)})Q$ is equal to 
\begin{align*}
& - Q(\lam^2\tM_U + N_2(\lam))^2 M_{f(\lam)} - \lam^4 \tG_4^{(v)})Q+  
\Og_{Q\Hg_2}(\lam^6(\log\lam)^3) 
\\ 
& = -  \lam^4 QM_{c_v^2 f(\lam)}Q+ \lam^6 QM_{\tilde{U}}^3Q +  F_4(\lam)  +
\Og_{Q\Hg_2}(\lam^6(\log\lam)^3).
\end{align*}
Including $- \lam^4 QM_{c_v^2 f(\lam)}Q + \lam^6 QM_{\tilde{U}}^3 Q$ in 
$\lam^2 A_Q(\lam)$, 
we obtain the lemma.  
\edpf 

\subsection{Resonances, regular and singular at zero} 

The coefficient of the leading term $\lam^2 g_1(\lam) Q \tG_2^{(v)} Q$ 
of \refeq(BF5) or \refeq(AQ)  is of rank two: 
\begin{gather} \lbeq(QG2vQ)
QG_2^{(v)}Q = \frac12 (\ph_1 \otimes \ph_1 + \ph_2 \otimes \ph_2), \\ 
\ph_j(x) = (Q M_{x_j}v)(x)= (x_j - c_j)v(x), \quad 
c_j= (\tv, x_j \tv)   \lbeq(phj) 
\end{gather}
and the behavior of $B(\lam)^{-1}$ as $\lam \to 0$ depends 
on the second term $T_0 = M_U + G_{2,l}^{(v)}$ on the space ${\Ker} QG_2^{(v)}Q$ 
in \refeq(AQ).   
Let $S_0$ be the projection onto ${\Ker} Q G_2^{(v)} Q\vert_{QL^2}$. 
The following lemma is evident.  

\bglm[\cite{LSY}]\lblm(S-0)  Let $\ph_1(x)$ and $\ph_2(x)$ be defined by \refeq(phj).
\begin{align} \lbeq(S-0)
{\Ker} Q G_2^{(v)} Q\vert_{L^2}& ={\Ker} Q G_2^{(v)} Q\vert_{QL^2}\oplus \C{v}, \\
{\Ker} Q G_2^{(v)} Q\vert_{QL^2}& 
= \{u \in QL^2 \colon  u \perp \ph_j , j=1,2\} \\
& = \{u \in \HL \colon \la M_{x^\a} v, u \ra=0, \ |\a| \leq 1 \}. \lbeq(S0-a) 
\end{align} 
\edlm 
By virtue of \reflm(S-0), $S_0^\perp=Q \ominus S_0$ is of rank two in $Q\Hs$:   
\[
S_0^\perp Q \Hs=\{a_1\ph_1+ a_2\ph_2\colon a_1, a_2 \in \C\}.
\]

Recall that we say $H$ is {regular} (or singular) at zero if $\Ng_\infty(H)=\{0\}$ 
(or $\Ng_\infty(H)\not=\{0\}$) (see \refeq(reso-space)).  
The following proposition is stated and proved in \S 5 of \cite{LSY}.  
We shall present a new proof in Appendix.  

\bgprop \lbprop(resonances)
{\rm (1)} The operator $H$ is {\it regular}  at zero if and only if 
$S_0 T_0 S_0 \vert_{S_0\HL}$ is non-singular.   
\vspace{0.2cm}

\noindent 
{\rm (2)}  If $H$ is {\it singular }  at zero and    
$f \in \Yg(H) \colon = \Ker (S_0 T_0 S_0)\vert_{S\Hs}\setminus \{0\}$. Then,  
for a unique $c_f \colon =(c_0,c_1,c_2)\in \C^3$  
\bqn \lbeq(8) 
(M_U + G_{2,l}^{(v)}) f = (c_0 +  c_1 x_1 + c_2 x_2)v(x)
\eqn
and 
\bqn \lbeq(9) 
\Phi(f)(x)\colon = G_{2,l}^{(v)}f(x) -(c_0 +  c_1 x_1 + c_2 x_2 )v(x)\in \Ng_\infty(H).
\eqn  
The map $\Phi \colon \Yg(H) \to \Ng_\infty(H)$ defined by \refeq(9) is an onto isomorphism 
and the inverse is given by $\Phi^{-1}\ph= M_U M_v \ph$. 
\edprop 

{In what follows we always assume $H$ is regular at zero}.  Let 
\bqn \lbeq(D0-0)
D_0 \colon = (S_0 T_0 S_0 \vert_{S_0\Hs})^{-1}. 
\eqn 
We compute $B(\lam)^{-1}$ for small $0<\lam<a$. In view of \refeq(BF5) 
we first do it for $A_Q(\lam)$. 
We decompose as 
\bqn \lbeq(de)
Q\Hs = S_0^\perp \Hs \oplus S_0 \Hs, \quad S_0^\perp= Q\ominus S_0
\eqn 
and apply Feshbach formula. 
In this decomposition, since  $Q\tilde{G}_{2 }^{(v)}Q S_0 = S_0 Q\tilde{G}_{2}^{(v)}Q=0$,  
${A}_Q(\lam) $ becomes the operator matrix:
\bqn \lbeq(N+)
{A}_Q(\lam) = 
\begin{pmatrix} a_{11} & a_{12} \\ a_{21} & a_{22} \end{pmatrix}
= \colon 
\begin{pmatrix} S_0^\perp A_Q (\lam)S_0^\perp 
& S_0^\perp T_1(\lam) S_0 \\
S_0 T_1(\lam)S_0^\perp & S_0 T_1(\lam) S_0  
\end{pmatrix}.
\eqn 
For $a_{22}=S_0 T_1(\lam) S_0$, we have the following lemma. 

\bgdf  For a Banach space $\Xg$ and $a>0$,  $C^\infty_{b,a}(\Xg)$  is 
the space of $\Xg$-valued  $C^\infty$ functions of $\lam\in [0,a)$   
which are bounded along with derivatives of any order.
\eddf 

\bglm  \lblm(tD1)  For $0<\lam<a$, 
$S_0T_1(\lam) S_0$ is invertible in $S_0 \HL$ and    
\bqn \lbeq(tD1)
(S_0T_1(\lam) S_0)^{-1} = S_0 M_{h(\lam)^{-1}}S_0 + X(\lam), 
\quad X(\lam) \in C^\infty_{b,a} (S_0 \Hg_2).
\eqn 
We denote 
\bqn \lbeq(D1)
D_1(\lam) \colon = (S_0T_1(\lam) S_0)^{-1}. 
\eqn 
\edlm 
\bgpf 
By the assumption $S_0\tT_0 S_0$ is invertible in 
$S_0 \Hs$ and clearly so is $S_0T_1(\lam)S_0$. 
We show \refeq(tD1). Let 
\bqn \lbeq(s0)
D_1 (\lam) \a = \b  \quad  \mbox{or}\quad    S_0T_1(\lam) S_0 \b = \a, \quad \a, \b \in S_0 Q\HL.
\eqn 
Let  $\Phi=P\oplus S_0^\perp = {\bf 1} \ominus S_0$  
and $\Psi(\lam)= \Phi T_1(\lam) S_0 D_1(\lam)$. Then, 
${\rm rank }\, \Psi(\lam) \leq 3$,  hence $\Psi(\lam)\in C^\infty_{b,a}(\Hg_2)$ 
and, \refeq(s0) implies 
\[
T_1(\lam) S_0 \b = \a+ \Phi T_1(\lam) S_0 \b
= \a + \Phi T_1(\lam) S_0 D_1(\lam) \a 
=\a + \Psi(\lam)\a. 
\] 
It follows that 
\[
M_{h(\lam)} S_0 \b 
= \a + \Psi(\lam) \a - \tG_{2,l}^{(v)}S_0 \b  
= \a + \Psi(\lam) \a - \tG_{2,l}^{(v)}S_0 D_1(\lam) \a .
\]
Let 
$X_1(\lam) 
=  M_{h(\lam)^{-1}}(\Psi(\lam)- \tG_{2,l}^{(v)}S_0 D_1(\lam))$.  
Then, $X_1(\lam) \in C^\infty_{b,a} (\Hg_2)$ and 
\[
S_0 \b= M_{h^{-1}(\lam)}\a+ X_1 (\lam) \a .
\]
Hence $\b= S_0 M_{h^{-1}(\lam)}S_0 \a+ X(\lam)\a$, 
$X(\lam)\colon =S_0 X_1(\lam)S_0 \in C^\infty_{b,a}(S_0\Hg_2)$ as desired. 
\edpf   

Since $a_{22}^{-1}= D_1(\lam)$,  $a_{11}- a_{12}a_{22}^{-1} a_{21} $ is equal to 
\begin{align}
& S_0^\perp (g_1(\lam)\tG_2^{(v)}+ T_1(\lam))S_0^\perp - 
S_0^\perp T_1(\lam) S_0 D_1(\lam) 
S_0 T_1(\lam) S_0^\perp \notag \\
& = g_1(\lam) (S_0^\perp \tG_2^{(v)} S_0^\perp + g_1(\lam)^{-1}F_3(\lam)), 
\lbeq(d-0)
\end{align} 
where $F_3(\lam) \in C^\infty_{b,a}(\Bb(S_0^\perp\Hs))$ is given by 
\bqn \lbeq(F3)
F_3(\lam) \colon = S_0^\perp (T_1(\lam)  - 
T_1(\lam) S_0 D_1(\lam) S_0 T_1(\lam) )S_0^\perp. 
\eqn 
By the definition of $S_0^\perp$,  it is clear that 
\bqn \lbeq(F2) 
(S_0^\perp \tG_2^{(v)} S_0^\perp)^{-1}=\colon F_2
\eqn 
exists in $S_0^\perp \Hs$ and 
$a_{11}- a_{12}a_{22}^{-1} a_{21} $ has the inverse  
\bqn 
d(\lam) = g_1(\lam)^{-1}F_2 (S_0^\perp + g_1(\lam)^{-1} F_3(\lam) F_2)^{-1}.
\lbeq(d-inv)
\eqn 
Let  $e(\lam): S_0^\perp \Hs \to S_0 \Hs$ and ${}^t e(\lam) \colon S_0 \Hs \to S_0^\perp \Hs$ 
be defined by 
\bqn  
e(\lam) \colon = D_1(\lam) S_0 T_1(\lam)S_0^\perp,  \ \  
{}^t e(\lam) \colon = S_0^\perp  T_1(\lam)S_0 D_1(\lam)   \lbeq(e)
\eqn 
and let, in the decomposition \refeq(de) of $Q\Hs$,   
\bqn 
F_{mat}(\lam)\colon  = \begin{pmatrix} d(\lam) &  - d(\lam){}^t e(\lam)  \\   
- e(\lam) d(\lam)    & e(\lam) d(\lam) {}^t e(\lam)  
\end{pmatrix}.   \lbeq(Fmat)
\eqn  
We have  
\bqn \lbeq(Fmat-rem)
{\rm rank}\, F_{mat}(\lam) \leq 4 \ \mbox{and} \  F_{mat}(\lam) = \Og((\log \lam)^{-1}).
\eqn  
The first part of the next lemma follows via Feshbach formula.  

\bglm \lblm(Alam-inv)  For $0<\lam<a$, ${A}_Q(\lam)$ is invertible in $Q\Hs$ and 
\bqn  \lbeq(Alam-inv)
A_Q(\lam)^{-1}= F_{mat}(\lam) + D_1(\lam); 
\eqn 
$B(\lam)$ is invertibe in $Q\Hs$ and   $B(\lam)^{-1} $ is equal to 
\bqn   
\lam^{-2} A_Q(\lam)^{-1}   - 
 \lam^{-4} A_Q(\lam)^{-1}F_4(\lam)A_Q(\lam)^{-1}  
+ \Og_{Q\Hg_2}^{(4)}(\lam^2(\log \lam)^4),  \lbeq(B-inv) 
\eqn 
where 
$F_4(\lam) \in \Og_{Q\Hg_2}^{(4)}(\lam^4(\log \lam)^2) $ is  defined  in \refeq(BF5a). 
\edlm 
\bgpf We prove the second statement only.  
Since $A_Q(\lam)^{-1}$ exists, we have from \refeq(BF5) that  
\[
B(\lam)= \lam^2 (Q+(F_4(\lam)+ \Og_{Q\Hg_2}(\lam^6(\log\lam)^3))
(\lam^2 A_Q(\lam))^{-1}) A_Q(\lam).
\]
Hence, $B(\lam)$ is invertible in $Q\HL$ and   $B(\lam)^{-1}$ is equal to 
\bqn \lbeq(B-inv-1)
 \lam^{-2}A_Q(\lam)^{-1}
(Q + \lam^{-2} F_4(\lam)A_Q(\lam)^{-1})^{-1} + \Og_{Q\Hg_2}(\lam^2 (\log\lam)^3).
\eqn 
Expanding the first term of \refeq(B-inv-1),  we obtain \refeq(B-inv) since 
\[
\lam^{-2} A_Q(\lam)^{-1}(\lam^{-2}F_4(\lam)A_Q(\lam)^{-1})^2 
(Q + \lam^{-2} F_4(\lam){A}_Q(\lam)^{-1})^{-1}
\]
is of class $\Og_{Q\Hg_2}^{(4)}(\lam^2(\log\lam)^4)$.
\edpf

\section{Simplifying $\Qg_v(\lam)$ modulo GPR}  
 
Since $B(\lam)^{-1}$ exists,  \reflm(JN) implies that $\tMg(\lam^4)^{-1}$is equal to 
\[ 
 (\tMg(\lam^4)+Q)^{-1}+(\tMg(\lam^4)+Q)^{-1}QB(\lam)^{-1}Q(\tMg(\lam^4)+Q)^{-1}
\] 
and $\Qg_v(\lam) = c_v \lam^2 M_v \tMg(\lam^4)^{-1}M_v$ to   
\begin{multline}  
\Qg_v(\lam) = c_v \lam^2 M_v (\tilde{\Mg}(\lam^4)+Q)^{-1}M_v \\
+ c_v \lam^2 M_v (\tilde{\Mg}(\lam^4)+ Q)^{-1}Q
B(\lam)^{-1}Q (\tilde{\Mg}(\lam^4)+Q)^{-1}M_v.  \lbeq(inv-M)
\end{multline}
We prove \refth(regular-case) by showing that \refeq(inv-M) is a GPR. For doing so, 
we first simplify the right of \refeq(inv-M) by removing terms which are ``easily'' shown to be 
GPRs.  In what follows  
\[
X(\lam)\equiv Y(\lam)  \ \mbox{means } \ X(\lam) - Y(\lam) \ \mbox{is a GPR (for small $\lam>0$)}.
\]   

By virtue of \reflm(JN-3) 
$c_v \lam^2 M_v (\tilde{\Mg}(\lam^4)+Q)^{-1} M_v$ is a GPR and we have 
$\Qg_v(\lam) \equiv \Cg(\lam)$ where, ignoring the unimportant constant $c_v$ 
on the right side,
\bqn \lbeq(step-1)
\Cg(\lam)\colon= \lam^2 
M_v (\tilde{\Mg}(\lam^4)+ Q)^{-1}Q B(\lam)^{-1}
Q (\tilde{\Mg}(\lam^4)+Q)^{-1} M_v .
\eqn 

\subsection{Simplification via \reflmb(T-K)}  We first remove GPRs 
from $\Cg(\lam)$ which satisfy conditions of \reflm(T-K) or \refcor(T-K).

We recall that $S_0^\perp \Hs={\rm span}\{\ph_1, \ph_2\}$ (cf. \refeq(phj)), hence, 
$\Bb(S_0^\perp \Hs)\subset \Hg_2$.   
We write $\Og_{S_0^\perp\Hs}$ for $\Og_{\Bb(S_0^\perp\Hs)}$ for shortening formulas.  
Let 
\begin{align} 
& F_{30}= 
S_0^\perp (\tT_0- \tT_0(S_0\tT_0 S_0)^{-1}\tT_0)S_0^\perp,  \lbeq(F3-detail) \\
& d_0(\lam)= g_1(\lam)^{-1} F_2 (S_0^\perp + g_1(\lam)^{-1}F_{30}F_2)^{-1}, \lbeq(d0-def) \\  
& e_0 = (S_0 T_0 S_0)^{-1}(S_0 T_0 S_0^\perp).   \lbeq(e-0) 
 \end{align}

\bglm  \lblm(d-e) 
For $d(\lam)$ and $e(\lam)$ we have 
\begin{align}
& d(\lam)=d_0(\lam) + \Og_{S_0^\perp\Hs}(\lam^2 (\log\lam)^{-2}) ,  \lbeq(d-e)\\ 
& e(\lam) = e_0+ \Og_{\Hg_2(S_0\Hs, S_0^\perp \Hs)}(\lam^2).   \lbeq(e-e)
\end{align} 
\edlm 
\bgpf We have $T_1(\lam)= \tT_0 + \Og_{\Bb(\Hs)}(\lam^2) $ and 
$D_1(\lam)  =(S_0\tT_0 S_0)^{-1} + \Og_{\Bb(S_0\Hs)}(\lam^2)$. 
It follows $F_3(\lam) = F_{30}+ \Og_{S_0^\perp \Hs}(\lam^2)$ and  
\begin{align} 
d(\lam) & = g_1 (\lam)^{-1}F_2 (S_0^\perp + g_1(\lam)^{-1}F_{30}F_2 + 
\Og_{S_0^\perp\Hs}(\lam^2 (\log\lam)^{-1}))^{-1}  \notag \\
&  = g_1 (\lam)^{-1}F_2 (S_0^\perp + g_1(\lam)^{-1}F_{30}F_2)^{-1} + 
\Og_{S_0^\perp\Hs}(\lam^2 (\log\lam)^{-2}) .  \lbeq(d-s)
\end{align}
 We omit the similar proof of  \refeq(e-e).  
\edpf 

We shall proceed step by step.  
\paragraph{ \bf Step 1}  We substitute \refeq(JN-a) for $(\tilde{\Mg}(\lam^4)+Q)^{-1}$'s 
in $\Cg(\lam)$ of \refeq(step-1). Then, since  
$B(\lam)^{-1}\in  \Og_{\Bb(Q\HL)}(\lam^{-2})$,  
$-\Ng_{4}(\lam)+ \Og_{\Hg_2}^{(4)}(\lam^6\log\lam)$ of \refeq(JN-a) 
produces $\Og_{\Lg^1}^{(4)}(\lam^4)$ for $\Cg(\lam)$ by \reflm(JN-3), 
which  is a GPR by \refcor(T-K) (3).  Hence,    
\bqn \lbeq(step-2)
{\Cg}(\lam) \equiv \lam^2M_v ({\bf 1}+ {A}_2(\lam))^{-1}QB(\lam)^{-1}Q 
({\bf 1}+ {A}_2 (\lam))^{-1}M_v.
\eqn 

\paragraph{ \bf Step 2} We next substitute \refeq(B-inv) for $B(\lam)^{-1}$ in \refeq(step-2). 
Then,  
\[
\lam^2M_v ({\bf 1}+ {A}_2(\lam))^{-1}\Og_{Q\Hg_2}^{(4)}(\lam^2 (\log \lam)^4)
({\bf 1}+ {A}_2 (\lam))^{-1}M_v \in \Og_{\Lg^1}^{(4)}(\lam^4 (\log \lam)^4).
\]
is a GPR as in Step 1 and  we may replace $B(\lam)^{-1}$ in \refeq(step-2) by 
\bqn  \lbeq(B2)
B_2(\lam)\colon =Q(A_Q(\lam)^{-1} - 
\lam^{-2} A_Q(\lam)^{-1}F_4(\lam)A_Q(\lam)^{-1})Q.
\eqn 
to produce
\bqn   \lbeq(step-3)
\Cg(\lam) \equiv M_v ({\bf 1}+ {A}_2 (\lam))^{-1}B_2(\lam)
({\bf 1}+ {A}_2(\lam))^{-1}M_v. 
\eqn 

\paragraph{ \bf Step 3}  We then replace $({\bf 1}+ {A}_2(\lam))^{-1}$'s  
in \refeq(step-3) by \refeq(A-inv-a).  
Since \refeq(step-3) with one of the $({\bf 1}+ {A}_2(\lam))^{-1}$ being replaced 
by $\Og_{\Hg_2}(\lam^4(\log\lam)^2)$ is equal to 
$\Og_{\Lg^1}(\lam^4 \log\lam)$, which is a GPR by \refcor(T-K) (3), this produces 
\bqn \lbeq(step-4-1)
\Cg(\lam)  \equiv M_v ({\bf 1}- A_{2}(\lam))
B_2(\lam)({\bf 1}- A_{2}(\lam))M_v . 
\eqn 

\paragraph{ \bf Step 4}  Expand the right side of \refeq(step-4-1).  Then, since 
$S_0 M_{h(\lam)^{-1}}S_0 - M_{h(\lam)^{-1}}$ is 
of finite rank,  we see by virtue of \refeqss(A2,tD1,Alam-inv) that 
$M_v A_{2} (\lam)B_2(\lam)A_{2}(\lam)M_v$ 
is equal to the sum of the multiplication with $c_v^{-2}\lam^4 h(\lam,x)^{-1} \in \Og_{L^1(\R^2)}(\lam^4)$ 
and $\Og_{\Lg^1}(\lam^4\log^2\lam)$ both of which are GPRs.  This reduces \refeq(step-4-1) to   
\bqn  \lbeq(step-4)
\Cg(\lam) \equiv  
M_v(B_2(\lam) - A_{2}(\lam)B_2(\lam)- B_2(\lam)A_{2}(\lam)) M_v .
\eqn  

\paragraph{ \bf Step 5}  We next substitute \refeq(B2) for last two $B_2(\lam)$s  
in \refeq(step-4). 
Then,   
$\lam^{-2} A_Q(\lam)^{-1}F_4(\lam)A_Q(\lam)^{-1}A_{2}(\lam)$ 
and 
$\lam^{-2} A_{2}(\lam)A_Q(\lam)^{-1}F_4(\lam)A_Q(\lam)^{-1}$,  
 when sandwiched by $M_v$,
become operators of class $\Og_{\Lg^1}(\lam^4(\log \lam)^4)$ which are GPRs. 
This simplifies \refeq(step-4) to    
\bqn \lbeq(step-4a)
\Cg(\lam)  \equiv 
M_v(B_2(\lam) - A_{2}(\lam)A_Q(\lam)^{-1}- A_Q(\lam)^{-1}A_{2}(\lam)) M_v.
\eqn 
\paragraph{ \bf Step 6}  Substitute \refeq(Alam-inv) for $A_Q(\lam)^{-1}$ in \refeq(step-4a) 
and expand the result.  The following lemma implies   
\bqn \lbeq(step-4b)
\Cg(\lam)  \equiv 
M_v(B_2(\lam) - A_{2}(\lam)D_1(\lam)- D_1(\lam)A_{2}(\lam)) M_v. 
\eqn 

\bglm \lblm(A2mF)  Both 
$M_v A_{2}(\lam)F_{mat}(\lam)M_v$ and $M_v F_{mat}(\lam) A_{2}(\lam))M_v$ 
are GPRs. 
\edlm 
\bgpf  We prove the lemma for 
$M_v A_{2}(\lam)F_{mat}(\lam)M_v$. The proof for the other is similar. 
By virtue of \reflm(d-e) 
\begin{align}
& F_{mat}(\lam)= F^{(0)}_{mat}(\lam)+\Og_{\Hg_2}(\lam^2),  \lbeq(F-simp)  \\
& F_{mat}^{(0)}(\lam) \colon = 
\begin{pmatrix}  d_0(\lam)    & - d_0(\lam) {}^te_0 \\
 -e_0 d_0(\lam)   &  -e_0 d_0(\lam) {}^t e_0 \end{pmatrix}  
\lbeq(F-simp-0)  
\end{align} 
and  $M_vA_{2}(\lam) \Og_{\Hg_2}(\lam^2) M_v\in \Og_{\Lg^1}(\lam^4 \log\lam)$ 
is a  GPR. 
From \refeq(d0-def)  we learn that $d_0(\lam)$ is a holomorphic function of  $g_1(\lam)^{-1}$ in 
a neighborhood of $\lam=0$, $d_0'(\lam) \in \Og_{S_0^\perp\Hs}( \lam^{-1}g(\lam)^{-2})$ 
and $d''_0(\lam)\in \Og_{S_0^\perp\Hs}( \lam^{-2}g(\lam)^{-2})$.   It follows that 
$M_v A_{2}(\lam)F_{mat}^{(0)}(\lam)M_v$ 
is also a GPR by \reflm(T-K) (2).  This proves the lemma.
\edpf

From \refeq(step-4b) we deduce the following final result of this subsection. 

\bglm \lblm(step-5)  Modulo GPR $\Cg(\lam)$ is equal to   
\begin{multline} \lbeq(step-5)
M_v(A_Q(\lam)^{-1} - 
	\lam^{-2}F_{mat}(\lam) F_4(\lam) D_1(\lam) - 
	\lam^{-2} D_1(\lam) F_4(\lam) F_{mat}(\lam)  \\
	- \lam^{-2} D_1(\lam) F_4(\lam)D_1 (\lam)
	- A_{2}(\lam)D_1(\lam)- D_1(\lam)A_{2}(\lam))M_v.
\end{multline}
\edlm 
\bgpf  We substitute \refeq(Alam-inv) for both $A_Q(\lam)^{-1}$'s  in 
the second term on the right of \refeq(B2), expand the resulting formula 
and, then substitute the result for $B_2(\lam)$ in \refeq(step-4).  We shall be 
done if  we show that 
\[
R(\lam) \colon = \lam^{-2} M_v F_{mat}(\lam)F_4(\lam)F_{mat}(\lam)M_v
\]
is a GPR. Since $F_4(\lam)\in \Og_{\Hg_2(Q\Hs)}(\lam^4(\log\lam)^2)$, 
\refeq(F-simp) implies that   
\bqn \lbeq(MvFm)
R(\lam) \equiv  \lam^{-2} M_v F_{mat}^{(0)}(\lam)F_4(\lam)F_{mat}^{(0)}(\lam)M_v .
\eqn 
By virtue of \refeq(n4) and \refeq(BF5a) 
\[
F_4(\lam)= \lam^4 H_0 + \lam^4 g_1 (\lam) H_1 + \lam^4 g_1(\lam)^2 H_2,  
\ \ H_0,H_1, H_2  \in Q\Hg_2 Q, 
\]
which we substitute in \refeq(MvFm). 
Then, thanks to the factor $g_1(\lam)^{-1}$ in $d_0(\lam)$ (see \refeqs(d0-def,F-simp-0)),   
\[
\lam^{2} M_v F_{mat}^{(0)}(\lam)(H_0 + g_1(\lam) H_1) F_{mat}^{(0)}(\lam)M_v 
\in \Og_{\Lg^1}(\lam^2(\log \lam)^{-1})
\]
and it is a GPR by \refcor(T-K) (3).  To show the same for 
\bqn \lbeq(lamMvg)
\lam^{2} g_1(\lam)^2 M_v  F_{mat}^{(0)}(\lam)  H_2 F_{mat}^{(0)}(\lam)M_v
\eqn 
we note that,  with another $\Bb(S_0^\perp\Hs)$-valued holomorphic function of 
$g_1(\lam)^{-1}$ which we denote by $F_3(\lam)$ 
\[
g_1(\lam) d_0(\lam) = F_2 (S_0^\perp + g_1(\lam)^{-1}F_{30}F_2)^{-1}= F_2+ g_1(\lam)^{-1} F_3(\lam)
\]
and  $F_{mat}^{(0)}(\lam)$ is equal to the sum 
\begin{align*}
& g_1(\lam)^{-1} \begin{pmatrix}  F_2   & - F_2 {}^t e_0 \\
 -e_0 F_2   &  - e_0 F_2 {}^t e_0 \end{pmatrix}  
+ g_1(\lam)^{-2}
\begin{pmatrix}  F_3(\lam)   & - F_3(\lam) {}^t e_0 \\
 -e_0 F_3(\lam)   &  - e_0 F_3(\lam) {}^t e_0 \end{pmatrix} \\
& = \colon I_0 + I_1(\lam). 
\end{align*}
Note that $I_0$ is $\lam$-independent  and is of finite rank.  
Thus,  the operator  \refeq(lamMvg) 
with two $F_{mat}^{(0)}(\lam)$'s being replaced  by $I_0$ 
is equal to $\lam^2 \Lg^1$ which is a GPR by \reflm(T-K) (1);  the ones 
with at least one of $F_{mat}^{(0)}(\lam)$ is replaced  
by $I_1(\lam)$ are of $\Og_{\Lg^1}(\lam^2 (\log\lam)^{-1})$ 
and they are GPRs by \refcor(T-K) (3).  Thus, $R(\lam)$ is a GPR and the proof is 
completed. 
\edpf

\subsection{Simplification via cancellations} 
We further simplify $\Cg(\lam)$ by using the cancellation property of $Q$ and $S_0$:
\bqn \lbeq(cancel) 
Q v=0, \quad S_0 v= S_0 (x_1v) = S_0 (x_2v)=0.
\eqn 
Let $R_j$, $j=1,2$, be Riesz transforms:   
\[
(R_j u)(x)= \frac1{(2\pi)^2} \int_{\R^2}e^{ix\xi}(i \xi_j/|\xi|) \hat{u}(\xi) 
d\xi, 
\] 
which are GOPs. The next lemma shows that 
the multiplications by  $Q$ and $S_0$ {\it from the left} to $M_v \Pi(\lam)$    
produce factors $\lam$ and $\lam^2$ respectively.

\bglm \lblm(cancel) Let $u \in \Dg_\ast$. Then: 
\begin{align} 
& Q M_v \Pi(\lam)u(x)= \lam \sum_{j=1,2} \int_0^1 
Q M_{x_j v} (\Pi(\lam)\tau_{-\th{x}}R_j u)(0)d\th, 
\lbeq(QMv)  \\
& S_0 M_v \Pi(\lam) u(x) \notag \\
& \hspace{1cm} = 
\lam^2 \sum_{j,k=1,2}  \int_0^1 (1-\th) S_0 M_{x_j x_k v} (\Pi(\lam)
\tau_{-\th{x}}R_j R_k u)(0)d\th. 
\lbeq(SQMv) 
\end{align} 
\edlm 
\bgpf By Taylor's formula     
\begin{align}
& \int_{{\mathbb S}}e^{i\lam{x}\w}\hat{u}(\lam\w) d\w 
= \int_{{\mathbb S}}\left(1+ i\lam{x}\w\int_0^1 
e^{i\th \lam{x}\w}d\th \right)\hat{u}(\lam\w)d\w  \lbeq(1st) \\
& = \int_{{\mathbb S}}
\left(1+ i\lam{x} \w + (i\lam{x\w})^2 
\int_0^1 (1-\th)e^{i\th\lam{x}\w}d\th\right)  \lbeq(2nd)
\hat{u}(\lam\w) d\w. 
\end{align}
Since $QM_v 1=0$ and $i \w_j \hat{u}(\lam\w)= ({\Fg}R_j u)(\lam\w)$ 
and, since 
\[
\int_{{\mathbb S}}e^{i\th \lam{x}\w}\hat{u}(\lam\w)d\w
=
\int_{{\mathbb S}}\Fg(\tau_{-\th {x}}{u})(\lam\w)d\w
= 2\pi\lam^2 \Pi(\lam)(\tau_{-\th{x}} u)(0), 
\] 
\refeq(1st) implies \refeq(QMv).  Eqn. \refeq(SQMv) follows from \refeq(2nd) 
by virtue of the cancellation properties \refeq(cancel)  and  
$- \w_i \w_j \hat{u}(\lam\w)= ({\Fg}R_i R_j  u)(\lam\w)$. 
\edpf 

Next lemma is a result of the combination of \reflms(WT-trans,cancel). Recall 
that $T^{(v)}= M_v T M_v$,  $T^{(v)}(\lam) =M_v T(\lam) M_v$ etc. 

\bglm \lblm(cancel-a)
Let $\k(\lam)$ be a GMU and $\m_j(\lam) = \lam^j \k(\m)$ for $j=1, 2$. 
Then, for $u \in \Dg_\ast$, we have the following statements:  

\noindent 
{\rm (1)} Suppose $T^{(v)}\in \Lg^1$ and $T= TQ$, then $\W(T^{(v)})\k(|D|)u(x)$ is equal to 
the superposition by 
$\sum_{j,k=1,2} \int_0^1  d\th $ of 
\bqn \lbeq(QS) 
\int_{\R^4} z_j T^{(v)} (y,z) 
\tau_y (K \m_1(|D|) \tau_{-\th{z}}R_j u)(x) dy dz.
\eqn  

\noindent 
{\rm (2)} Suppose  $T^{(v)}\in \Lg^1$ and $T= TS_0$, then  
$\W(T^{(v)})\k(|D|)u(x)$ is equal to the superposition by 
$\sum_{j,k=1,2} \int_0^1 (1-\th) d\th $ of 
\bqn \lbeq(QSa) 
\int_{\R^4} z_j z_k T^{(v)}(y,z) 
\tau_y (K \m_2(|D|) \tau_{-\th{z}}R_j R_k u)(x) dy dz .  
\eqn  

\noindent
{\rm (3)}  For $\tW(T^{(v)}(\lam))$  extensions of statements {\rm (1) and (2)}  
hold with obvious modifications via \refeq(int-parts) in the proof of 
\reflm(T-K) {\rm (2)}. 
\edlm

\bgrm  \lbrm(cancel)  
{\rm (1)}  
\reflm(cancel-a) implies that, if 
$T(\lam)=T(\lam)Q$ or $T(\lam)=T(\lam)S_0$, 
then  we may deal with $T^{(v)}(\lam) \in \Og^{(j)}_{\Lg^1}(f(\lam))$ in $\tW(T^{(v)}(\lam))$ 
as if it belongs to $\Og^{(j)}_{\Lg^1}(\lam f(\lam))$ or to $\Og^{(j)}_{\Lg^1}(\lam^{2}f(\lam))$ 
respectively, $j=0,1, \dots$.  

\noindent 
{\rm (2)} As is remarked in the proof below the integrals \refeq(QS), 
\refeq(QSa) and the ones in statement {\rm (3)}  
converge absolutely for a.e. $x \in \R^2$.  
\edrm 

\bgpf 
If $T= TQ$, then $T^{(v)}= M_v T Q M_v$ and \refeq(QMv) implies that  
\begin{align}
&\W(T^{(v)})\k(|D|)u(x)
= \int_0^\infty R_0^{+}(\lam^4)T^{(v)}(\lam) \Pi(\lam)u(x) \k(\lam) \lam^3 d\lam 	
\notag \\
&  = \sum_{j=1}^2 \int   
	\Rg(\lam,x-y)z_j T^{(v)}(y,z)(\Pi(\lam)\tau_{-\th{z}}R_j u)(0) \m_1(\lam) \lam^3 dX ,\lbeq(X-int)
\end{align}
where \refeq(X-int) is the iterative integral with respect to 
$dX= d\th dz dy d\lam$  over 
$(\th, z,y, \lam) \in (0, 1)_\th \times \R^2_z \times \R^2_y \times (0,\infty)_\lam$  
in this order, viz. $d\th$ first, $dz$ next  etc.   However, the proof of \reflm(meaning) 
shows that, for a.e.  $x \in \R^2$, 
the integral with respect to the $6$-dimensional measure $dX$ converges absolutely 
and the order of integrals may be changed freely. Hence,   
the right side of \refeq(X-int) is equal to $\sum_{j=1}^2 \int_0^1d\th $ of 
\[
\int_{\R^4} z_j T^{(v)}(y,z) 
\tau_y \left(\int_0^\infty \Rg(\lam,x)(\Pi(\lam)\tau_{-\th{z}}R_j \m_1(|D|)u)(0) 
\lam^3 d\lam\right) 	dy dz . 
\]
The function inside the parentheses is equal to 
$(K \tau_{-\th{z}}R_j \m_1(|D|)u)(x)$ and \refeq(QS) follows.  
Similar argument by using \refeq(SQMv) in place of 
\refeq(QMv) implies \refeq(QSa).  We omit the repetitive 
proof of statement (3). 
\edpf

\bglm \lblm(final) Modulo GPR we have that  
\bqn  
\Cg(\lam) \equiv 
M_v d(\lam)M_v - M_v e(\lam)d(\lam)M_v. 
\lbeq(step-8)
\eqn 
\edlm 
\bgpf We shall examine each term of \refeq(step-5) separately.    

(1) We recall that $F_4(\lam) = \Og_{Q\Hg_2}(\lam^4(\log\lam)^2)$, 
$F_{mat}(\lam)=F_{mat}(\lam)Q$ and $D_1(\lam)=D_1(\lam)S_0$. 
Then,  by \refrm(cancel)  we may consider 
\[
\lam^{-2}M_v (F_{mat}(\lam) F_4(\lam) D_1(\lam)+ 
D_1(\lam) F_4(\lam) F_{mat}(\lam) + D_1(\lam) F_4(\lam)D_1 (\lam))M_v
\]
as a member of $\Og_{\Lg^1}(\lam^3(\log\lam)^2)$ and, hence, 
as a GPR by statement (3) of \refcor(T-K).  

(2) Next we show $M_v A_{2}(\lam)D_1(\lam) M_v$ is also a GPR. 
By \refeq(A2) and \refeq(tD1),  it is equal to  
\bqn \lbeq(eqlm)
\lam^2 M_v (\tM_{U}+ \tG_{2,l}^{(v)}+ g_1(\lam)\tG_2^{(v)}) 
(S_0 M_{h(\lam)^{-1}}S_0 + X(\lam))M_v 
\eqn 
with $X(\lam) \in C_{b,a}^\infty(S_0\Hg_2)$.  We expand \refeq(eqlm). 
Then,  thanks to the factor $S_0 M_v$ on the right end,  
all terms may be considered as members of $\Og_{\Lg^1}(\lam^4\log\lam)$ 
which are GPRs except    
\begin{multline} \lbeq(mvs0)
\lam^2 M_v \tM_{U}S_0 M_{h(\lam)^{-1}}S_0M_v \\
= \lam^2 M_{c_v h(\lam,x)^{-1}v(x)U(x)}S_0 M_v   + \lam^2 M_{c_v  U(x)} S_0^\perp 
M_{h(\lam)^{-1}} S_0 M_v. 
\end{multline}
However, by virtue of $S_0 M_v$ on the right ends, \refeq(mvs0) may  also be 
considered as a GPR (recall that $\dim S^\perp \Hs=3$). 
Thus, $M_v A_{2}(\lam)D_1(\lam) M_v$ is a GPR. 

(3)  
Reversing the order of 
$A_{2}(\lam)$ and $D_1(\lam)$, we have  
\begin{multline} \lbeq(mvs0-reverse) 
M_v D_1(\lam)A_{2}(\lam)M_v \\
= \lam^2 M_v 
(S_0 M_{h(\lam)^{-1}}S_0 + X(\lam)) (\tM_{U}+ \tG_{2,l}^{(v)}+ g_1(\lam)\tG_2^{(v)})M_v.
\end{multline} 
Here there is no $S_0$ in front of the right-most $M_v$. Nevertheless 
\begin{align*}
& \lam^2 M_v (S_0 M_{h(\lam)^{-1}}S_0 + X(\lam)) (\tM_{U}+ \tG_{2,l}^{(v)})M_v \\
& \hspace{1cm} = \lam^2 M_{c_v h(\lam)^{-1}V}+ \lam^2 C^2([0,a). \Lg^1)
\end{align*}
is a GPR by virtue of \refcor(T-K) (2). We shall show that  the term 
$\lam^2 g_1(\lam)M_vD_1(\lam)\tG_2^{(v)}M_v$ which is produced by 
$g_1(\lam)\tG_2^{(v)}$ is also a GPR in the next lemma. 
Hence, 
\bqn \lbeq(step-6)
{\Bg}(\lam) \equiv M_v A_Q(\lam)^{-1}M_v.
\eqn 
(4) 
We  substitute \refeq(Alam-inv) for $A_Q(\lam)^{-1}$ in \refeq(step-6). 
Then, $D_1(\lam)$ and the second column of $F_{mat}(\lam)$ of \refeq(Fmat) 
produce GPRs by the same reason as above, since they carry $S_0$ on the right end. 
This proves the lemma. 
\edpf 

The following lemma finishes the  proof of \reflm(final). 

\bglm \lblm(kapp)
For $a>0$ small enough, the operator-valued function  
$\lam^2 g_1(\lam) M_v D_1(\lam) G_2^{(v)}M_v \chi_{\leq a}(\lam)$  is a GPR.    
\edlm 
\bgpf Ignoring the constant $c_v$, we consider $\lam^2 g_1(\lam) I(\lam)$ 
where $I(\lam) = M_v D_1(\lam)G_2^{(v)}M_v$.  Since $D_1(\lam)= S_0 D_1(\lam) S_0$ 
and $S_0 G_2^{(v)}= 4^{-1} S_0(x^2 v)\otimes v $,  we have 
$I(\lam) = m(\lam) \otimes  k$ where   
\bqn \lbeq(Ilam)
m(\lam,x)\colon =4^{-1} M_v D_1(\lam) (x^2 v), \ \ 
k(x)\colon = |V(x)|.
\eqn
By virtue of  \refeq(condi-1), \reflm(tD1) and \refeq(cancel),  we have 
\bqn \lbeq(kappa) 
m(\lam,x)\in C_{b,a}^\infty(\ax^{-4}L^1(\R^2)), \quad 
\int_{\R^2} m(\lam,x)dx =0 .
\eqn
Hence, we also have $\la x \ra m''(\lam,x)\in C_{b,a}^\infty(\ax^{-3}L^1(\R^2))$.

We shall show that 
\bqn 
W_{l}u \colon 
= \int_0^\infty R_0^{+}(\lam^4)  (m(\lam) \otimes k)\Pi(\lam)u 
\lam^5 g_1(\lam) \chi_{\leq a}(\lam)d\lam
\eqn 
is a GOP.   Substitute $\Pi(\lam)= \lam^{-2}\Pi_2(\lam)$ and 
$R_0^{+}(\lam^4) = (2\lam^2)^{-1}(G_0(\lam^2) - G_0(-\lam^2))$. 
We have   
$W_{l}u= (1/2)(W_{l}^{(1)}u- W_{l}^{(2)}u)$, where 
\begin{gather}  
W_{l}^{(1)}u
= \int_0^\infty G_0(\lam^2)(m(\lam)\otimes k) \Pi_2(\lam)u 
\lam g_1(\lam) \chi_{\leq a}(\lam)d\lam,  \lbeq(W1u)\\
W_{l}^{(2)}u = \int_0^\infty G_0(-\lam^2)
(m(\lam) \otimes k) \Pi_2(\lam)u 
\lam g_1(\lam) \chi_{\leq a}(\lam)d\lam.   \lbeq(W2u) 
\end{gather} 
We replace $m(\lam)$ by $m_{\leq a}(\lam)= \chi_{\leq {2a}}(\lam)m(\lam)$, 
which does not change $W_l^{(j)} u$, $j=1,2$, and denote $I_a(\lam) = m_{\leq a}(\lam) \otimes k$.   
By integration by parts 
\bqn \lbeq(Ilam-int)
I_a (\lam)= \int_{0}^\infty (\r-\lam)^{+} I_a''(\r)d\r.
\eqn

We first prove that $W_{l}^{(1)}$ is GOP.  
On substituting  \refeq(Ilam-int) in \refeq(W1u) and 
by changing the order of integration we have 
\[
W_{l}^{(1)}u= 
\int_0^\infty \left(\int_0^\infty G_0(\lam^2)(\r-\lam)_{+} I_a''(\r)\Pi_2(\lam)u 
\lam g_1(\lam) \chi_{\leq a}(\lam)d\lam \right) d\r.
\]
Since $(\r-\lam)_{+} \Pi_2(\lam)u= \r \Pi_2(\lam)(1-|D|/\r)_{+}u$ by \refeq(mult),  
the inner integral becomes 
\[
\Wg(\r)u \colon = \r \int_0^\infty G_0(\lam^2) I_a''(\r)\Pi_2(\lam)(1-|D|/\r)_{+}u 
\lam g_1(\lam) \chi_{\leq a}(\lam)d\lam .
\]
Here the Fourier transform of $(1-|\xi|)_{+}$ is integrable on $\R^2$ and 
\[
\|(1-|D|/\r)_{+}u\|_p\leq C_p \|u\|_p, \quad 1\leq p\leq \infty
\] 
with $C_p$ independent of $\r>0$;  
$I_a''(\r)= m_{\leq a}''(\r)\otimes k$ and by virtue of \refeq(kappa), 
\bqn  \lbeq(k)
\int_{\R^2} m_{\leq a}''(\r,x)dx =0 \ \ \mbox{and} \ \ \ax m_{\leq a}''(\r, x)\in L^1(\R^2).
\eqn  
It follows from  Proposition 1.4 (7) of 
\cite{Ya-2dim-new} that the logarithmic singularity of $g_1(\lam)$ 
at $\lam=0$ can be canceled and  
\[
\|\Wg(\r)u\|_p \leq C_p \chi_{\leq 2a}(\r) 
\|\ax{m}_{\leq a}''(\r)\|_1 \|k\|_1 \|u\|_p, \quad 1<p<\infty, 
\]
where we have used ${m}_{\leq a}''(\r)=0$ for $\r\geq 2a$ to insert $\chi_{\leq 2a}(\r) $ in 
the front of the right side.  We should remark the notation here  slightly differs 
from that in \cite{Ya-2dim-new}, e.g. $G_0(\lam)$ there is denoted by 
$G_0(\lam^2)$ here. Thus,  
\bqn \lbeq(W1-proof)
\|W_{l}^{(1)}u\|_p \leq C_p \left(\int_0^\infty 
\|\ax{m}_{\leq a}''(\r)\|_1 \|k\|_1\chi_{\leq 2a}(\r) d\r\right) \|u\|_p
\eqn 
and $W_l^{(1)}$ is a GOP. 

We next prove that 
$\chi_{\geq 4a}(|D|)W_{l}^{(2)}$ and $\chi_{\leq 4a}(|D|)W_{l}^{(2)}$ 
are GOPs, which of course implies that $W_{l}^{(2)}$ is a GOP.  
The argument above implies that it suffices to show this 
when $m(\lam)$ is independent of $\lam$.  

We omit the proof for $\chi_{\geq 4a}(|D|)W_{l}^{(2)}$ which is 
the repetition of the one of the first part of Lemma 3.8 of \cite{Ya-2dim-new} 
with $G_0(-\lam)$ and $K$ replacing $G_0(-\lam^2)$ and $\tK_2$ respectively   
and which uses the cancellation property \refeq(k) and \reflm(K-GOP) (1). 
We refer to \cite{Ya-2dim-new} for the details.

The following proof for $\chi_{\leq 4a}(|D|)W_{l}^{(2)}$ is  a modification of 
the one of the second part of Lemma 3.8 of \cite{Ya-2dim-new}.   
Since $m \in M_v S_0 \Hs$,  we have  
\bqn 
\hat{m}(\xi) = - \sum_{j,k=1}^2 \xi_j \xi_k \int_0^1 
(1-\th) \left(\int_{\R^2}  e^{-i\th{x}\xi} x_j x_k m(x) dx\right) d\th 
\eqn 
and $\chi_{\leq 4a}(|D|)G_0(-\lam^2)m(x)$ is equal to 
\begin{align*}
&  \sum_{j,k=1}^2 \frac{-1}{2\pi}\int_{\R^2}
\left(\int_{\R^2}\int_0^1 (1-\th) \frac{\chi_{\leq 4a}(|\xi|)e^{i\xi(x-\th{y})}\xi_j\xi_k y_j y_k  m(y)}{|\xi|^2+ \lam^2}dy d\th\right)
d\xi   \notag \\
& = \sum_{j,k=1}^2 \int_0^1 (1-\th) d\th 
\int_{\R^2} y_j y_k {m}(y)\t_{\th{y}}
\left(\frac{-1}{2\pi} \int_{\R^2} 
\frac{\chi_{\leq 4a}(|\xi|)e^{i\xi{x}}\xi_j\xi_k }{|\xi|^2+ \lam^2}d\xi \right)dy.   
\end{align*}
Using the obvious identity 
$|\xi|^2 (|\xi|^2+ \lam^2)^{-1}= 1- \lam^2(|\xi|^2+ \lam^2)^{-1}$ 
we write the innermost integral in the form:   
\[
- R_j R_k \left(\lam^2 \chi_{\leq 4a}(|D|) \Gg({i\lam},x) - (\Fg\chi_{\leq 4a}(|\xi|))(x)\right) .
\]
Thus,  $\chi_{\leq 4a}(|D|)W_l^{(2)}u(x)$ is equal to the integral 
\[ 
- \sum_{j,k=1}^2  \int_0^1 (1-\th) \left( \int_{\R^2} y_j y_k m(y) \t_{\th{y}} 
 R_j R_k(I_1 u(x)- I_2u(x))dy \right)d\th, 
\] 
where   
\begin{gather}  
I_1 u(x) = \int_0^\infty 
\lam^2 \chi_{\leq 4a}(|D|) \Gg({i\lam},x) 
\la  k , \Pi_2(\lam) u\ra \lam g_1(\lam) \chi_{\leq a}(\lam) d\lam; 
\lbeq(f-term)  \\
I_2 u(x) = (\Fg\chi_{\leq 4a}(|\xi|))(x)  \int_0^\infty 
\lam^2  
\la  k , \Pi_2(\lam) u\ra \lam g_1(\lam) \chi_{\leq a}(\lam) d\lam .
\lbeq(s-term) 
\end{gather}  
It is sufficient to show that $I_1$ and $I_2$ are GOPs. 

It is evident that 
$\Fg(\chi_{\leq 4a})(x)\in L^p(\R^2)$  for all $1\leq p \leq \infty$ 
and the integral of \refeq(s-term) is equal to 
\begin{align*}  
\ell(u)\colon & = \int_0^\infty 
\left(\int_{{\mathbb S}^1}\hat{k}(\lam\w), \hat{u}(\lam\w) d\w
\right) \lam g_1(\lam)\chi_{\leq a}(\lam)d\lam \\
& = \int_{\R^2} \hat{k}(\xi)\hat{u}(\xi) g_1(|\xi|)\chi_{\leq a}(|\xi|)d\xi 
= (u,  |V|\ast\Fg(g_1(|\xi|) \chi_{\leq a}(|\xi|))).
\end{align*}
Here $\Fg(g_1(|\xi|) \chi_{\leq a}(|\xi|)) (x)\in L^p(\R^2)$ 
as was shown in the proof of Lemma 3.6, \cite{Ya-2dim-new} 
and $\ell$ is a bounded linear functional on $L^p(\R^2)$ for all $1<p<\infty$. 
Thus, $I_2$ is a GOP. 

Removing $\chi_{\leq 4a}(|D|)$ from $I_1 $, we let  
\[  
\tI_1 u(x)= \int_0^\infty \Gg_{i\lam}(x) \la k | \Pi_2(\lam)u\ra  
\lam^3 g_1(\lam) \chi_{\leq a}(\lam)d\lam  .
\]
We shall prove that $\tI_1$ is a GOP, which clearly implies the same 
for $I_1 u(x)$.  By changing the order of integrals and recalling \refeq(tk2), we have 
\begin{align*}
\tI_1 u(x) & = \int_{\R^2} k(y) dy 
\int_0^\infty \Gg({i\lam}, x) \left( \int_{\mathbb S^1} \Fg_{\t_{-y}u}(\lam\w) d\w\right) \lam^3  g_1(\lam) 
\chi_{\leq a}(\lam)d\lam  \\
 & =  \int_{\R^2} k(y) (\tK_2 \tau_{-y} \mu(|D|)u)(x)  dy , 
\quad \m(\lam) \colon = \lam^2 g_1(\lam), 
 \end{align*} 
Since $\m(\lam)$ is a GMU and $\tK_2$ is a GOP by \reflm(K-GOP), $\tI_1$ is a GOP.  
This completes the proof. 
\edpf 

\subsection{Further simplification}  
We shall further simplify $\Cg(\lam)$. 
We continue to omit the phrase ``for small $\lam>0$''.
\bglm \lblm(final-2) 
We have that $\Cg(\lam)\equiv M_v d_0(\lam) M_v$. 
\edlm 
\bgpf    
We shalI prove $J_2(\lam)\colon = M_v e(\lam) d(\lam)M_v $ is a GPR. 
This will imply together with \reflms(d-e,final)  that 
\[
\Cg(\lam) \equiv  M_v d(\lam)M= M_v d_0(\lam) M_v+ \Og_{\Lg^1}(\lam^{2} (\log\lam)^{-1})
\]
and $\Cg(\lam)\equiv M_v d_0(\lam) M_v$ by \refcor(T-K).    

By \reflm(d-e) and \refcor(T-K) again we have  
\[
J_2(\lam) \equiv  M_v e_0 S_0^\perp d_0(\lam)M_v  .  
\]
We orthonormalize $\{\ph_1, \ph_2\}$ of \refeq(phj) 
and denote the result again by $\{\ph_1, \ph_2\}$  
so that $S_0^\perp Q= \ph_1\otimes \ph_1 + \ph_2 \otimes \ph_2$. 
Let 
\[
\mbox{$\p_j = e_0 \ph_j$ and  $\s_j(\lam) ={d}_0(\lam)^\ast \ph_j$ for $j=1,2$,}
\] 
where $d_0(\lam)^\ast$ is the adjoint of $d(\lam)$, 
so that $\p_j \in S_0 \Hs$, $\s_j(\lam) \in S_0^\perp \Hs \subset Q\Hs$ and 
\[
M_v e_0 S_0^\perp d_0(\lam)M_v = \sum_{j=1}^2 M_v \p_j \otimes M_v \s_j(\lam) = \colon 
\sum_{j=1}^2 J_{2j}(\lam).
\]
We prove only that $J_{21}(\lam)$ is a GPR or  
\bqn \lbeq(Wauc)
W_{a}u\colon = \int_0^\infty R_0^{+}(\lam^4)  J_{21}(\lam) 
\Pi(\lam)u \lam^3 \chi_{\leq a}(\lam)d\lam
\eqn 
is a GOP.   Proof for $J_{22}(\lam) $ is similar.  Replace $\chi_{\leq a}(\lam)$ by 
$\chi_{\leq a}(\lam)\chi_{\leq 2a}(\lam)$ and 
denote $J_a(\lam) = J_{21}(\lam) \chi_{\leq a}(\lam)$ so that \refeq(Wauc) becomes 
\bqn \lbeq(Wa-a)
W_{a}u\colon = \int_0^\infty R_0^{+}(\lam^4)  J_a (\lam) 
\Pi(\lam)u \lam^3 \chi_{\leq 2a}(\lam)d\lam. 
\eqn 
Express $J_a (\lam) = \int_0^{2a}  (\r-\lam)_{+} J_a''(\r) d\r$ as before and repeat the argument 
after \refeq(Ilam-int). We obtain   
\bqn \lbeq(Wa-b)
W_{a}u=\int_0^{2a} \left( \int_0^\infty R_0^{+}(\lam^4)  J''_a (\r)  
\Pi(\lam) u(\r) \lam^3 \chi_{\leq 2a}(\lam)d\lam \right) \r d\r \,.
\eqn 
where $u(\r)\colon = (1-|D|/\r)_{+}u$ satisfies $\|u(\r)\|_p \leq C_p\|u\|_p$ 
for $1\leq p \leq \infty$ 
with $\r$-independent constant $C_p$.  We have 
\bqn \lbeq(inR0)
R_0^{+}(\lam^4)  J''_a (\r)  \Pi(\lam)u(\r)
= (M_v \tilde{\s}_1(\r) , \Pi(\lam)u(\r)) R_0^{+} (\lam^4) M_v \p_1  
\eqn 
where $\tilde{\s}_1(\r) =(\chi_{\leq a}(\r)\s_1(\r))''$. 
Since $\p_1 \in S_0 \Hs$,  
\[
R_0^{+}(\lam^4)  (M_v\p_1) (x)= \frac1{2\pi} 
\int_{\R^2}\frac{e^{ix\xi} \Fg(M_v\p_1)(\xi)}{\xi^4-\lam^4-i0}d\xi 
\]
is, as in the proof of \reflm(final),  equal to 
\begin{align}
& \sum_{|\a|=2} \frac{R^\a}{(2\pi)^2}\int_0^1\int_{\R^2} (1-\th)  y^\a (M\p_1)(y) \tau_{\th{y}}
\left(\lim_{z\to \lam+i0}\int_{\R^2}\frac{e^{ix\xi} |\xi|^2 d\xi}{|\xi|^4-z^4}\right) dy  d\th \notag \\
& =  \sum_{|a|=2} \frac{R^\a}{2}
\int_0^1\int_{\R^2} (1-\th)  y^\a (M_v\p_1)(y) 
\tau_{\th{y}}
\left(\Gg(\lam,x)+\Gg({i\lam},x) \right) dy  d\th.  \lbeq(Last-1)
\end{align} 
Since $\tilde{\s}_1(\r)\in Q\Hs$,  $(M_v \tilde{\s}_1(\r), \Pi(\lam) u)$ is equal to 
\begin{align}
& \sum_{j=1}^2 \frac{\lam}{2\pi\lam^2} 
\int_{\R^2} w_j v(w) \tilde{\s}_1(\r,w) 
\left(\int_0^1 \int_{{\mathbb S}^1}
e^{i\th\lam{w\w}}\widehat{R_j u}(\lam\w) 
d\w{d\th}\right) dw \notag  \\
& =  \sum_{j=1}^2 \frac{1}{\lam} 
\int_0^1 \int_{\R^2} w_j v(w)\tilde{\s}_1(\r,w) (\Pi_2(\lam) R_j \tau_{-\th{w}}u)(0) d\th {dw}.
\lbeq(Last-2)
\end{align}
Substitute \refeqs(Last-1,Last-2) in \refeq(inR0) and  
combine the result with \refeq(Wa-b). We obtain that $W_a u(x)$ is equal to  
$\sum_{|\a|=2}\sum_{l=1}^2$ of 
\begin{align*} 
& 
 \int_0^1 (1-\th) d\th \int_0^1 d\tilde{\th}\int_0^{2a} 
\r d\r \int_{\R^2} y^{\a} (v\p_1)(y) dy \int_{\R^2} w_l v(w) \tilde{\s}_1(\r,w) dw \\
& \times \tau_{\th{y}} R^\a \left(
\int_0^\infty (\Gg(\lam,x)+\Gg(i\lam,x))\Pi_2(\lam) \left(R_l \tau_{-\tilde{\th}{w}}u\right)(0) 
\lam^2 \chi_{\leq 2a}(\lam)d\lam \right) .
\end{align*} 
Let $\k(\lam) = \lam \chi_{2a}(\lam)$. Then,  $\k$ is a GMU and 
the function inside the parentheses is equal to 
$(\tK_1 + \tK_2)R_l \tau_{-\tilde{\th}{w}}\k(|D|)u$, which is bounded by $C\|u\|_p$ in the norm 
of $L^p(\R^2)$, $1<p<\infty$, 
with a constant independent of $\tilde{\th}w$.   
Minkowski's 
inequality proves that $W_a$ is a GOP.     
\edpf

\section{Completion of the proof of \refth(regular-case)} 
By virtue of \reflm(final-2) the next proposition finishes the proof of \refth(regular-case).

\bgprop  \lbprop(final) 
The operator $M_v d_0 (\lam)M_v $ is GPR. 
\edprop 

We have to prove that  
\bqn \lbeq(J1) 
W_{J}u \colon = \int_0^\infty R_0^{+}(\lam^4)M_v d_0 (\lam)M_v \Pi(\lam)u \lam^3 
\chi_{\leq {a}}(\lam)d\lam, 
\quad u \in \Dg_\ast
\eqn 
is a GOP.  Let $\{\ph_1, \ph_2\}$ be the orthonormal basis of $S_0^\perp \HL$  used 
in the proof of \reflm(final-2) above. Then,  by \refeq(d0-def), 
\bqn \lbeq(d1-exp)
d_0(\lam) = \sum_{j,k=1}^2 g_1(\lam)^{-1} a_{jk}(\lam) \ph_j \otimes \ph_k 
\eqn 
via analytic functions $a_{jk}(\lam)=A_{jk}(g_1(\lam)^{-1})$ of $g_1(\lam)^{-1}$, $j,k=1,2$, 
and, it suffices to prove that $W_J$ is a GOP when $d_0(\lam)$ is replaced by 
$g_1(\lam)^{-1} a_{jk}(\lam) \ph_j \otimes \ph_k$  for each $j,k=1,2$. 
We shall do this only for $j=k=1$ 
denoting the resulting operator by $W_J$ again and omitting the 
indices. The proof for other $j,k$ is similar.  
Let 
\bqn \lbeq(mujk)
\tm(\lam)=g_1(\lam)^{-1}a(\lam)\chi_{\leq {a}}(\lam), \quad   
\m(\lam) = \lam^2 \tm(\lam). 
\eqn 
By virtue of \reflm(F) in Appendix 
$\Fg(\tm(|\xi|))(x)$ and $\Fg(\m(|\xi|))(x)$ are in $L^p(\R^2)$ 
for all $1\leq p \leq \infty$. 

We split $W_J$ into the high and the low energy parts:
\bqn 
W_J = \chi_{\geq 4a}(|D|)W_J + \chi_{\leq 4a}(|D|)W_J = \colon 
 W_{J,\leq{4a}}   + W_{J, \geq{4a}}. \lbeq(split-g-l) 
\eqn 

\bglm   For any $a>0$,  $W_{J,\geq{4a}}$ is a GOP.   
\edlm 
\bgpf  Since $\chi_{\geq 4a}(|D|)$ is bounded, $W_{J,\geq{4a}}u $ is equal to 
\bqn 
\int_0^\infty \chi_{\geq 4a}(|D|)
R_0^{+}(\lam^4)M_v (\ph \otimes \ph)M_v \Pi(\lam)u \lam^3 
\tm(\lam)d\lam.  \lbeq(J1-geq) 
\eqn  
Via Fourier transform we have   
\bqn  \lbeq(geq)
\chi_{\geq 4a}(|D|)R_0^{+}(\lam^4)= 
\frac{\chi_{\geq 4a}(|D|)}{|D|^4} +  
\frac{\chi_{\geq 4a}(|D|)}{|D|^4} \lam^4 R_0^{+}(\lam^4) , 
\eqn 
which we substitute in  \refeq(J1-geq). Then, \refeq(J1-geq) with   
${\chi_{\geq 4a}(|D|)}/{|D|^4}$ in place of $\chi_{\geq 4a}(|D|)R_0^{+}(\lam^4)$ 
produces  
\begin{multline}  \lbeq(first-r)
\left(\frac{\chi_{\geq 4a}(|D|)}{|D|^4} (v\ph)(x) \right)
\left\la  v\ph, 
\frac1{2\pi}\int_0^\infty \int_{\mathbb S^1}e^{i\lam x\w}\tm(\lam)\hat{u}(\lam\w)d\w \lam d\lam 
\right\ra   \\
= \left(\frac{\chi_{\geq 4a}(|D|)}{|D|^4} (v\ph)(x) \right)
\left\la v\ph, \tm(|D|){u} \right\ra. 
\end{multline}
Since $\Fg(\chi_{\geq 4a}(|\xi|)|\xi|^{-4})(x) $ and $\Fg(\tm(|\xi|))(x)$ are 
both in $L^p(\R^2)$ for all $	1\leq p \leq \infty$ and $v(x)\ph(x) \in L^1(\R^2)$, 
we have 
\[
\|\refeq(first-r)\|_p \leq C_p \|u\|_p, \quad 1\leq p \leq \infty.
\]

On substituting the second term on the right of \refeq(geq) for 
$\chi_{\geq 4a}(|D|)R_0^{+}(\lam^4)$, 
\refeq(J1-geq) becomes 
\[
\frac{\chi_{\geq 4a}(|D|)}{|D|^4} \int_0^\infty 
R_0^{+}(\lam^4)M_v (\ph \otimes \ph)M_v \Pi(\lam)u \lam^7 \tm(\lam)d\lam, 
\] 
which is also a GOP by  \reflm(T-K) (1).   This proves the lemma. 
\edpf 

We next study the low energy part: 
\bqn 
W_{J,\leq{4a}}=\int_0^\infty \chi_{\leq 4a}(|D|)
R_0^{+}(\lam^4)M_v (\ph \otimes \ph)M_v \Pi(\lam)u \lam^3 
\tm(\lam)d\lam.  \lbeq(J1-leq) 
\eqn  
The following argument 
patterns after that in \S 5.6.2 of \cite{Ya-2dim-new}.   
We split it into several steps:

\paragraph{\bf Step 1.  Good and bad parts.} 
Take the inner product of $v(x)\ph(x)$ and  Taylor's formula \refeq(2nd).  
Since $\ph \in (Q\ominus S_0)\Hs$ and $\Pi(\lam) = \lam^{-2}\Pi_2(\lam)$,  
we have $\la \ph | M_v \Pi(\lam)u \ra= \b(\lam) + \c(\lam) $, where 
\begin{align}
& \b(\lam)=  \lam^{-1} \sum_{l=1}^2 \la \ph, x_l v \ra (\Pi_2(\lam) R_l u)(0), \lbeq(f-1st) \\
& \c(\lam)=-\int_0^1 (1-\th) \sum_{|\a|=2} 
\left(\int_{\R^2} z^\a {\ph}(z) v(z) (\Pi_2 (\lam) \tau_{-{\th}z} R^\a u)(0) dz\right)  d\th . 
\lbeq(f-2nd) 
\end{align} 
It follows that  
\bqn \lbeq(subs)
W_{J,\leq{4a}} u(x) = W^{(b)}_{J,\leq{4a}} u(x) + W^{(g)}_{J,\leq{4a}} u(x)
\eqn 
where  the ``bad'' part $W^{(b)}_{J,\leq{4a}} u(x)$ and the ``good'' part $ W^{(g)}_{J,\leq{4a}} u(x)$ 
are defined respectively by    
\begin{align}
& W^{(b)}_{J,\leq{4a}} u(x)= 
\int_0^\infty \chi_{\leq 4a}(|D|)(R_0^{+}(\lam^4)M_v \ph)(x) \b(\lam) \lam^3 \tm(\lam) d\lam,  \lbeq(Wb) \\
& W^{(g)}_{J,\leq{4a}} u(x)= 
\int_0^\infty \chi_{\leq 4a}(|D|)(R_0^{+}(\lam^4)M_v \ph)(x)  \c(\lam) \lam^3 \tm(\lam) d\lam.   \lbeq(Wg)
\end{align}

\bglm  The good part $W^{(g)}_{J,\leq{4a}}$ is a GOP. 
\edlm 
\bgpf  Let $u \in \Dg_\ast$. Substitute \refeq(f-2nd) for $\c(\lam)$ in \refeq(Wg) 
and change  the order of integrals. We obtain  
\begin{multline*}
W^{(g)}_{J,\leq{4a}} u(x)= -\sum_{|\a|=2}\int_0^1 (1-\th) \left(\int_{\R^2 \times \R^2} 
v(y) z^{\a}v(z) \ph (y)\ph(z) \right. \\
\left. \times \tau_{y}\chi_{\leq 4a}(|D|)\left(\int_0^\infty \Rg({\lam},x)
(\Pi_2(\lam) \tau_{-{\th}z} R^\a u)(0) \lam^3 \tm(\lam) d\lam \right) dy dz \right) d\th.
\end{multline*}
By \refeqss(res-ker,tk1,tk2) the $\lam$-integral becomes   
\begin{align*}
F_{\th{z}}u(x)\colon & = \frac12 \int_0^\infty  (\Gg({\lam},x)-\Gg({i\lam},x))
(\Pi_2 (\lam) \tau_{-{\th}z} R^\a u)(0) \lam \tm(\lam) d\lam \\
& = \frac12(\tilde{K}_1  - \tilde{K}_2)\tau_{-{\th}z} R^\a \tm(|D|) u(x) 
\end{align*}
and \reflm(K-GOP) implies 
$\|F_{\th{z}}u\|_p \leq C_p \|u\|_p$ for $1<p<\infty$ with $C_p$ being a constant 
independent of $\th{z}$.   Hence 
\[
\|W^{(g)}_{J,\leq{4a}} u\|_p  \leq C_p \|v \ph\|_1 \|\ax^2v \ph \|_1 \|u\|_p, \quad 1 <p<\infty
\]
and the lemma follows. 
\edpf 

\paragraph{\bf Step 2. Splitting the bad part}     
We are left with $W_{J,\leq{4a}}^{(b)}$.  We write $M_v \ph(x)= f (x)$. Since 
$\ph \in Q\Hs$, $\int_{\R^2} f(x)dx=0$ and, by Taylor's formula,   
\[
\hat{f}(\xi)= \xi \cdot \pa_\xi  \hat{f}(0) - \frac1{2\pi} \sum_{|\a|=2}
\int_0^1(1-\th) \xi^\a \left(\int_{\R^2}e^{-i\th{y}\xi}y^\a f(y) dy \right) d\th .  
\]
Then, via Fourier transform, 
\begin{align}
& \chi_{\leq 4a}(|D|)R_0^{+}(\lam)f (x)= \frac{1}{2\pi}
\sum_{m=1}^2 i (\pa_m \hat{f})(0)
R_m \int_{\R^2}\frac{e^{ix\xi} \chi_{\leq 4a}(\xi)d\xi}{|\xi|^4 -\lam^4-i0}  \lbeq(front-1)  
\\
& + 
\sum_{|a|=2} R^\a
\int_0^1(1-\th)d\th 
\int_{\R^2}  y^\a f(y) \tau_{\th{y}}  dy
\left( \frac1{4\pi^2}
\int_{\R^2} \frac{e^{ix\xi}|\xi|^2 \chi_{\leq 4a}(|\xi|)d\xi}
{|\xi|^4 -\lam^4-i0}
\right), 
\lbeq(front)
\end{align}
which we substitute for  $(\chi_{\leq 4a}(|D|)R_0^{+}(\lam) M_v \ph) (x)$ in \refeq(Wb). 
We obtain 
\bqn \lbeq(b-1-2)
W_{J,\leq{4a}}^{(b)}=W_{J,\leq{4a}}^{(b),1}+ W_{J,\leq{4a}}^{(b),2}
\eqn 
where $W_{J,\leq{4a}}^{(b),1}$ and $W_{J,\leq{4a}}^{(b),2}$ are produced by \refeq(front-1) and \refeq(front) respectively.    

\bglm  The operator $W_{J,\leq{4a}}^{(b),2}$ is a GOP. 
\edlm 
\bgpf  We have as previously 
\[
\frac1{4\pi^2}
\int_{\R^2} \frac{e^{ix\xi}|\xi|^2 \chi_{\leq 4a}(|\xi|)d\xi}
{|\xi|^4 -\lam^4-i0}=\frac{1}{2}\chi_{\leq 4a}(|D|)(\Gg(\lam,x) + \Gg_({i\lam},x))
\]
and $W_{J,\leq{4a}}^{(b),2}$ is equal to the superposition  
\begin{align*}
& \sum_{|a|=2} \sum_{l=1}^2 \la \ph, x_l v\ra R^\a
\int_0^1(1-\th)d\th 
\int_{\R^2}  y^\a f(y) \tau_{\th{y}}  dy \notag \\
& \times \left(\frac12\int_0^\infty \chi_{\leq 4a}(|D|)(\Gg(\lam,x) + \Gg({i\lam},x))
(\Pi_2(\lam)R_l u)(0) \tm(\lam) \lam^2 d\lam\right).  
\end{align*}
The second line is a GOP  by \reflm(K-GOP) (1) and the first line is an ``integrable''
factor. Minkowski's inequality implies the lemma. 
 \edpf 

Thus, we shall be done if we prove that $W_{J, \leq 4a}^{(b),1}$ is a GOP.  
In \refeq(Wb) we substitute \refeq(front-1) and \refeq(f-1st) for 
$\chi_{\leq 4a}(|D|)R_0^{+}(\lam) M_v \ph$  and $\b(\lam)$  respectively. 
Then, ignoring harmless constants and 
Riesz transforms $R_m$ and $R_l$ which are GOPs, 
we see that it suffices to prove that    
\bqn  \lbeq(before-final)
\tilde{W}_{J, \leq 4a}^{(b),1}u(x)= 
\int_0^\infty 
\left(\frac1{2\pi}
\int_{\R^2}\frac{e^{ix\xi}|\xi| \chi_{\leq 4a}(\xi)d\xi}{|\xi|^4 -\lam^4-i0} \right) 
  (\Pi_2(\lam)u)(0) \tm(\lam) \lam^2 d\lam
\eqn 
is a GOP.  We substitute in \refeq(before-final)
\bqn \lbeq(sum) 
\frac{|\xi|}{|\xi|^4 -\lam^4-i0} = 
\frac{\lam}{|\xi|^4 -\lam^4-i0} + 
\frac{1}{(|\xi|^2+ \lam^2)(|\xi|+ \lam)}
\eqn 
and split $\tilde{W}_{J, \leq 4a}^{(b),1}u(x)$ into the sum $L_1 u(x) + L_2 u(x)$, where  
\begin{align*}
L_1 u(x)& = \int_0^\infty 
\left(\frac1{2\pi}
\int_{\R^2}\frac{e^{ix\xi}\lam \chi_{\leq 4a}(\xi)d\xi}{|\xi|^4 -\lam^4-i0} \right) 
  (\Pi_2(\lam)u)(0) \tm(\lam) \lam^2 d\lam, \\
L_2u(x) & =\int_0^\infty 
\left(\frac1{2\pi}
\int_{\R^2}\frac{e^{ix\xi}\chi_{\leq 4a}(\xi)d\xi}
{(|\xi|^2 +\lam^2)(|\xi|+ \lam)} \right) 
  (\Pi_2(\lam)u)(0) \tm(\lam) \lam^2 d\lam .
\end{align*} 
Thanks to the factor $\lam$,  
\[
L_1 u(x)= \frac{1}{2}\int_0^\infty 
\chi_{\leq 4a}(|D|)(\Gg({\lam},x)-\Gg({i\lam}, x)) 
  (\Pi_2(\lam)u)(0) \tm(\lam) \lam d\lam 
\]
and $L_1$ is a GOP by virtue of \reflm(K-GOP) (1).   Recalling \refeq(mujk) for $\tm(\lam)$, 
we express 
\[ 
L_2 u(x) 
= \int_{\R^2}
\left(
\int_{\R^2}\frac{e^{ix\xi-iy\eta}\chi_{\leq 4a}(|\xi|)\chi_{\leq a}(|\eta|)|\eta| d\xi d\eta}
{(|\xi|^2 +|\eta|^2)(|\xi|+ |\eta|)} \right)\frac{a(|D|)}{g_1(D)} u(y) dy . 
\]
Since $a(\lam)$ is an analytic function of  $g_1(\lam)^{-1}$, 
$g_1(|D|)^{-1}a(|D|)$ is a GOP and we have only to consider instead of $L_2u(x)$ 
the integral operator  $Lu(x)$ with the integral kernel 
\[ 
L(x,y) = \frac1{(2\pi)^2}\int_{\R^2\times \R^2}\frac{e^{ix\xi-iy\eta}\chi_{\leq 4a}(|\xi|)\chi_{\leq a}(|\eta|)|\eta| d\xi d\eta} {(|\xi|^2 +|\eta|^2)(|\xi|+ |\eta|)} .
\]
The following proposition will 
complete the  proof of \refth(regular-case).
 
\bgprop \lbprop(reg)
The integral operator $L(x,y)$ is a GOP. 
\edprop 

\paragraph{\bf Proof of \refpropb(reg)} 
The proof is lengthy and it will be given by a series of lemmas. 
Using polar coordinates for $\xi$ and $\eta$ and integrating out with respect to the 
spherical variables,  we obtain  
\bqn 
L(x,y) = \int_{0}^{8a}dr \int_0^{2a}d\r 
\frac{J_0(r|x|)J_0(\r|y|)\chi_{\leq 4a}(r)\chi_{\leq a}(\rho)r\r^2}
{(r^2 +\r^2)(r+ \r)}.  \lbeq(Lxy) 
\eqn  

Statement (1) of the next lemma is  the special case of 
\reflm(preparation), (2) is well known (cf. \cite{DLMF} 10.13.7) and  
(3) is a result of simple computations.  
 
\bglm  \lblm(GOP) We have following statements: 
\ben 
\item[(1)]  Suppose that  
$T(x,y) \absleq C\la \log(|x|/|y|)\ra^l (x^2+ y^2)^{-1}$ for a $l=0,1, \dots$.  
Then, $T(x,y)$ is a GOP.   
\item[(2)]  The Bessel 
function $J_0(\lam)$ satisfies as $\lam \to \infty$   
\begin{align} 
J_0(\lam)& =\sqrt{\frac{2}{\pi}}\left(
\frac{\cos(\lam-\frac{\pi}{4})}{\lam^\frac12} + 
\frac{\sin(\lam-\frac{\pi}{4})}{8\lam^\frac32} \right)
+ O(\lam^{-\frac52}),  \lbeq(asymp)  \\
& =\sqrt{\frac{2}{\pi}}
\frac{\cos(\lam-\frac{\pi}{4})}{\lam^\frac12} + O(\lam^{-\frac32}) .
\lbeq(asymp-a)
\end{align}
\item[{\rm (3)}]  For the derivatives, we have the following identities:
\bqn  \lbeq(first-deri)
\left(\frac{C}{AB}\right)'=- \frac{C}{AB}\left(\frac{A'}{A}+\frac{B'}{B}-\frac{C'}{C}\right) .
\eqn 
\begin{multline} \lbeq(2nd-deri) 
\left(\frac{C}{AB}\right)''=\frac{C}{AB}\left( \frac{A'}{A}+\frac{B'}{B}-\frac{C'}{C}\right) ^2 \\
- \frac{C}{AB}\left( \frac{A''}{A}+ \frac{B''}{B}-\frac{C''}{C}- \frac{(A')^2}{A^2} - \frac{(B')^2}{B^2}
+\frac{(C')^2}{C^2}\right)\,.
\end{multline}
\een 
\edlm 

\bgdf Let $D\subset \R^2 \times \R^2$. We say that $L(x,y)$ is a GOP on 
$D$ if  $\chi_D(x,y) L(x,y)$ is a GOP, where $\chi_D(x,y)$ is the characteristic function 
of $D$.
\eddf  

 Let $D_j$, $j=1,2,3,4$ be domains of $\R^2 \times \R^2$ defined by 
\begin{gather*}
D_1=\{|x|\leq R, |y|\leq R\}, \ \ 
D_2=\{|x|\leq R , |y|>R\}, \\ 
D_3=\{|x| > R, |y|\leq R\},  \ \ 
D_4=\{|x| > R, |y> R\}, 
\end{gather*}
for a large $R$ so that $\R^2 \times \R^2 = \cup_{j=1}^4 D_j$.  
We shall show 
\bqn \lbeq(L-log-a)
L(x,y)\chi_{D_j}(x,y) \absleq C \la \log(|x|/|y|) \ra \left\{\br{l}
|x|^{-2}\ \ \mbox{if $|x|\geq |y|$ in $D_j$} , \\
|y|^{-2}\ \ \mbox{if $|y|\geq  |x|$ in $D_j$}  \er \right.
\eqn 
for $1\leq j \leq 4$.  It is evident that  \refeq(L-log-a) implies   
\bqn  
L(x,y) \absleq C \la \log(|x|/|y|) \ra (|x|^2+ |y|^2)^{-1}
\eqn 
and $L$ is a GOP by \reflm(GOP) (1).  

It is evident that $L(x,y)$ is smooth on 
$\R^2 \times \R^2$.  Since $D_1$ is compact, the following lemma is obvious.   

\bglm For any $R>0$, $L(x,y)\chi_{D_1}(x,y)$ is a GOP. 
\edlm 

To prove \refeq(L-log-a) for $j=2,3,4$, we split 
the domain of integration $(r,\r) \in (0,8a) \times (0, 2a)$ of \refeq(Lxy) into 
four regions where $r$ satisfies either $r|x|\geq 1$ or $r|x| \leq 1$ 
and $\r$ either $\r|y|\geq 1$ or $\r|y|\leq 1$. We estimate 
$J_0(r|x|)\absleq C$ or $J_0(\r|y|)\absleq C$ when $r|x|\leq 1$ or 
$\r|y|\leq 1$ respectively; if  $r|x|\geq 1$ or $\r|y|\geq 1$ we substitute 
the asymptotic formula \refeq(asymp) or \refeq(asymp-a)  with $\lam= r|x|$ or 
$\lam= \r|y|$ respectively for $J_0(r|x|)$ or $J_0(\r|y|)$  and 
estimate the resulting oscillatory integral via integration by parts. 
Thus, the arguments will be to some 
extent similar and the proof will be repetitive, hence, 
we shall be sketchy in several places.  We assume $R>10/a$ in what follows. 

We record here some integration by parts formulas and their immediate consequences 
which will be used in the proof.  In the next lemma 
$c(\pi)= \cos(1-\frac{\pi}{4})$ and $s(\pi)= \sin(1-\frac{\pi}{4})$ . 

\bglm 
\lblm(int-part) We have the following equations and estimates: 
\begin{align*} 
& \hspace{-0.5cm}
 {\rm (F1)} \ \  
\frac{1}{|y|^\frac32 }
\int_{1/|y|}^{2a} \frac{\sin(\r|y|-\frac{\pi}{4})\chi_{\leq a}(\rho)\r^{\frac12} d\r}
{ (r^2 +\r^2)(r+ \r)}  = \frac{c(\pi)}
{ (|y|^2 r^2 +1)(r|y|+ 1)}    \\
&  \hspace{1cm}   + 
\frac{1}{|y|^\frac52 }\int_{1/|y|}^{2a} \cos(\r|y|-\frac{\pi}{4})
\left( \frac{\r^{\frac12}\chi_{\leq a}(\rho) }{ (r^2 +\r^2)(r+ \r)} \right)' d\r   \\
& \hspace{1cm}    \absleq \frac{C}{(|y|^2 r^2+ 1)(|y| r + 1)} + \frac{C}{|y|^\frac52}
\left(1 + \int_{\frac1{|y|}}^{2a} \frac{\r^{-\frac12}d\r}{(r^2+ \r^2)(\r + r)}  \right).  \\
& \hspace{-0.5cm}
 {\rm (F2)} \ \ \int_{1/|y|}^{2a} \frac{\cos(\r|y|-\frac{\pi}{4})\chi_{\leq a}(\rho)\r^{\frac32} d\r}
{|y|^\frac12 (r^2 +\r^2)(r+ \r)}   \\
&   \hspace{1cm}  = - 
\frac{s(\pi)}
{ (|y|^2 r^2 +1)(r|y|+ 1)}   - \left. \frac{c(\pi)}{|y|^\frac52 }
\left( 
\frac{\r^{\frac32}  \chi_{\leq {a}}(\r)}
{ (r^2 +\r^2)(r+ \r)} 
\right)' \right\vert_{\r=\frac1{|y|}}^{2a}  \\
&  \hspace{1cm} - 
\frac{1}{|y|^\frac52 }\int_{1/|y|}^{2a} \cos(\r|y|-\frac{\pi}{4})
\left( \frac{\r^{\frac32}\chi_{\leq a}(\rho) }{ (r^2 +\r^2)(r+ \r)} \right)'' d\r.  \\
& \hspace{1cm} \absleq 
\frac{C} { (|y|^2 r^2 +1)(r|y|+ 1)}  + 
\frac{C}{|y|^\frac52} \left(1+\int_{\frac1{|y|}}^{2a}
\frac{\r^{-\frac12}d\r}{(r^2 +\r^2)(r+ \r)} \right). \\
& \hspace{-0.5cm}
 {\rm (F3)} \ \ 
\int_{1/|x|}^{8a} 
\frac{\sin(r|x|-\frac{\pi}{4})}{8r^\frac32 |x|^{\frac32}}
\frac{\chi_{\leq 4a}(r)r dr}
{(r^2 +\r^2)(r+ \r)} \\
& \hspace{1cm} = 
\frac{1}{8}\frac{|x|c(\pi)}{(|x|^2\r^2+1)(|x|\r+ 1)}  \\
&  \hspace{1cm}- \frac{1}{8|x|^{\frac52}}
\int_{1/|x|}^{8a} 
\cos(r|x|-\tfrac{\pi}{4})
\left(
\frac{\chi_{\leq 4a}(r)}
{(r^2 +\r^2)(r+ \r)r^\frac12} \right)' dr \\
& \hspace{1cm} 
\absleq 
\frac{C|x|}{(|x|^2\r^2+1)(|x|\r+ 1)}
+ \frac{C}{|x|^{\frac52}}
\left(1+ 
\int_{1/|x|}^{8a} 
\frac{ r^{-\frac32}dr}{(r^2 +\r^2)(r+ \r)} 
\right) .
\\
& \hspace{-0.5cm}
 {\rm (F4)} \ \ \int_{\frac{1}{|x|}}^{8a}
\frac{\cos(r|x|-\frac{\pi}{4})\chi_{\leq 4a}(r)r dr}{(r|x|)^{\frac12}(r^2 +\r^2)(r+ \r)} \\
& \hspace{1cm} = 
\left. \frac{s(\pi)r^{\frac12}}{|x|^{\frac32}(r^2 +\r^2)(r+ \r)} \right|_{\frac{1}{|x|}}
- \left. \left(\frac{c(\pi)r^{\frac12}}{|x|^{\frac52}(r^2 +\r^2)(r+ \r)}\right)' \right|_{\frac{1}{|x|}}  
\\
 & \hspace{1cm} - \int_{\frac{1}{|x|}}^{8a}
\frac{\cos(r|x|-\frac{\pi}{4})}{|x|^\frac52} 
\left(\frac{\chi_{\leq 4a}(r)r^{\frac12}}{(r^2 +\r^2)(r+ \r)} \right)''dr \\
& \hspace{1cm} 
\absleq \frac{C |x|}{(1+|x|^2\r^2)(1+ |x|\r)} + 
\frac{C}{|x|^\frac52}\left(1+ \int_{\frac1{|x|}}^{8a} \frac{r^{-\frac32}dr}{(r^2+\r^2)(r+\r)}\right) .
\end{align*} 
Notice that the right sides of (F1)  and (F3) coincide with (F2) and (F4) respectively.  
\edlm 
\bgpf 
 Equations are results of integration by parts. 
Estimates for the boundary terms are evident.  The derivatives 
inside the integrals may be written in the form either 
$\chi_{\leq a}'(\r)F(r,\r) + \chi_{\leq a}(\r)\pa_{\r} F(r, \r)$ or 
$\chi_{\leq 4a}'(r)F(r,\r)+ \chi_{\leq 4a}(r)\pa_r F(r,\r)$ 
and similarly for the second derivatives. 
Since derivatives of cut-off functions vanish outside compact 
intervals of $(0, 2a)$ or $(0,8a)$, we have  for $j=1,2$ that 
$\chi_{\leq a}^{(j)}(\r)F(r,\r)\absleq C$  and 
$\chi_{\leq 4a}^{(j)}(r)F(r,\r)\absleq C$;  for derivatives of the other part 
we use \refeq(first-deri) or \refeq(2nd-deri).  
Results follow immediately.
\edpf

\bglm  For $R>10/a$, $L(x,y)\chi_{D_2}(x,y)$ is a GOP.  
\lblm(L2-s)
\edlm 
\bgpf 
We shall prove $L(x,y)\chi_{D_2}(x,y) \absleq C |y|^{-2} $ which evidently 
implies \refeq(L-log-a). 
Let $(x,y) \in D_2$.  Estimate $J_0(r|x|)\absleq C$ for $0<r<8a$; 
split the interval of $\r$-integration as $(0,2a)= (0,1/|y|) \cup [1/|y|, 2a)$ 
and estimate $J_0(\r|y|) \absleq C$ for $0 < \r <1/|y|$. 
Then, $L(x,y)  \leq  L_{21}(x,y) + L_{22}(x,y)$ where 
\begin{align} 
L_{21}(x,y)= C \int_0^{8a} r \left(
\int_0^{\frac1{|y|}} 
\frac{\r^2 d\r}{(r^2 +\r^2)(r+ \r)} \right) dr,  \lbeq(L21) \\
L_{22}(x,y)= C \int_{0}^{8a} r \left|
\left(\int_{\frac1{|y|}}^{2a}
 \frac{J_0(\r|y|)\chi_{\leq a}(\rho)\r^2 d\r}
{(r^2 +\r^2)(r+ \r)} \right) \right| dr .  \lbeq(L22)
\end{align}
Change the order of integrals and  change variable $r$ to $t\r$. Then,  
\bqn 
L_{21}(x,y) = C \int_{0}^{1/|y|} {\r}  \left(\int_0^{8a/\r}
 \frac{t dt}{(t^2 +1)(t+ 1)} \right) d\r 
\leq  \frac{C }{|y|^2}.  
\lbeq(L1-est)
\eqn 
We substitute \refeq(asymp) with 
$\lam=\r|y|$ for $J_0(\r|y|)$ in \refeq(L22) and estimate the result by the sum:   
\[
L_{22}(x,y)\leq X_1(x,y)+ X_2(x,y) + X_3(x,y),
\]
where   
\begin{align} 
& X_1(x,y)= C \int_{0}^{8a} r \left|
\left(\int_{\frac1{|y|}}^{2a}
 \frac{\cos(\r|y|-\frac{\pi}{4})\chi_{\leq a}(\rho)\r^{\frac32} d\r}
{|y|^{\frac12} (r^2 +\r^2)(r+ \r)} \right) \right| dr ,   \lbeq(X1) \\
& X_2(x,y)= 
C \int_{0}^{8a} r \left|
\left(\int_{\frac1{|y|}}^{2a}
 \frac{\sin(\r|y|- \frac{\pi}{4})\chi_{\leq a}(\rho)\r^{\frac12} d\r}
{|y|^{\frac32} (r^2 +\r^2)(r+ \r)} \right) \right| dr,  \\
& X_3(x,y) \leq C \int_{0}^{8a} r \left(
\int_{1/|y|}^{2a}  \frac{(\r|y|)^{-\frac52}\r^2 d\r} 
{(r^2 +\r^2)(r+ \r)} 
\right) dr  .
\end{align} 
By using (F2) we estimate $X_1(x, y)$ by the sum of 
\[
\int_{0}^{8a} \frac{C rdr}{((r|y|)^2 + 1)(r|y|+1)} 
\leq \frac{C}{|y|^2}\int_0^{8a|y|}\frac{tdt}{(t^2+1)(t+1)} 
\leq \frac{C}{|y|^2}
\] 
and 
\begin{align} 
& \frac{C}{|y|^{\frac52}} \int_0^{8a} r \left(1+\int_{\frac1{|y|}}^{2a}
\frac{\r^{-\frac12}d\r}{(r^2 +\r^2)(r+ \r)} \right) dr  \notag \\
& \leq \frac{C}{|y|^{\frac52}}+ \frac{C}{|y|^\frac52} \int_{\frac1{|y|}}^{2a}
\left( \int_0^{\frac{8a}{\r}}\frac{tdt}{(t^2+1)(t+1)}\right) \frac{d\r}{\r^{\frac32}} 
\leq \frac{C}{|y|^{2}},  \lbeq(forX1)
\end{align}
where in the last step we changed the order of integrals and 
the variable $r\to  t\r$.  Thus,  $X_1(x,y) \absleq C |y|^{-2}$. 

We estimate  $X_2(x,y)$ by using (F1). Then, since the right sides of (F1) and (F2) are the same, 
the argument used for $X_1(x,y)$ implies $X_2(x,y)\absleq {C}|y|^{-2}$.   
 
Similarly we estimate $X_3(x,y)$ by   
\[ 
C |y|^{-\frac52}\int_{1/|y|}^{2a}  {\r^{-\frac32}} 
 \left(
 \int_{0}^{8a/\r} \frac{t dt}{(t^2 +1)(t+ 1)}\right)  d\r  \leq \frac{C}{|y|^2}.
\]
Summing up, we obtain  
$|L(x,y)\chi_{D_2}(x,y)| \leq C |y|^{-2}$ and $L$ is a GOP on $D_2$. 
\edpf

\bglm  For $R>0$ large enough, $L(x,y)\chi_{D_3}(x,y)$ is a GOP.   \lblm(D3)
\edlm 
\bgpf  Let $|x|>R$ and $|y|\leq R$ and $R>1/a$. We show 
\bqn   \lbeq(D3-e)
L(x,y)\chi_{D_3}(x,y) \leq C\frac{\la \log |x| \ra}{|x|^2}
\eqn 
which evidently implies \refeq(L-log-a).  Estimating $|J_0(\r|y|)|\leq C$,  we have  
\begin{align}
L(x,y) & \absleq C \int_{0}^{2a} \r^2 \left| 
\left(\int_0^{1/|x|}+ \int_{1/|x|}^{8a} \right) 
\frac{J_0(r|x|)\chi_{\leq 4a}(r)r dr}
{(r^2 +\r^2)(r+ \r)} \right| d\r \notag \\
& \leq  L_{31}(x,y)+ L_{32}(x,y), \lbeq(L34)
\end{align}
where the definition of $L_{31}(x,y)$ of $L_{32}(x,y)$ should be obvious. 
For $0<r<1/|x|$, we estimate $|J_0(r|x|)|\leq C$, change the order of integrals and  
the variable as $\r=tr$. Then,    
\begin{align}
L_{31}(x,y)& \leq  C\int_0^{1/|x|}
 r  \left(\int_{0}^{2a/r} \frac{t^2 dt}{(t^2+1)(t+1)} \right) dr  \notag \\
 &\absleq C\int_0^{1/|x|}  r(1+ |\log{r}|) dr    
\leq \frac{C\la \log x \ra}{|x|^2} .  \lbeq(L31)
\end{align}

Substitute  \refeq(asymp) for $J_0(r|x|)$ in $L_{32}(x,y)$ and denote 
by $Y_j (x,y)$ the function produced by the $j$-th term of \refeq(asymp) so that
\bqn 
L_{32}(x,y)\leq Y_1(x,y)+ Y_{2}(x,y)+ Y_{3}(x,y). 
\eqn
By applying  (F4) we estimate   
\[
Y_1(x,y) = C \int_{0}^{2a}\r^2 \left| 
\int_{\frac{1}{|x|}}^{8a}
\frac{\cos(r|x|-\frac{\pi}{4})\chi_{\leq 4a}(r)r dr}{(r|x|)^{\frac12}(r^2 +\r^2)(r+ \r)}\right| d\r 
\]
by the sum of 
\bqn 
\int_{0}^{2a}
\frac{C |x|\r^2d\r}{(1+|x|^2 \r^2)(1+ |x|\r)  }
\leq  \frac{C}{|x|^2}\int_0^{2a|x|}\frac{dt}{ 1+t}  
\leq \frac{C\la \log|x| \ra}{|x|^2}  \lbeq(Bdry-contri-1)
\eqn 
and 
\begin{align}
& \frac{C}{|x|^\frac52} \int_{0}^{2a} \r^2
\left(1+ 
\int_{\frac{1}{|x|}}^{8a} \frac{\chi_{\leq 4a}(r)r^{-\frac32} dr}
{(r^2+\r^2)(r+\r)}  \right) d\r    \lbeq(4-2-1) \\
& \leq \frac{C}{|x|^\frac52}  + \frac1{|x|^\frac52} \int_{\frac{1}{|x|}}^{8a} r^{-\frac32}\left( 
\int_{0}^{\frac{2a}{r}}\frac{t^2 dt}{(t^2+1)(t+1)}\right) dr . 
\lbeq(main-3)
\end{align} 
Split as $(1/|x|, 8a)= (1/|x|, 2a] \cup  (2a,8a)$ for the $r$-integral. Since 
$1/4 \leq 2a/r\leq 1$ when $r \in (2a,8a)$ and $1 \leq 2a/r\leq 2a|x|$ 
when $r\in (1/|x|, 2a]$,  the integral of \refeq(main-3) is estimated by 
\[
\frac{C}{|x|^\frac52} \left(C+ 
\int_{\frac{1}{|x|}}^{2a} r^{-\frac32}
\left(
\int^{2a|x|}_0 \frac{dt}{t+1}\right) dr \right)
\leq \frac{C\la \log{|x|}\ra}{|x|^2}. 
\]
Thus, $Y_1(x,y)\absleq  C \la \log {|x|} \ra |x|^{-2}$ on $D_3$.   

We apply (F3) for 
\bqn \lbeq(43)
Y_2 (x,y)= C \int_{0}^{2a} {\r^2} \left|\int_{1/|x|}^{8a} 
\frac{\sin(r|x|-\frac{\pi}{4})}{8r^\frac32 |x|^{\frac32}}
\frac{\chi_{\leq 4a}(r)r dr}
{(r^2 +\r^2)(r+ \r)} \right| d\r. 
\eqn 
Then,  since the rights sides of (F3) and (F4) are the same,  the 
argument used for $Y_1(x,y)$ above implies 
$Y_2(x,y) \absleq C\la \log |x|\ra {|x|^{-2}}$.

By changing the order of integration and the variable we estimate   
\begin{align*}
Y_3(x,y)& \absleq C \int_{0}^{2a} \r^2 \left(
\int_{1/|x|}^{8a}  \frac{(r|x|)^{-\frac52}r dr} 
{(r^2 +\r^2)(r+ \r)} \right) d\r   \\
& \leq 
\frac{C}{|x|^{\frac52}} \int_{1/|x|}^{8a}  \frac{(1 + |\log r|) dr}{r^{\frac32}}
\leq \frac{C\la \log |x|\ra}{|x|^2}.    
\end{align*} 
Combining these estimates, we obtain \refeq(D3-e). 
\edpf 

We proceed to estimate $L(x,y)\chi_{D_4}(x,y)$. 
 
\bglm \lblm(D4)
For large enough $R>0$, $L(x,y) \chi_{D_4}(x,y)$ is a GOP. 
\edlm 

We prove the lemma by a series of lemmas. 
Let $(x, y) \in D_4$ and $R>1/a$ be large enough. 
We split the integral \refeq(Lxy) as follows: 
\begin{align} 
& \left(\int_{0}^{\frac{1}{|x|}}+ \int_{\frac{1}{|x|}}^{8a}\right)dr
\left(\int_{0}^{\frac{1}{|y|}} + \int_{\frac{1}{|y|}}^{2a} \right)d\r
\frac{J_0(r|x|)J_0(\r|y|)\chi_{\leq 4a}(r)\chi_{\leq a}(\rho) r\r^2}
{(r^2 +\r^2)(r+ \r)}  \notag \\ 
& = L_{in,in}(x,y) + L_{in,out}(x,y) + L_{out,in}(x,y) + L_{out,out}(x,y),   \lbeq(split-a)
\end{align} 
where $L_{in,in}(x,y)$,  $L_{in,out}(x,y)$ etc. are integrals with respect to $dr\otimes d\r$ 
over the domains $[0,1/|x|] \times [0,1/|y|]$, 
$[0,1/|x|] \times [1/|y|, 2a]$ etc.
and we shall show they are GOPs on $D_4$ separately. 

\bglm \lblm(in-in)
For $R>0$ large enough $L_{in,in}(x,y)$ is a GOP on $D_4$.
\edlm 
\bgpf  Since $J_0(\lam)\absleq C$, we have    
\bqn \lbeq(in-in) 
L_{in,in}(x,y)\absleq  C\int_{0}^{\frac{1}{|x|}} \left( \int_{0}^{\frac{1}{|y|}}
\frac{r \r^2 }{(r^2 +\r^2)(r+ \r)} d\r \right) dr.
\eqn 
If $|x|\geq |y|$, then  $r<1/|x|<1/|y|$ and the integral on the right  is bounded by 
\bqn  
\int_{0}^{\frac{1}{|x|}} \left( \int_{0}^{r} d\r+  
 \int_{r}^{\frac{1}{|y|}} \frac{r d\r}{\r} \right) dr  
= \frac{3}{4|x|^2}+ \frac12\left(\frac{1}{|x|}\right)^2\log \frac{|x|}{|y|};
\lbeq(in-in1)
\eqn 
when $|y|\geq |x|$,  then, changing the role of $x$ and $y$, we estimate it by 
\bqn 
\int_0^{\frac{1}{|y|}} 
\left(\int_0^{\r} \frac{r}{\r} dr + \int_{\r}^{\frac{1}{|x|}} \frac{\r^2}{r^2} dr\right) d\r 
\leq \frac{5}{12}\left(\frac{1}{|y|}\right)^2.  \lbeq(in-in2)
\eqn
It follows that  
 \bqn 
L_{in,in}(x,y) \absleq 
\frac{C}{|x|^2+|y|^2}\left\la \log\left(\frac{|x|}{|y|}\right) \right\ra
\eqn 
and $L_{in,in}(x,y)$ is a GOP on $D_4$. 
\edpf

\bglm \lblm(in-out)  For $R>0$ large enough $L_{in,out}(x,y)$ is a GOP on $D_4$.
 \edlm 
\bgpf Since  $|J_0(r|x|)|\leq C$,  we have 
\bqn  
L_{in,out}(x,y) \absleq  C 
\int_{0}^{\frac{1}{|x|}}r  \left|
\int_{\frac{1}{|y|}}^{2a} 
\frac{J_0(\r|y|)\chi_{\leq a}(\rho) \r^2 d\r}
{(r^2 +\r^2)(r+ \r)} 
\right| dr\,.   \lbeq(in-out-1)
\eqn 
(1) Let $|x|>|y|$ first.  
We substitute \refeq(asymp-a) with $\lam= \r|y|$ for $J_0(\r|y|)$  
in \refeq(in-out-1) so that  
$L_{in,out}(x,y)\absleq  L_{in,out}^{(m)}(x,y) +  L^{(r)}_{in,out}(x,y)$,  
where 
\bqn
L_{in,out}^{(m)}(x,y)= \sqrt{\frac{2}{\pi}}\int_{0}^{\frac{1}{|x|}}r  \left|
\int_{\frac{1}{|y|}}^{2a} 
\frac{\cos(\r|y|-\frac{\pi}{4})\chi_{\leq a}(\rho) \r^\frac32 d\r}
{|y|^{\frac12}(r^2 +\r^2)(r+ \r)} 
\right| dr  \lbeq(184)
\eqn
and 
\bqn 
L^{(r)}_{in,out}(x,y) = C \int_{0}^{\frac{1}{|x|}}r  \left(
\int_{\frac{1}{|y|}}^{2a} 
\frac{(\r|y|)^{-\frac32} \r^2 \chi_{\leq a}(\r)d\r}{(r^2 +\r^2)(r+ \r)} \right) dr .   
\eqn
By applying (F2) to  the inner integral, we estimate \refeq(184) by the sum of 
\begin{align}
& C \int_0^{\frac{1}{|x|}}\frac{r dr }{(r^2 |y|^2+1)(r|y|+ 1)}  \notag  \\
& = \frac{C}{|y|^2} 
\int_0^{\frac{|y|}{|x|}}
\frac{t dt }{(t^2 +1)^2(t+1)}  
\leq \min
\left\{
\frac{C}{|x|^2}, 
\frac{C}{|y|^2} 
\right\}  \lbeq(contri-bdry3a) 
\end{align} 
and
\bqn 
\frac{C}{|y|^{\frac{5}{2}}} 
\int_{0}^{\frac{1}{|x|}}  r \left( 1+ 
\int_{\frac{1}{|y|}}^{2a} \frac{\r^\frac12 d\r  }
{(r^2 +\r^2)(r+ \r)} \right) dr    \leq \frac{C}{|x|^2}, \lbeq(bdry-3-2)
\eqn 
where in the left of \refeq(bdry-3-2) we estimated the inner integral by 
$C |y|^{\frac52}$ by using $r<1/|x|<1/|y|<\r$. 
 
It is immediate to see that 
\bqn 
L^{(r)}_{in,out}(x,y) \absleq  
C |y|^{-\frac32}
\int_{0}^{\frac{1}{|x|}}r  \left(
\int_{\frac{1}{|y|}}^{2a} \r^{-\frac52} d\r \right) dr
\leq \frac{C}{|x|^2} .
\lbeq(V2-2)
\eqn
Summing up \refeq(contri-bdry3a),  \refeq(bdry-3-2) and \refeq(V2-2), 
we obtain $L_{in,out}(x,y)\absleq C|x|^{-2}$ on $\{(x,y) \in D_4: |x|>|y|\}$. 

(2) Let $|y|>|x|$ next.  We now substitute \refeq(asymp) 
with $\lam= \r|y|$ in \refeq(in-out-1).  Then, $O((\r|y|)^{-\frac52})$ produces   
\begin{align} 
 L^{(r)}_{in,out}(x,y) & \absleq C \int_{\frac{1}{|y|}}^{2a} 
\left(\int_{0}^{\frac{1}{|x|}}\frac{r (\r|y|)^{-\frac52} \r^2 dr}{(r^2 +\r^2)(r+ \r)} \right) d\r 
\notag \\
& \leq \frac{C}{|y|^\frac52} \int_{\frac{1}{|y|}}^{2a}
\left(\int_{0}^{\frac{1}{\r |x|}}\frac{t dt}{(t^2 +1)(t+1)} \right) \frac{d\r}{\r^\frac32} 
\leq \frac{C}{|y|^2} ,
\lbeq(V2-3)
\end{align}
where we have estimated the $t$-integral by a constant. 
The first term $(\r|y|)^{-\frac12}\cos(\r|y|-\frac{\pi}{4})$ of \refeq(asymp) 
produces 
\refeq(184) which, as was shown above, is bounded by the sum of \refeq(contri-bdry3a)
which is now bounded by  $C|y|^{-2}$ and the left side of \refeq(bdry-3-2) which is 
bounded by 
\[
\frac{C}{|y|^{\frac{5}{2}}}\left(\frac1{|x|^2} + 
\left(\int_{0}^{\frac{1}{|x|}}  dr\right) \left(\int_{\frac{1}{|y|}}^{2a} \frac{d\r  }{\r^\frac32}\right) 
\right) 
\leq \frac{C}{|y|^2}.
\]
The contribution of $(4\r|y|)^{-\frac32}\sin(\r|y|-\frac{\pi}{4})$ to 
$L_{in,out}(x,y)$ is bounded by 
\bqn \lbeq(pre-f)
\frac{C}{|y|^{\frac32}}
\int_{0}^{\frac{1}{|x|}}r  \left|
\int_{\frac{1}{|y|}}^{2a} 
\frac{\sin(\r|y|-\frac{\pi}{4})\chi_{\leq a}(\rho) \r^{\frac12} d\r}
{(r^2 +\r^2)(r+ \r)} \right| dr .
\eqn 
This is less than $X_2(x,y)$ of the proof of \reflm(L2-s) and is bounded by $C|y|^{-2}$. Thus, 
$L_{in,out}(x,y)$ is also a GOP on $\{(x,y)\in D_4: |y|>|x|\}$.   The lemma is proved.
\edpf 

\bglm \lblm(out-in)  For $R>0$ large enough, $L_{out,in}(x,y)$ is a GOP on $D_4$.
\edlm 
\bgpf  Since $J_0(\r|y|)\absleq C$, we have 
\bqn 
L_{out,in}(x,y) 
\absleq 
C \int_{0}^{\frac{1}{|y|}}  \r^2
\left|
\int_{\frac{1}{|x|}}^{8a}
\frac{J_0(r|x|)\chi_{\leq 4a}(r)r dr}
{(r^2 +\r^2)(r+ \r)}  \right| d\r.    \lbeq(4-1)
\eqn 
(1) Let $|y|>|x|$ first. Since we also have $J_0(r|x|)\absleq C$, 
\[ 
L_{out,in}(x,y) \absleq  
C \int_{0}^{\frac{\c}{|y|}}{\r^2}
\left(
\int_{\frac{\c}{|x|}}^{8a} \frac{dr}{r^2}  \right) d\r
= \frac{C|x|}{|y|^3 } \leq \frac{C}{|y|^2}.
\]
(2)  Let $|x|>|y|$ next.  The right side of \refeq(4-1) is equal to 
$L_{32}(x,y)$ of the proof of \reflm(D3) and we have shown 
that it is bounded by $C\la \log(|x|/|y) \ra |x|^{-2}$.  Hence 
$L_{out,in}(x,y) $ is a GOP also on $\{(x,y)\in D_4: |x|>|y|\}$. 
This proves the lemma. 
\edpf 

We finally study $L_{out,out}(x,y)$ which is equal to 
\bqn 
\lbeq(oo)
\int_{\frac{1}{|x|}}^{8a} J_0(r|x|)\chi_{\leq 4a}(r)r
\left(\int_{\frac{1}{|y|}}^{2a}
\frac{J_0(\r|y|)\chi_{\leq a}(\rho) \r^2 }
{(r^2 +\r^2)(r+ \r)} d\r\right) dr.
\eqn 

\bglm  For $R>0$ large enough, $L_{out,out}(x,y)$ is a GOP on $D_4$. 
\edlm 
\bgpf 
(1) Let $|x|>|y|$ first. We substitute \refeq(asymp) with $\lam= r|x|$ for $J_0(r|x|)$ in \refeq(oo) 
so that 
 \[
L_{out,out}(x,y)= Z_{1}(x,y) + Z_{2}(x,y) + Z_{3}(x,y),
\]
where $Z_{j}(x,y)$ is produced by the $j$-th term of \refeq(asymp).

(1a)  If we change the order of integrals, then $Z_{1}(x,y)$ becomes    
\bqn \lbeq(Z1)
C \int_{\frac{1}{|y|}}^{2a}  
{J_0(\r|y|)\r^2\chi_{\leq{a}}(\r)}
\left(\int_{\frac{1}{|x|}}^{4a}
\frac{\cos(r|x|-\frac{\pi}{4})\chi_{\leq 4a}(r) r^\frac12}
{|x|^\frac12 (r^2 +\r^2)(r+ \r)} dr\right) d\r. 
\eqn 
We estimate $J_0(\r|y|)\absleq C (\r|y|)^{-\frac12}$ and 
substitute (F4) for the inner $dr$-integral. Then, $Z_{1}(x,y)$ is bounded in modulus by 
the sum of 
\[
C \int_{\frac{1}{|y|}}^{2a}  
\frac{|x|\r^{\frac32} d\r}{|y|^\frac12 (|x|^2\r^2 +1)(|x|\r+1)}  
\leq 
\frac{C}{|x|^2} \left(\frac{|x|}{|y|}\right)^\frac12 
\int_{\frac{|x|}{|y|}}^{2a|x|}  \frac{dt}{t^\frac32}\leq \frac{C}{|x|^2},  
\] 
and 
\bqn \lbeq(thatof) 
\frac{C}{|x|^\frac52}\left(1+ \int_{\frac{1}{|x|}}^{4a}
r^{-\frac32}dr \left( \int_{\frac{1}{|y|}}^{2a}  
\frac{|J_0(\r|y|)|\r^2}{(r^2 +\r^2)(r+ \r)}d\r \right) \right).
\eqn
We estimate the inner $d\r$-integral by  
\[
\int_{\frac{1}{|y|}}^{2a}    
\frac{C |y|^{-\frac12}\r^\frac32 d\r} {(r^2+\r^2)(r+\r)} 
\leq 
\frac{C}{|y|^\frac12} \int_{\frac{1}{|y|}}^{2a}  \frac{d\r}{\r^\frac32} \leq C
\]
and \refeq(thatof) by  $C |x|^{-2}$. This implies  $Z_{1}(x,y)\absleq C |x|^{-2}$. 

\noindent 
(1b)  We change the order of integrals and use 
$J_0(\r|y|)\absleq C(\r|y|)^{-1/2}$ for $\r|y|>1$.  
Then,  
\bqn \lbeq(Z2)
Z_2(x,y) \absleq C \int_{\frac{1}{|y|}}^{2a} \frac{\r^{\frac32}}{|y|^\frac12}\left|
\int_{\frac{1}{|x|}}^{8a}
\frac{\sin(r|x|-\frac{\pi}{4})\chi_{\leq 4a}(r) r }
{(r|x|)^{\frac32}(r^2 +\r^2)(r+ \r)} dr\right| d\r  
\eqn 
We substitute (F3) for the inner integral in the right of \refeq(Z2);  
$|Z_2(x,y)|$ is estimated by the sum of  
\[ 
\frac{C}{|x|^2} \int_{\frac{1}{|y|}}^{2a}  
\frac{|x|^3 \r^{\frac32} d\r}{|y|^{\frac12}(|x|^2\r^2 +1)(|x|\r+ 1)}  
\leq \frac{1}{|x|^2} 
\left(\frac{|x|}{|y|}\right)^\frac12 \int_{\frac{|x|}{|y|}}^{2a|x|}  t^{-\frac32}dt 
\leq \frac{C}{|x|^2}
\]
and 
\[
\frac{C}{|x|^\frac52|y|^3}+ \frac{C}{|x|^\frac52} 
\int_{\frac{1}{|x|}}^{8a}\frac{1}{r^{\frac32}}
\left(\int_{\frac{1}{|y|}}^{2a}  \frac{\r^{\frac32} d\r}
{|y|^{\frac12}(r^2 +\r^2)(r+ \r)}\right)  dr  
\leq \frac{C}{|x|^2}
\]
where we estimated the inner integral in the last step by a constant.

\noindent 
(1c) Let $Z_3(x,y)$ be \refeq(oo) with $O((r|x|)^{-\frac52})$ replacing $J_0(r|x|)$.  Then, 
using $J_0(\r|y|)\absleq C(\r|y|)^{-\frac12}$ for $\r>1/|y|$, we obtain  
\begin{align}
& Z_3(x,y) \absleq C
\int_{\frac{1}{|x|}}^{8a} (r|x|)^{-\frac52} r \left(
|y|^{-\frac12} \int_{\frac{1}{|y|}}^{2a}
\frac{\r^{\frac 32} d\r }
{(r^2 +\r^2)(r+ \r)}\right) dr  \notag \\
& \leq \frac{C}{|x|^\frac52}
\int_{\frac{1}{|x|}}^{8a} r^{-\frac32} \left (
|y|^{-\frac12} \int_{\frac{1}{|y|}}^{2a} \r^{-\frac 32} d\r 
\right) dr 
\leq \frac{C}{|x|^2}. 
\end{align}
Thus,  $L_{out,out}(x,y)$ is GOP on $\{(x,y) \in D_4 \colon |x|>|y\}|$. 

(2)  Let  $|y|>|x|$ next. 
We substitute  \refeq(asymp) with $\lam=\r|y|$ for $J_0(\r|y|)$ in \refeq(oo) 
and denote the function produced by the $j$-th term of 
\refeq(asymp) by  $A_{j}(x,y)$  so that 
$L_{out,out}(x,y)= A_1(x,y) + A_{2}(x,y) + A_{3}(x,y)$ 
We shall prove $A_j (x,y) \absleq C |y|^{-2}$ for $j=1,2,3$. 

\noindent 
(2a)  In the right side of 
\[
A_1(x,y)= C \int_{\frac{1}{|x|}}^{8a}J_0(r|x|)\chi_{\leq 4a}(r) r 
\left( 
\int_{\frac{1}{|y|}}^{2a}   \frac{\cos(\r|y|-\frac{\pi}{4})    
\r^{\frac32}\chi_{\leq a}(\r)}
{|y|^\frac12 (r^2 +\r^2)(r+ \r)} d\r
\right) dr
\]
we estimate $|J_0(r|x|)|\leq C$ and substitute (F2)  for the inner $d\r$-integral.  
Then, $A_1(x,y)$ is bounded in modulus by the sum of 
\[
C\int_{\frac{1}{|x|}}^{8a}  \left(\frac{ rdr }
{(|y|^2 r^2 +1)(|y|r+1)} \right) dr  
\leq \frac{C}{|y|^2}, 
\] 
and   
\[ 
\frac{C}{|y|^{\frac52}}\left(1+   
\int_{\frac{1}{|y|}}^{2a}
\r^{-\frac12}
\left( 
\int_{\frac{1}{|x|}}^{8a}
\frac{r dr}{(r^2 +\r^2)(r+ \r)} \right) d\r \right) 
\leq \frac{C}{|y|^2}.
\] 
(2b)  For 
\[
A_2 (x,y)= C \int_{\frac{1}{|x|}}^{8a}J_0(r|x|)\chi_{\leq 4a}(r) r \left( 
\int_{\frac{1}{|y|}}^{2a}   \frac{\sin(\r|y|-\frac{\pi}{4})
\r^{\frac12}\chi_{\leq a}(\r)}{|y|^\frac32 (r^2 +\r^2)(r+ \r)} d\r\right) dr
\]
we use $|J_0(r|x|)|\leq C$ and substitute (F1)  for the inner $d\r$-integral.
Then, $A_2(x,y)$ is bounded in modulus by the sum of   
\bqn 
C \int_{\frac{1}{|x|}}^{8a}
\frac{r  dr}{(|y|^2r^2 +1)(|y|r+ 1)}   
\leq \frac{C}{|y|^2}
\eqn 
and 
\[
\frac{C}{|y|^\frac52}\left(1+ \int_{\frac{1}{|y|}}^{2a}\r^{-\frac12} 
\left(\int_{\frac{1}{|x|}}^{8a}\frac{r dr}{(r^2 +\r^2)(r+ \r)}\right) d\r \right) 
\leq \frac{C}{|y|^2}. 
\]
(2c) Finally for $A_3(x,y)$ which is obtained by substituting $O((\r|y|)^{-\frac52})$ for 
$J_0(\r|y|)$ we estimate $J_0(r|x|)\absleq C$ and integrate with respect to $r$ first. 
Then,   
\begin{align*}
A_3(x,y)& \absleq C \int_{\frac{\c}{|y|}}^{2a}   
(\r|y|)^{-\frac52} 
\r^2\chi_{\leq a}(\rho)
\left|\int_{\frac{\c}{|x|}}^{8a}\frac{J_0(r|x|)\chi_{\leq 4a}(r) r }
{(r^2 +\r^2)(r+ \r)} dr\right| d\r  \\
& \absleq \frac{C}{|y|^{\frac52}}
\int_{\frac{\c}{|y|}}^{2a}   \r^{-\frac12}
\left(\int_{\frac{\c}{|x|}}^{8a}\frac{ r dr }
{(r^2 +\r^2)(r+ \r)} \right) d\r  \leq \frac{C}{|y|^2}. 
\end{align*} 
Thus, $L_{out,out}(x,y)$ is a GOP also on $\{(x,y)\in D_4 \colon |y|>|x|\}$. 
 
Combining all these together, we have proved that \refeq(oo) is a GOP on $D_4$ 
and this completes the proof of 
the proposition and, hence, of \refth(regular-case). 
\edpf

\section{Appendix}

\subsection{Integral inequality}   
In \S 6, we employed the following lemma 
which is a simple modification of the well-known theorem 
on integral operators with homogeneous kernels.

\begin{lemma} \lblm(preparation)
Let $1<p< \infty$, $\sup_{x\in \R^2} |F(x, |x|y)|\leq f(|y|)$ and       
\bqn 
F_p \colon = \int_{\R^2} \frac{f(|y|)}{(1+|y|^2)|y|^{\frac{2}{p}}}dy<\infty .
\eqn 
Then, the integral operator 
\bqn 
Tu(x) = \int_{\R^2} \frac{F(x,y)u(y)}{|x|^2+ |y|^2}dy 
\eqn 
satisfies  $\|T u \|_p \leq F_p \|u\|_p$ for $u \in \Dg_\ast$. 
\edlm 
\bgpf  Let $Mu(\r)$ be the spherical average of $u$ on the sphere $|x|=\r$ (see\refeq(sphe-mean)). 
Then, changing the variables $y$ to $|x|y$ and integrating on the sphere first, we obtain 
\[ 
Tu(x)  \absleq 
\int_{\R^2}\frac{|f(|y|)u(|x|y)|}{1+ |y|^2}dy  
\leq 2\pi \int_0^\infty \frac{|f(r)| M{|u|}(|x|r)}{1+r^2}r dr .
\]  
H\"older's inequality implies  
\[
\int_{\R^2} M{|u|}(|x|r)^p dx = 2\pi \int_0^\infty M{|u|}(sr)^p s ds 
\leq \frac1{r^2}\int_{\R^2}|u(x)|^p dx .
\]
It follows by Minkowski's inequality that 
\[
\|Tu\|_ p \leq 2\pi \left(\int_0^\infty \frac{|f(r)|r^{1-\frac{2}{p}}dr}{1+r^2} \right)\|u\|_p 
= F_p \|u\|_p.
\]
This proves the lemma.
\edpf 

\subsection{Proof of \reflmb(H02y)} 

We use Peral's theorem (\cite{Peral}  p.139, where the 
end points $p=1$ and $p=\infty$ should be excluded): 

\bglm \lblm(peral) Let $\psi(\xi) \in C^\infty(\R^n)$ be such that 
$\psi(\xi)=0$  in a neighbourhood of $\xi= 0$ and 
$\psi(\xi)=1$ for $|\xi|>a$ for an $a>0$. Then, 
\[
\Phi u(x)=\frac{1}{(2\pi)^{n/2}}\int_{\R^n} e^{ix\xi+i|\xi|}\frac{\psi(\xi)}{|\xi|^b} 
\hat{f}(\xi)d\xi
\]
is bounded in $L^p(\R^n)$, $1<p<\infty$, if and only if
\[
\left|\frac1{p}-\frac12\right| < \frac{b}{n-1}\,.
\]
In particular, it is a GOP if $n=2$ and $b=1/2$. 
\edlm

Since $\chi_{\geq {a}}(\lam)= \chi_{\geq {a/2}}(\lam)
\chi_{\geq {a}}(\lam)$,  we have from \refeq(res-ker) and \refeq(Green) that  
\[
\Rg(|D|, y)\chi_{\geq {a}}(|D|)=
\frac{i\chi_{\geq {a/2}}(|D|)}{8 |D|^{2}}
(H_0^{(1)}(|D||y|)- H_0^{(1)}(i|D||y|))\chi_{\geq {a}}(|D|).
\]
It is evident that 
$i\chi_{\geq {a}}(|D|/2)|D|^{-2}$ is a GOP and it suffices to show that 
\begin{gather} 
\|H_0^{(1)}(|D||y|)\chi_{\geq {a}}(|D|)\|_{\Bb(L^p)} 
\leq C_{a,p}(1+ |\log |y||), 
\lbeq(a) \\
\|H_0^{(1)}(i|D||y|)\chi_{\geq {a}}(|D|)\|_{\Bb(L^p)} 
\leq C_{a,p}(1+ |\log |y||).
\lbeq(ia)
\end{gather}
We shall prove \refeq(a) only. 
The proof for \refeq(ia) is similar.  Let  
\[
\Hg_{\geq{a}}(\lam)= H_0^{(1)}(\lam)\chi_{\geq a}(\lam), \quad 
\Hg_{\leq{a}}(\lam)= H_0^{(1)}(\lam)\chi_{\leq a}(\lam).
\] 
By using $N(\lam)$ of the  proof of \reflm(Green) we have that  
\[
\Hg_{\geq{a}}(\lam)= \frac{e^{i\lam}}{i\pi{\lam}^\frac12}\chi_{\geq {a/2}}(\lam) 
\cdot \lam^{\frac12}N(\lam)\chi_{\geq {a}}(\lam).
\]
By Peral's theorem 
$e^{i|D|}\chi_{\geq {a/2}}(|D|)/({i\pi|D|^\frac12})$
is a GOP;  by virtue of  
\refeq(F-estimate) $|D|^{\frac12}N(|D|)\chi_{\geq {a}}(|D|)$ 
satisfies the H\"ormander condition and is also a GOP. It follows that 
$\Hg_{\geq{a}}(|D|)$ is a GOP and, by the scaling invariance  
of $L^p$-bounds of Fourier multipliers, that 
$\|\Hg_{\geq{a}}(|D|y)\|_{\Bb(L^p)}\leq C_p$ 
with a constant $C_p$ independent of $y\in \R$. 
Hence  
\bqn \lbeq(high-H)
\|\Hg_{\geq{a}}(|D|y)\chi_{\geq {a}}(|D|)\|_{\Bb(L^p)}\leq C_p.
\eqn 

We next show that
\bqn \lbeq(low-H)
\|\Hg_{\leq{a}}(|D|y)\chi_{\geq a}(|D|)\|_{\Bb(L^p)}\leq C_p(1+ |\log |y|),
\eqn 
which will finish the proof of \reflm(H02y).  
The series expansion of the Hankel function (see \refeqs(Green-b,g)), 
\bqn 
\frac{i}{4} H^{(1)}_{0}(z)= \sum_{n=0}^\infty 
\left( g(z) + \frac{c_n}{2\pi} \right) 
\frac{(-z^2/4)^n}{(n!)^2},  \lbeq(Hankel) \\
\eqn 
and \refeq(g) imply that for $j=0,1, \dots$ 
\[
|\pa_\lam^j \left(\Hg_{\leq {a}}(\lam) 
- g(\lam)\chi_{\leq {a}}(\lam)\right)|
\leq C_j \lam^{2-j}|\log \lam| \quad (\lam \to 0)\,.
\]
Hence, the Fourier transform of 
$\Hg_{\leq {a}}(|\xi|) -g(|\xi|)\chi_{\leq{a}}(|\xi|)$ 
is integrable on $\R^2$ and 
\[
\|\Hg_{\leq{a}}(|D||y|)  - 
g(|y||D|)\chi_{\leq{a}}(|D||y|)\|_{\Bb(L^p)}
\leq C_p, \quad 1\leq p\leq \infty
\] 
with a constant $C_p$ independent of $y\in \R^2$. Thus,   
\[ 
\|
\Hg_{\leq{a}}(|D||y|)\chi_{\geq a}(|D|) - 
g(|y||D|)\chi_{\leq{a}}(|D||y|)\chi_{\geq{a}}(|D|)
\|_{\Bb(L^p)}
\leq C_p \,.
\]
But \refeq(g) implies $g(|y||D|)= g(|y|)-(\log |D|)/(2\pi)$ 
and  
\[
|\pa_{\lam}^j \{(\log \lam)\chi_{\leq{a}}(\lam|y|)\chi_{\geq{a}}(\lam)\}| 
\leq (1+ |\log |y||), \quad j=0,1,2
\]
This implies 
\[
\|g(|y||D|)\chi_{\leq{a}}(|D||y|)\chi_{\geq{a}}(|D|)\|_{\Bb(L^p)}
\leq  C_p (1+ |\log |y||).
\]
The desired estimate \refeq(low-H) follows by adding last two estimates for multipliers.  
\qed

\subsection{Proof of \refpropb(resonances)} 

(1) Suppose that $X\colon = S_0 T_0 S_0 \vert_{S_0\Hs}$ is singular.  Then, $0$ must be 
an isolated eigenvalue. Indeed, since  $X$ is selfadjoint in $S_0 \Hs$, if $0\in \s(X)$ is not an isolated 
eigenvalue,  there must exists an series $\{f_n\} \subset S_0 \Hs$ such that $\|f_n\|=1$, 
$f_n \to 0$ weakly and $X f_n = S_0 (U + G_{2,l}^{(v)})f_n \to 0$ as $n \to \infty$. Then, since 
$G_{2,l}^{(v)}\in \Hg_2$, $G_{2,l}^{(v)}f_n \to 0$ and $S_0 U f_n \to 0$. But, we have 
$S_0^\perp  U\in \Hg_2$ and $S_0^\perp U f_n  \to 0$ simultaneously . 
Hence  $U f_n \to  0$ and $f_n \to 0$.  This is a contradiction.

Thus, if $X\colon = S_0 T_0 S_0 \vert_{S_0\Hs}$  is singular, then, 
there exists $f \in  S_0 \Hs\setminus\{0\}$ such that $ T_0 f \in S_0^\perp \Hs$, viz. 
\bqn \lbeq(8a) 
Uf+ v G_{2,l} v f = v(c_1 x_1+ c_2 x_2+c_0)
\eqn 
for a unique set of scalars $(c_1, c_2, c_3)\in \C^3$ 
and, multiplying both sides by $Uv$, we obtain 
$vf+ V (G_{2,l} v f - (c_1 x_1+ c_2 x_2+c_0))=0$.  Let  
\bqn \lbeq(forward)
\ph= G_{2,l} v f - (c_1 x_1+ c_2 x_2+c_0).
\eqn 
Then $\lap^2 \ph = vf = - V\ph$ or $(\lap^2 + V)\ph=0$;  
$\ph \not=0$ since we have $Uf= -v\ph$ by \refeqs(8a,forward) and $\ph =0$ 
would imply $Uf=0$ and $f=0$, which is a contradiction. 

We next show  that 
$\ax^{-1}\ph \in L^\infty(\R^2)$, which will imply  $\ph \in \Ng_\infty\setminus\{0\}$. 
To see this  it suffices to show that for $|\a|\leq 1$ 
\bqn \lbeq(log)
Y_\a(x)\colon = \int_{\R^2}(\log |x-y| )y^\a (vf)(y) dy \absleq C\ax^{-1} ,
\eqn 
which clearly implies  
\begin{align*}
& G_{2,l} (v f)(x) = C \int_{\R^2} |x-y|^2 \log |x-y| (vf)(y) dy \\ 
& = C x^2 \int_{\R^2} \log |x-y| (vf)(y) dy 
- 2 C x \cdot \int_{\R^2} \log |x-y| (y (vf)(y)) dy   \\
& \qquad +C \int_{\R^2} \log |x-y| (y^2 (vf)(y)) dy \in \ax L^\infty(\R^2)  
\end{align*}
and hence $\ax^{-1}\ph \in L^\infty(\R^2)$. We shall show \refeq(log) for $|\a|\leq 1$.  

Let $R>1$ and $|x|\leq R$ first. Then, 
H\"older's inequality implies for $q'=q/(q-1)$ that  
\begin{multline}
\int_{|x-y|<R}(\log |x-y| )y^\a (vf)(y) dy \\ 
\absleq \|\log |y|\|_{L^{2q'}(|y|\leq R)} 
\|y^\a v\|_{L^{2q}(|y|\leq 2R)} \|f\|_{2} \leq C.  \lbeq(HoR)
\end{multline}
If  $|x-y|\geq R$,  then $\log |x-y|\leq \log(1+|x|)+ \log (1+|y|)\leq C\ay$ for $|x|\leq R$ and 
\[
\int_{|x-y|\geq R}(\log |x-y| )y^\a (vf)(y) dy \absleq 
C \|\ay^{|{\a}|+1}(vf)\|_{1} \leq \|\la y\ra^{4}V\|_1^\frac12 \|f\|_2.
\]
Together with \refeq(HoR), this proves \refeq(log) for $|x|\leq R$.  

Let $|x| \geq R$ next.  Since  $\la x^\a v, f\ra=0$,
\[
Y_\a (x)= \int_{\R^2}(\log |x-y|-\log|x|)y^\a (vf)(y) dy .
\]
If $2|y|\leq |x|$, then $|x-\th{y}|\geq |x|/2$ and 
\[
\log |x-y|-\log|x|= \int_0^1 \frac{-2(x-\th{y})\cdot{y}}{|x-\th{y}|^2}d\th
\absleq  4\frac{|y|}{|x|}.
\]
It follows that 
\begin{align}
& \int_{2|y|\leq |x|}|(\log |x-y|-\log|x|) y^\a (vf)(y)| dy \leq \frac{2}{|x|}\|\ay^2 (vf)\|_1 
\notag \\
&\qquad \leq \frac{4}{|x|}\|\ay^{4} v \|_2\|f\|_2. \lbeq(A)
\end{align}
If $2|y|\geq |x|\geq R$, then $0<\log |x|\leq C|y|^\ep $ and  Schwarz' inequality implies  
\begin{multline}
\int_{2|y|\geq |x|}|(\log|x|) y^\a (vf)(y)| dy 
\leq \int_{2|y|\geq |x|}||y^\a |y|^\ep (vf)(y)| dy 
\\
\leq \frac{C}{\ax} \|\ay^{2+\ep} (vf)\|_1 \leq C 
\leq \frac{C}{\ax} \|\ay^{2+\ep}v\|_2 \|f\|_2<\infty .   \lbeq(B)
\end{multline}
By interpolating \refeqs(1-1,condi-1), we have 
$\sup_{x}\|\ay^4 V\|_{\frac{N}{N-1}(|x-y|\leq R)}<\infty$ 
for a sufficiently large $N$ and, via H\"older's inequality, that   
\begin{multline}
\int_{2|y|\geq |x|, |x-y|\leq R} 
|(\log |x-y|) y^\a (vf)(y)| dy \\ 
\leq \frac{C}{|x|}\|\log |y|\|_{L^{N}(|y|\leq R)}
\sup_{x}\|\ay^2 v\|_{L^{\frac{2N}{N-1}}(|x-y|\leq R)}\|f\|_2<\infty.    \lbeq(C)
\end{multline}
When $|x-y|\geq R$, then $0< \log |x-y| \leq \log \ax + \log \ay \leq C \ay^\ep$ and 
\begin{multline}
\int_{2|y|\geq |x|, |x-y|\geq R}|(\log |x-y|) y^\a (vf)(y)| dy 
\leq \frac{C}{\ax}\|\ay^{2+\ep}v\|_2 \|f\|_2.  \lbeq(D)
\end{multline}
Summing up \refeqss(B,C,D) and combining the result with \refeq(A), 
we have \refeq(log) for $|x|>R$, which completes the proof of \refeq(log).
Thus, $H$ is singular at zero if $X$ is singular.  

Next we suppose that $H$ is singular at zero, viz.  
$(\lap^2+ V(x))\ph(x)=0$  for an $\ph \in \ax L^\infty(\R^2)\setminus\{0\}$. 
Let 
$f= Uv \ph$ so that $vf = V \ph  = -\lap^2 \ph$. Then, $vf \in \ax^{-8-\ep}L^1(\R^2)$ 
and,  as was shown by \cite{LSY}, 
\[
\int_{\R^2} v(x) f(x) dx =0, \ \mbox{hence} \ f \in  Q \Hs.  
\]
 (The last line of (5.5) of \cite{LSY} must be replaced by 
$ \d \|\ay^{-1}\ph(y)\|_{L^\infty} \|\eta \|_{L^1}$ and we were not 
able to follow the argument ``Similarly, we also have 
$\la x_1v, \ph \ra =  \la x_2v, \ph\ra =0$'' after (5.5).)  Let 
$u= G_{2,l}v f $. Then $\lap^2 u = v f = V\ph= -\lap^2 \ph$, 
$\lap^2(G_{2,l}v f + \ph)=0$ and the Fourier transform and 
the distribution theory imply 
\[
G_{2,l}vf  + \ph = \sum_{|\a|\leq 3}c_\a x^\a. 
\] 
However,  $c_\a=0$ if $|\a|\geq 2$ because $\ph(x)\absleq C \ax$ and 
because 
$f \in  Q \Hs$ implies $G_{2,l}vf\absleq C\ax \log |x|$ as $|x|\to \infty$ 
by the proof of the if part above. Thus, 
\bqn \lbeq(backward)
G_{2,l}vf  + \ph = c_1 x_1+ c_2 x_2+c_0
\eqn 
and this implies $\la y_j v, f \ra =0$ for $j=1,2$, since, otherwise, 
\[
x_j \int_{\R^2} \log |x-y| (y_j (vf)(y)) dy = 
x_j \log |x| \cdot \int_{\R^2} (y_j (vf)(y)) dy + \Og(1)
\]
as $|x|\to \infty$ 
and  \refeq(backward) cannot happen.   Thus, 
\[
\mbox{$f \in S_0\Hs$ and 
$v\ph+ vG_{2,l}v f = T_0 f = (c_1x_1+ c_2 x_2+c_0)v\in S_0^\perp \Hs $},
\]
which implies $S_0 T_0 S_0 f = 0$ and $X$ is singular.

(2) The first part of the proof of statement (1) already proves the first statement of (2) 
and that the map $\Phi$  is well defined from $\Yg(H)$ to $\Ng_\infty(H)$. 
If $\Phi(f) = \Phi(g)$ for $f,g \in \Yg(H)$, then $0=\lap^2( \Phi(f) - \Phi(g)) = f-g$ 
and $\Phi$ is one-to-one. To show it is map onto $\Ng_\infty(H)$, we set $f = M_U M_v \ph$ 
for $\ph\in \Ng_\infty(H)$. Then, the second part of the proof of statement (1) 
shows that $f \in S_0 \Hs$ and the pair $(\ph, f)$ satisfies \refeq(backward) for 
a set scalars $(c_0, c_1, c_2)$. But it implies, since $M_v \phi = M_U f$, 
\[
M_U f + G_{2,l}^{(v)} f = (c_1x_1+ c_2 x_2+c_0)v
\]
and $(c_0,c_1,c_2)$ in the \refeq(backward) must be equal to $c_f$ defined by \refeq(8), 
which implies  $\ph = \Phi(f)$. Thus, $\Phi$ is a onto map. This completes the proof.

\subsection{A Fourier transform} 
We used the following lemma in \S 6. 
\bglm  \lblm(F) 
Let $A(z)$ be analytic in a neighborhood of the origin. 
Let $\m(\lam) = A(g_1(\lam)^{-1})\chi_{\leq a}(\lam)$. 
Then,  for $a>0$ small enough, $\Fg(\m(|\xi|))(x)$ is a bounded $C^\infty$-function 
of $x\in \R^2$ and, for any $\ep>0$,
\bqn \lbeq(F)
|\Fg(\m(|\xi|))(x)|\leq C_\ep \ax^{-2} \la \log |x| \ra^{-2+\ep}, \quad 
|x|\geq 1.
\eqn 
\edlm 
\bgpf  Since $\m(|\xi|)$ is compactly supported, rotationary invariant 
and integrable,  $\Fg(\m(|\xi|))(x)$ is rotationary invariant, smooth 
and  bounded. Thus, we have only to prove \refeq(F).   Let 
$a(\lam)= A(g_1(\lam)^{-1})$; $a(\lam)$ is well defined and continuous  
around $\lam=0$ and  
$\lap_{\xi}\big(a(|\xi|)\chi_{\leq {a}}(|\xi|)\big) - (\lap_{\xi}a(|\xi|)) \chi_{\leq {a}}(|\xi|)
\in C_0^\infty(\R^2)$. It follows by integration by parts that  
\begin{align}
& \Fg(\m(|\xi|))(x) = 
\frac{-1}{2\pi |x|^2}\int_{\R^2} e^{ix\xi}\lap_{\xi}\big(a(|\xi|)
\chi_{\leq {a}}(|\xi|)\big)d\xi   \notag \\ 
& \equiv  
\frac{-1}{2\pi |x|^2}\int_{\R^2} e^{ix\xi}\lap_{\xi}\big(a(|\xi|))
\chi_{\leq {a}}(|\xi|)d\xi  \lbeq(evi)
\end{align}
modulo a rapidly decreasing function and we shall show \refeq(F) for \refeq(evi). We shall often write  
\[
|\xi|=r \ \mbox{and}\  g_1(r)^{-1}=\r.
\]
By the chain rule we have 
\begin{align}
& \nabla_{\xi} g_1(|\xi|)^{-1}= -g_1(|\xi|)^{-2}\nabla_\xi g_1(|\xi|)=
\frac{1}{2\pi}g_1(|\xi|)^{-2}\frac{\xi}{|\xi|^2},   \lbeq(nabla-g-1) \\ 
& \lap_{\xi} g_1(|\xi|)^{-1} =  2g_1(|\xi|)^{-3}(\nabla g_1(|\xi|))^2 
 -g_1(|\xi|)^{-2}\lap_{\xi} g_1(|\xi|). \lbeq(g1-lap)
\end{align}
Here we have $g_1(|\xi|)^{-2}\lap_{\xi} g_1(|\xi|)=0$ since, as is well known, 
\[
\lap_{\xi} g_1(|\xi|)= -\frac1{2\pi} \lap \log |\xi| = \d_0(\xi), 
\]
$g_1(|\xi|)^{-2}$ is continuous at $\xi=0$ and 
$g_1(0)^{-2}=0$. Thus, by \refeq(g1-lap), 
\bqn \lbeq(lap-g-1)
 \lap_{\xi} g_1(|\xi|)^{-1} =  2g_1(|\xi|)^{-3}(\nabla g_1(|\xi|))^2 
= \frac{2\r^3}{(2\pi r)^2}.
\eqn 
Differentiating $\nabla_{\xi} a(|\xi|)= A'(\r)\nabla_{\xi} g_1(|\xi|)^{-1}$ and 
using \refeq(lap-g-1), we have 
\bqn  \lbeq(lap-g-3)
\lap_\xi a(|\xi|)= A'(\r)\frac{2\r^3}{(2\pi r)^2} + A''(\r)\frac{\r^4}{(2\pi r)^2}.
\eqn 
Write $\lap_\xi\big( a(|\xi|)\big) \chi_{\leq a}(|\xi|)= f(r)$ so that 
\begin{gather} 
f(r)= \big(
A'(\r)\frac{2\r^3}{(2\pi r)^2} + A''(\r)\frac{\r^4}{(2\pi r)^2}\big) \chi_{\leq a}(r); 
\lbeq(gath-1) \\
\int_{\R^2}e^{ix\xi}\lap_\xi\big(a(|\xi|)\big)\chi_{\leq a}(|\xi|)d\xi 
= \int_0^\infty  J_0(|x|r) f(r) r dr .  \lbeq(gath-2)
\end{gather} 
We estimate \refeq(gath-2) for large $|x|>1$. From \refeq(gath-1)  
\bqn \lbeq(lap-g) 
|f(r)| \leq C r^{-2}\la \log r \ra^{-3} \chi_{\leq a}(r)
\eqn  
and,  for any $0<\ep<1$, 
\begin{align}
& \int_{0}^{\frac1{|x|}} J_0(|x|r) f(r) r dr \absleq  C \int_0^1 \la \log (r/|x|) \ra^{-3}r^{-1}dr  
\notag \\
& \leq C_\ep \int_0^1 r^{-1}\la \log r \ra^{-1-\ep} \la \log |x| \ra^{-2+\ep}dr 
\leq C \la \log |x| \ra^{-2+\ep}.  \lbeq(e-1)
\end{align}  
For  $r \geq 1/|x|$, we use the asymptotic formula for $J_0(|x|r)$. Then,    
\begin{align} 
& \int_{\frac1{|x|}}^{2a} J_0(|x|r) f(r) r dr
=\int_{1}^{2a|x|} J_0(r) f(r/|x|) |x|^{-2} rdr  \notag \\
& = (2/ \pi)^{1/2}\int_1^{2a|x|} {(\cos(r-\frac{\pi}{4})+ O(r^{-1}))} {r^{-\frac12}} f(r/|x|) |x|^{-2} r dr.
\end{align} 
By \refeq(lap-g) the  integral produced by $O(r^{-1})$ is bounded in modulus  by 
a constant times   
\begin{align}
\int_1^{2a|x|}  {r^{-\frac52}}\la \log (|x|/r)\ra^{-3}  dr
& \leq C \int_1^{|x|^\ep}\la \log |x|\ra^{-3}r^{-\frac52}dr + 
C\int_{|x|^\ep}^{2a|x|} r^{-\frac52} dr \notag  \\
& \leq C(\la \log |x|\ra^{-3}+ |x|^{-\frac{3\ep}2}). \lbeq(e-2) 
\end{align}
Let $\t= r-\pi/4$. Then, by integration by parts 
\begin{multline}  \lbeq(e-3a) 
\int_1^{2a|x|} \cos(r-\frac{\pi}{4}) {r^{-\frac12}} f(r/|x|) |x|^{-2} r dr  
= \left. (\sin \t) {r^{\frac12}} f(r/|x|) |x|^{-2} \right\vert^{2a|x|}_1   \\
- \int_1^{2a|x|} (\sin\t)\big(r^{\frac12} f(r/|x|) |x|^{-2}\big)' dr.
\end{multline}
Since $f(2a)=0$, the first term on the right is equal  to   
\bqn \lbeq(e-3)
-\sin(1-\pi/4)f(1/|x|)|x|^{-2} \absleq C \la \log |x| \ra^{-3}. 
\eqn
To estimate last term on the right of \refeq(e-3a) we compute 
\[
\big(r^{\frac12} f(r/|x|) |x|^{-2}\big)' 
= \tfrac12 r^{-\frac12}f(r/|x|)|x|^{-2} + r^{\frac12}f'(r/|x|)|x|^{-3} .
\]
Then,  by \refeq(lap-g) once more, 
\begin{multline}  \lbeq(e-4)
\int_1^{2a|x|} (\sin\t) r^{-\frac12}f(r/|x|)|x|^{-2} dr \\
\absleq \left(\int_1^{|x|^\ep} + \int_{|x|^\ep}^ {2a|x|}\right) 
r^{-\frac52}\la \log r/|x|\ra^{-3}dr  
\leq 
C(\la \log |x|\ra^{-3} + |x|^{-3\ep/2}).
\end{multline}
Differentiating \refeq(gath-1) by $r$ by using $d\r/dr=\r^2/(2\pi r)$, it is easy to 
see that $f'(r) \absleq C r^{-3}\la \log r\ra^{-3}$.  Thus, 
\begin{multline} \lbeq(e-5)
\int_1^{2a|x|} (\sin\r)r^{\frac12} f'(r/|x|) |x|^{-3} dr \\
\absleq C \int_1^{2a|x|} r^{-\frac52} \la \log(r/|x|) \ra^{-3} dr 
\leq C (\log |x|)^{-3}
\end{multline}
Combining estimates \refeqsss(e-1,e-2,e-4,e-5), we obtain  
\[
\Fg(\m(|\xi|))(x) \absleq C_\ep \la x \ra^{-2}\la \log |x| \ra^{-2+\ep}
\]  
for any $\ep>0$ and the lemma is proved. 
\edpf

\end{document}